%% file: sample-acmsmall.tex
\documentclass[acmsmall]{acmart}
\usepackage{amsmath,amssymb,amsfonts}
\usepackage{algorithmic}
\usepackage{graphicx}
\usepackage{textcomp}
\usepackage{xcolor}
\usepackage{url}

\usepackage{hyperref}

\AtBeginDocument{%
  \providecommand\BibTeX{{%
    \normalfont B\kern-0.5em{\scshape i\kern-0.25em b}\kern-0.8em\TeX}}}

\input{macros}

\begin{document}
\title{Mapping the Trust Terrain: LLMs in Software Engineering - Insights and Perspectives}

\author{Dipin Khati}
\orcid{0009-0008-4489-7733}
\affiliation{%
 \institution{William \& Mary}
 \city{Williamsburg}
 \state{Virginia}
 \country{USA}
 \postcode{23188}
 }
 \email{dkhati@wm.edu}

\author{Yijin Liu}
\affiliation{%
 \institution{William \& Mary}
 \city{Williamsburg}
 \state{Virginia}
 \country{USA}
 \postcode{23188}
 }
 \email{yliu85@wm.edu}

\author{David N. Palacio}
\email{danaderpalacio@wm.edu}
\orcid{0001-6166-7595}
\affiliation{%
  \institution{William \& Mary}
  \city{Williamsburg}
  \state{Virginia}
  \country{USA}
  \postcode{23185}
}

 \author{Yixuan Zhang}
\affiliation{%
 \institution{William \& Mary}
 \city{Williamsburg}
 \state{Virginia}
 \country{USA}
 \postcode{23188}
 }
 \email{yzhang104@wm.edu}
 
\author{Denys Poshyvanyk}
\affiliation{%
  \institution{William \& Mary}
  \city{Williamsburg}
  \state{Virginia}
  \country{USA}
  \postcode{23185}}
\email{dposhyvanyk@wm.edu}

\begin{abstract}
The application of Large Language Models (LLMs) in Software Engineering (SE) continues to grow rapidly across both industry and academia. As these models become integral to critical SE processes, ensuring their reliability and trustworthiness becomes essential. Achieving this requires a balanced approach to trust: excessive trust can introduce security vulnerabilities, while insufficient trust may hinder innovation. However, the conceptual landscape of trust in LLMs for SE(LLM4SE) remains unclear. Key concepts such as trust, distrust, and trustworthiness lack precise definitions, factors that shape trust formation remain underexplored, and metrics for trust in LLMs remain undeveloped. To clarify the current research landscape and identify future directions, we conducted a comprehensive review of $88$ articles: a systematic review of $18$ studies on LLMs in SE, supplemented by an analysis of $70$ articles from the broader trust literature. Furthermore, we surveyed $25$ domain experts to gather practitioners' perspectives on trust and identify gaps between their experiences and the existing literature. Our findings provide a structured overview of trust-related concepts in LLM4SE, outlining key areas for future research. This study contributes to building more trustworthy LLM-assisted software engineering processes, ultimately supporting safer and more effective adoption of LLMs in SE.
\end{abstract}




\maketitle

\input{text/1_introduction}

\input{text/2_motivation_background}

\input{text/3_research_questions}

\input{text/4_methodology}


\input{text/6_RQ1}

\input{text/7_RQ2}

\input{text/8_RQ3}

\input{text/9_threats}

\input{text/12_conclusion}

\appendix
\clearpage
\bibliographystyle{ACM-Reference-Format}
\bibliography{sample-acmsmall}

\end{document}

%% file: macros.tex
\usepackage{caption}
\usepackage{blindtext}
\usepackage{tcolorbox}
\usepackage[final]{pdfpages}
\usepackage{lipsum,multicol}
\usepackage{xcolor}
\usepackage{tikz}
\usepackage{listings}
\usepackage{enumitem}
\usepackage{hyperref}
\usepackage{amsfonts}
\usepackage{wrapfig}
\usepackage{booktabs}
\usepackage{multirow}
\usepackage{array}
\usepackage{colortbl}

\usepackage[edges]{forest}
\usepackage{xcolor}
\usepackage{geometry}

\newcommand{\llms}{LLMs\xspace}

\newcommand{\ie}{\textit{i.e.,}\xspace}
\newcommand{\eg}{\textit{e.g.,}\xspace}

\newtcolorbox{boxK}{
    fontupper = \small,
    sharpish corners, 
    boxrule = 0pt,
    toprule = 1.0pt, 
    enhanced,
    fuzzy shadow = {0pt}{-2pt}{-0.5pt}{0.5pt}{black!35} 
}

\newcommand{\figref}[1]{Fig.~\ref{#1}\xspace}


%% file: text/1_introduction.tex
\section{Introduction}

The use of LLMs in SE is becoming increasingly prevalent, with applications spanning bug fixing ~\cite{bug_fixing_tufano}, defect prediction ~\cite{defect_prediction}, input generation ~\cite{input_generation}, and various other tasks ~\cite{deep_code_search, traceability_2017, feature_location, feature_envy_2018, watson2021systematic}. These models are now widely adopted by software developers, researchers, and students, demonstrating their potential to enhance productivity and automate complex SE tasks.

However, the successful integration of LLMs into SE workflows depends not only on their technical capabilities, but also on how practitioners perceive and \textit{trust} them~\cite{roychoudhury2025aisoftwareengineerprogramming}. For example, misaligned trust between the LLM and the practitioner, either excessive (\ie overtrust) or insufficient (\ie undertrust), can significantly impact the effectiveness and security of these LLMs. A recent empirical study on ChatGPT and Gemini in Java projects found that up to 7\% of automatic refactorings broke functionality or introduced syntax errors\cite{liu2024empiricalstudypotentialllms}, demonstrating risks of overtrust. Conversely, Peng et al.’s controlled study found that developers using GitHub Copilot completed a JavaScript HTTP server task 55.8\% faster\cite{peng2023impactaideveloperproductivity}, highlighting significant productivity gains achievable with appropriate trust. Hence, achieving the right alignment is crucial, as both extremes entail undesirable consequences: overtrust in LLMs can lead to security vulnerabilities, data integrity risks, and erroneous decision-making ~\cite{Thorne_2024}, while undertrust can hinder adoption, reducing the potential benefits of LLMs ~\cite{balayn2024empiricalexplorationtrustdynamics}.



The alignment of trust between LLMs and practitioners shapes the range of possible trust relationships in software engineering settings. These relationships involve not only the trustor, the party that decides whether to trust and the trustee, the entity that is trusted~\cite{trustor_trustee}, but also the specific activity in which trust occurs~\cite{balayn2024empiricalexplorationtrustdynamics}. For instance, a practitioner (i.e., a trustor) may choose to accept or reject a code snippet generated by an LLM such as GitHub Copilot (i.e., trustee). In addition, trust perceptions vary according to the trustor, the SE task, and the domain~\cite{balayn2024empiricalexplorationtrustdynamics}. A developer, for example, might confidently rely on an LLM to generate boilerplate code but hesitate to use it for test case generation due to concerns about domain-specific accuracy. Understanding these core trust concepts is essential to unlock the complexities of trust relationships in LLM-assisted software engineering.


However, core concepts such as \textit{trust, distrust, and trustworthiness} are often used interchangeably, obscuring their precise meanings ~\cite{sharma2024suggestthathumantrust}. As a result, fundamental questions about \textit{trust} in LLM4SE remain unresolved. What factors shape trust and do these factors vary across experience levels, such as between expert and novice developers, or across different SE tasks? Is distrust simply the absence of trust or does it represent a distinct state? Addressing these unresolved questions is crucial for successfully adopting LLMs in SE workflows, as demonstrated by research findings in the field of Human-Computer Interaction (HCI)~\cite{Choung_2022, Glikson2020HumanTI}. These HCI studies have shown that trust is a crucial factor that influences the willingness of stakeholders to adopt LLMs. Unfortunately, existing studies provide a limited foundational understanding of how trust is defined, the factors that influence it, and the metrics used to assess it.  


To establish this foundational understanding, we must first assess the current state of research on trust in LLM4SE. Although some existing work has explored aspects of trust in LLMs and even examined factors influencing trust in specific LLM tools within SE, to the best of our knowledge, no comprehensive review has yet been conducted. This absence of a systematic synthesis makes it difficult to determine what aspects of trust have been explored and where significant gaps remain. Previous reviews in broader Artificial Intelligence (AI) contexts provide valuable conceptual insights, but often lack the necessary focus on SE-specific tasks, developer interactions, and the unique challenges posed by LLMs in this domain~\cite{trustLLM, aljohani2025comprehensivesurveytrustworthinesslarge, liu2024trustworthyllmssurveyguideline}. Consequently, the absence of a structured overview of trust concepts hinders efforts to define key trust-related concepts, identify influencing factors, and develop reliable evaluation methodologies. Furthermore, the perspective of SE practitioners on the definitions, factors, and metrics of trust remains largely underexplored, resulting in a disconnect between academic research and real-world needs. Without a clear understanding of the existing landscape, advancing trust-aware LLM systems for SE remains a significant challenge.

To address these challenges, we have taken a comprehensive approach that combines a literature review with practitioner insights. First, we review $18$ SE-focused articles to examine how trust-related concepts are defined, what factors influence trust in LLM4SE, and what metrics are used for evaluation. Second, we conducted a survey study with $25$ SE practitioners to capture real-world perspectives on trust in LLM4SE, identify task-specific trust factors, and explore variations in trust perceptions at different levels of expertise. In addition, we carried out a complementary analysis of $70$ articles from disciplines such as Deep Learning (DL), HCI, and Automation to enrich and complement the trust conceptualization of other fields.

By integrating findings from the software engineering literature and practitioner experience, we offer a comprehensive perspective on the current state of trust research in LLM4SE. Beyond identifying gaps in definitions, influencing factors, and evaluation methodologies, our study states the limitations of existing approaches and highlights the urgent need for standardized trust assessment frameworks. Furthermore, we examine the disconnect between academic research and practitioner expectations, revealing key areas where trust calibration -- the alignment of user trust level with actual reliability or trustworthiness of a system~\cite{Turner2024} --  poses challenges in real-world SE workflows. To improve these aspects, we propose concrete steps to validate trust definitions, improve trust factors, and establish trust metrics. By systematically synthesizing these findings, this work lays the foundations for future advances in trust-based LLM integration, offering concrete directions to improve the reliability, interpretability, and user confidence in these systems.


We structure our investigation around three key Research Questions (RQs) to map current understanding of trust concepts in LLM4SE. Through a systematic review of the literature and a survey study with SE practitioners, we examine how trust, distrust, and trustworthiness are defined in the context of LLM4SE (\ref{rq:definition}), identify the factors that shape trust perceptions (\ref{rq:factors}), and analyze existing approaches to measuring trust in LLM (\ref{rq:metrics}). Our findings reveal that while LLMs are increasingly adopted in SE, trust remains a loosely defined and inconsistently measured construct. Existing research neglects domain-specific definitions of trust-related concepts, often borrowing from other disciplines without adaptation to SE. In addition, trust factors, such as precision, interpretability, robustness, and workflow integration, are not consistently considered in different studies, leading to a fragmented understanding of the factors driving trust in LLM4SE. Finally, trust evaluation remains an open challenge, with limited efforts to establish standardized multidimensional metrics that account for the complexity of trust in SE workflows.

{To our knowledge, this is the first study to systematically examine trust in LLM4SE through a literature review and a practitioner survey. Our contributions include:}

\begin{enumerate}
\item We demonstrate a current understanding of trust definitions, factors that influence trust in LLM4SE, and the metrics used to evaluate trust in LLM4SE, through an extensive review of the literature, including $18$ articles focused on SE, and analysis $70$ articles from the broader trust literature. 

\item We complement our review with a survey study of $25$ participants to map current research and gather the perspectives of practitioners on trust in \llms in SE. We highlight the gaps between the literature and practical scenarios and provide a roadmap for future research.

\item We propose precise definitions for trust, distrust, and trustworthiness in LLM4SE, integrating insights from both existing literature and practitioners' perspectives. We also identify key factors influencing trust in LLM4SE, grouping them into model-specific and user-centric factors. Furthermore, we highlight the limitations of single-item metrics and advocate for comprehensive evaluation frameworks that better capture the multifaceted nature of trust in LLM4SE.

\item  We provide an online appendix with all of our data and results to facilitate reproducibility and encourage contributions from the community to promote trust in \llms in SE~\cite{AnonymousRepoTrustSurvey}.

\end{enumerate}

%% file: text/2_motivation_background.tex
\section{Why are Trust, Distrust, and Trustworthiness relevant for researching LLMs in Software Engineering?} \label{sec:motivation}

Misplaced trust, whether excessive or insufficient, can lead to poor decisions and unintended consequences, such as security vulnerabilities, code smells, and data leaks~\cite{Thorne_2024,mohsin2024trustlargelanguagemodels, balayn2024empiricalexplorationtrustdynamics}. For example, researchers recently discovered that AI-powered code assistants, including GitHub Copilot and ChatGPT, often generate insecure code snippets that developers often do not recognize as flawed~\cite{sandoval2024security, pearce2022copilot}. The study also found that when LLM-generated suggestions contained subtle security vulnerabilities, developers were more likely to trust and use them without adequate scrutiny, leading to increased security risks \cite{pearce2022copilot}. This tendency to over-rely on LLM-generated code without thorough verification goes beyond individual mistakes: it reflects a growing shift in how developers interact with LLMs and make critical coding decisions. As LLMs become integral to software development, they do more than suggest code; \textit{they influence workflows, assumptions, and even risk perception}. The growing reliance on LLMs in SE requires a structured approach to understanding trust, distrust, and trustworthiness~\cite{trustLLM,lo2023trustworthysynergisticartificialintelligence}.

Structured understanding of trust in LLM4SE can draw inspiration from research in other fields, such as the ethics of HCI and AI. Studies in these fields highlight key factors that influence trust, including accuracy, interpretability, and transparency \cite{Glikson2020HumanTI, sharma2024suggestthathumantrust}. However, trust in SE is particularly complex due to the technical nature of the domain, the potential for hidden errors in generated code, and the need to align with best practices \cite{cui2024fftharmlessnessevaluationanalysis, balayn2024empiricalexplorationtrustdynamics}. Unlike traditional software tools, LLMs produce nondeterministic outputs, which means that the same prompt can produce different responses, making trust assessment even more challenging \cite{zhu2024promptrobustevaluatingrobustnesslarge, ye2023assessinghiddenrisksllms}. Recognizing these challenges, researchers have called for a deeper investigation of trust in SE, \eg D. Lo~\cite{lo2023trustworthysynergisticartificialintelligence} emphasizes that as software development moves toward Software 2.0, where LLMs play a central role in coding, ensuring trustworthiness in AI-assisted development is imperative.

Although these calls for research highlight the urgency of understanding trust in LLM4SE, existing studies have not yet fully addressed this issue. Despite the significance of trust in AI-powered SE tools, existing research on LLM4SE has largely overlooked definitions, factors, and metrics of trust. Our review identifies three key challenges related to trust in software engineering research. First challenge ($CH_1$) is, only a small fraction of SE studies explicitly define trust, and the distinctions between trust, distrust, and trustworthiness remain underexplored, making it difficult to assess their role in real-world SE workflows. Second challenge ($CH_2$), while the broader trust literature provides valuable insight into the factors that influence trust, these perspectives have not been adequately integrated into SE research, leaving a gap in understanding what affects trust in developer interactions with LLMs. Third challenge ($CH_3$), there is a noticeable absence of standardized metrics to measure trust, and the perspectives of software practitioners on how trust should be quantified and evaluated are not well represented.

To address these challenges, we conduct a structured literature review, analyze trust research beyond SE, and complement our findings with a practitioner survey. In doing so, our goal is to provide a comprehensive understanding of trust in LLM4SE and lay the groundwork for future research and the development of reliable and trustworthy LLM-powered coding assistants aligned with developers.

%% file: text/3_research_questions.tex
\section{Research Questions} \label{sec:rqs}

This study investigates the trust in LLM4SE through three RQs. Each question is addressed using a systematic review of the trust literature in LLM4SE and a survey with practitioners. To provide additional context, each research question is complemented with an analysis of the underlying trust concept from the broader trust literature. 

\begin{figure}[htbp]
    \centering
    \includegraphics[width=\linewidth,height=0.8\textheight,keepaspectratio]{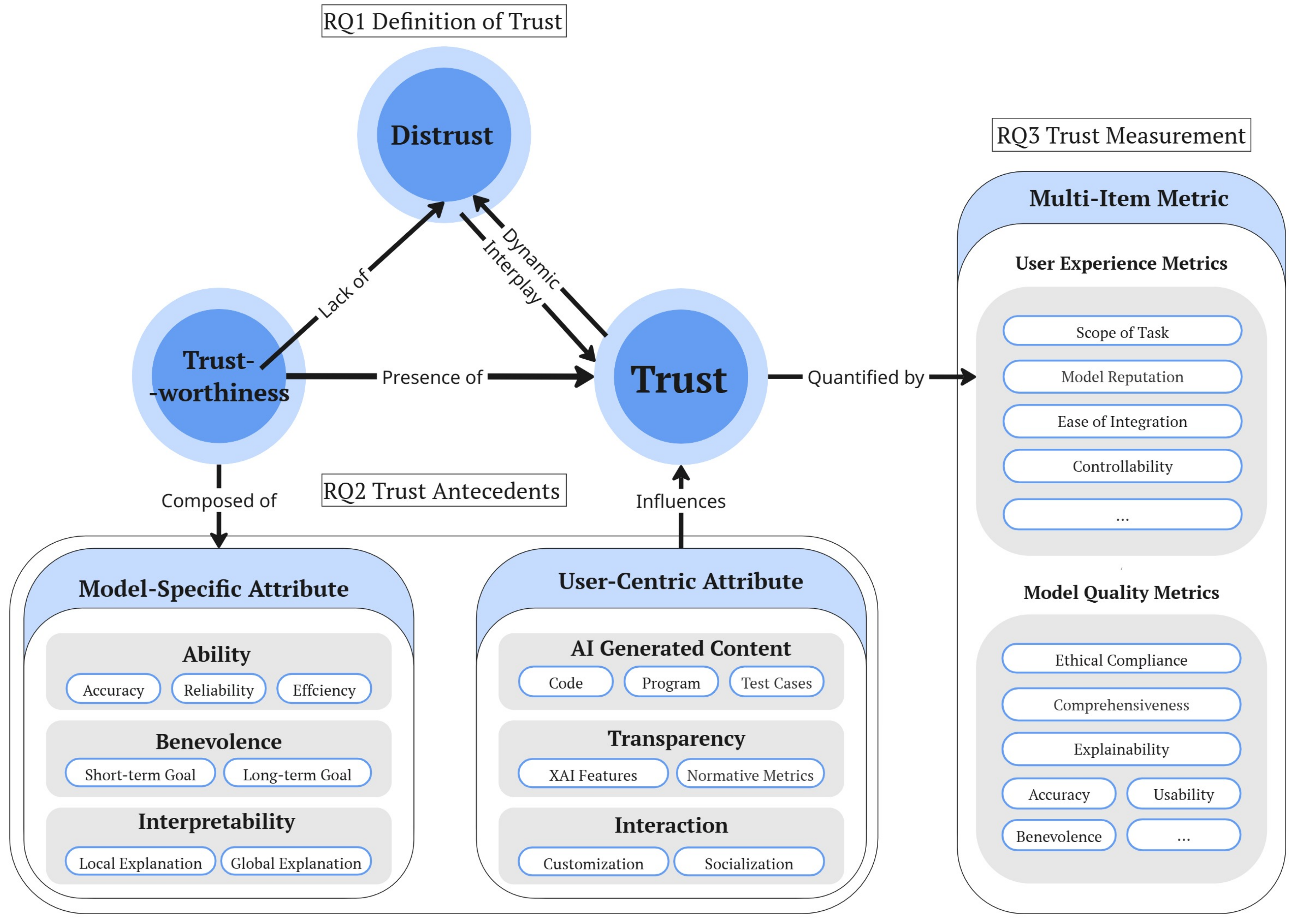}
    \caption{An Integrated Conceptual Model for Trust in LLMs for Software Engineering}
    \label{fig:overall_framework}
\end{figure}

\begin{enumerate} [label=\textbf{RQ$_{\arabic*}$}, ref=\textbf{RQ$_{\arabic*}$}, wide, labelindent=5pt]\setlength{\itemsep}{0.2em}
    \item \label{rq:definition}
     Definition of Trust: \textit {How is trust in LLM defined and conceptualized in the context of SE? }  This question aims to explore how \textbf{trust}, \textbf{distrust}, and \textbf{trustworthiness} are defined by analyzing the existing literature on LLM4SE and the perspectives of practitioners and addressing $CH_1$. These findings are contextualized through definitions found in the wider trust literature to highlight key similarities, differences, and gaps. By analyzing all these perspectives, we provide refined, evidence-based definitions that align with academic and practitioner expectations.

    \item \label{rq:factors}
     Antecedents of Trust: \textit{What are the antecedents of trust (\ie factors that can lead to trust) in LLM in the context of SE?} This question investigates the key factors that lead to the formation of trust in LLM4SE, drawing from both the literature of LLM4SE and practitioner experiences and addressing $CH_2$. An additional context is provided by analyzing trust factors in established trust models and related domains. The analysis provides actionable recommendations for fostering trust in LLM4SE.

    \item \label{rq:metrics}
     Trust Metrics: \textit{How is trust in \llms measured in the context of SE?} This question evaluates the existing metrics used to assess trust in LLM4SE to address $CH_3$. We investigate existing trust assessment methods in LLM4SE, examine how professionals evaluate trust in real world settings, and compare these with broader trust evaluation frameworks. By integrating these perspectives, we propose a road map for developing a comprehensive and domain-specific trust metric for LLM4SE.

\end{enumerate}

\figref{fig:overall_framework} presents an overall conceptual framework of our work, illustrating the interrelationships between trust definitions, antecedents, and metrics in LLM-assisted software engineering.

%% file: text/4_methodology.tex
\section{Methodology} \label{sec:methodology}

Our approach consists of two phases: a systematic review of the literature and a survey study. To provide more context and information, we also conducted a complementary analysis of the broader trust literature.

\subsection{Phase\texorpdfstring{$_1$}{1}: Systematic Literature Review}
\label{subsec:slr}
\begin{figure}[ht]
  \centering
  \includegraphics[width=.9\linewidth]{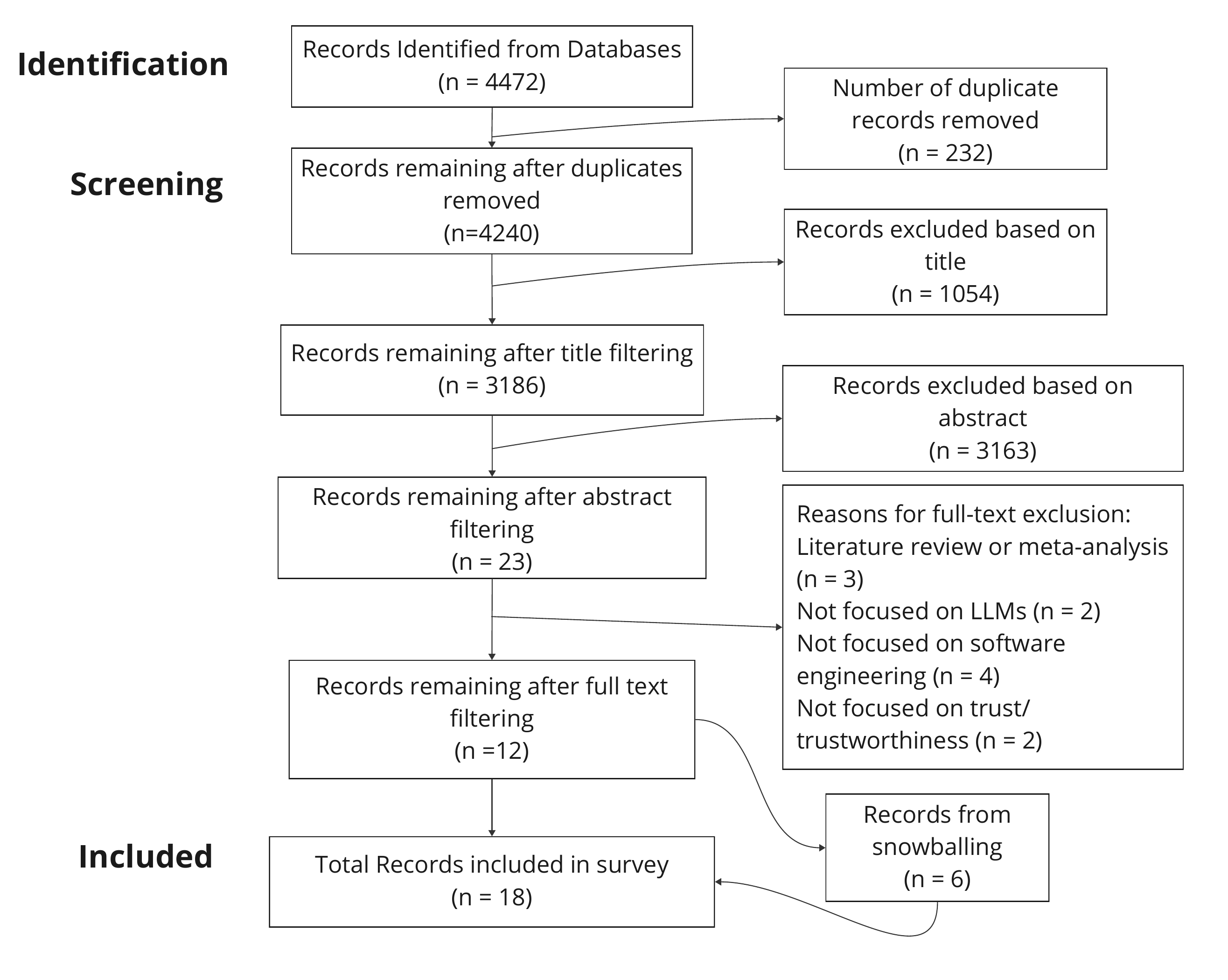}

  \Description[PRISMA flow diagram]{A PRISMA ~\cite{Prisma} flow diagram describing the stages of our review with the number of records at each stage}
   \caption[PRISMA flow diagram]{A PRISMA ~\cite{Prisma} flow diagram that describes our review process, labeled with the number of records at each stage}
   \label{fig:methodology}  
\end{figure}

We conducted a Systematic Literature Review (SLR) following the approach of Kitchenman et al. ~\cite{barabaraCit}. Before starting our review process, we developed RQs according to Kitchenman's guidelines. This helped us to systematically find papers that are related to our research goals. We then performed the following steps: 1) Search for Primary Studies 2) Screening 2a) Filter by title 2b) Filter by abstract 2c) Full-text Screening 3) Snowballing 4) Manual addition 5) Data Extraction and Analysis. Fig.~\ref{fig:methodology} shows a visual representation of our complete pipeline. Each step in this process was conducted independently by the two authors, who met after each step to ensure alignment and resolve any differences. The third author was consulted as needed to help settle any disputes.  

\subsubsection{Search for Primary Studies}
\label{subsec_search}

We focus on the period from January 1, 2002, to November 1, 2024. We chose 2002 as our starting period because this was when the method for modeling language using feedforward neural networks was published by Bengio et al. ~\cite{NIPS2000_728f206c}.

\textbf{Databases:} We selected the following databases, including ACM Digital Library, IEEE Xplore, Springer Link, Google Scholar, and DBLP. We then formulate a search string to query these databases. These databases were chosen because they were the main venues in SE, Machine Learning (ML), HCI, and DL that encompass our research goals.

\textbf{Search strings:} We used various combinations of search strings to get the best results.  
The search string \textit{(“trust” OR “distrust” OR “trustworthiness”) AND (“SE” OR “Software Engineering” OR “Software Development” OR “Code Generation” OR “Program Repair” OR "Automatic Program Repair" OR “Software Testing” OR “Test Generation”) AND (“LLM” OR “Large Language Model” OR “LLMs” OR “Large Language Models” OR “Deep Learning” OR “Machine Learning”)} provided the most comprehensive results. Alongside "SE" and "Software Engineering," we included terms related to the three SE-related activities of interest (i.e., Code Generation, Program Repair, and Software Testing) to ensure the inclusion of papers from SE and its subfields.

\textbf{Search strategies:} We used different mechanisms to query each database. For IEEEXplore and Springer Link, we used their advanced search feature to filter by date and venue and used their website to export the first 1,000 most relevant results as an Excel sheet. However, DBLP, Google Scholar, and ACM Digital Library do not have an export feature; thus, we wrote a simple Python program to query using a REST API and appropriate filters. After searching the databases using the specified search strings, we retrieved 2,000 papers from ACM, 1,000 from IEEE, 460 from Google Scholar, 1,000 from Springer, and 12 from DBLP, resulting in a total of 4,472 papers. The comparatively low number of results from DBLP is expected due to its search functionality, which performs a literal match on all search terms and does not support the phrase or proximity searches available in the other databases we queried. After eliminating 232 duplicates, we had 4,240 potentially relevant papers from this step. The script and the breakdown of the number of articles retrieved from each database are available in our repository ~\cite{AnonymousRepoTrustSurvey}. 

\subsubsection{Filtering by inclusion criteria}
We had two iterative inclusion criteria to filter out articles from the primary search.
\begin{itemize}
    \item Filter by title:
We selected all search results by reviewing the title first. In this step, any papers with titles that contained keywords {\textit{"trust" OR "distrust" OR "trustworthiness" OR "SE" OR "Software Engineering" OR "Software Development" OR "Code Generation" OR "Program Repair" OR "Automatic Program Repair" OR "Software Testing" OR "Test Generation" OR "LLM" OR "Large Language Model" OR "LLMs" OR "Large Language Models" OR "Deep Learning" OR "Machine Learning"}} were considered for further screening. We kept all these keywords to ensure that we did not lose any relevant papers. We were left with $3,186$ papers for further screening from this step.

\item Filter by abstract:
{Filtering by abstract was a manual effort of the two authors of this paper. The abstract was systematically analyzed to confirm its relevance to our research goals. In this step, papers that contained some indication of using trust or concepts related to trust were included. Specifically, we searched for specific indicators in the abstracts, such as explicit mentions of trust, trustworthiness, or related concepts such as reliability, confidence, or user acceptance in the context of LLMs applied to SE tasks. Papers that merely mentioned LLMs or SE without a clear focus on trust were excluded. Furthermore, abstracts that discussed trust in general AI or machine learning contexts without specific application to LLM4SE were also filtered out. From this step, we filtered out 3,163 papers, leaving us with 23 papers. The high number of excluded articles reflects the novelty of trust research, specifically in the LLM4SE domain. Many articles may have touched on related topics, but did not have the specific intersection of trust, LLM, and software engineering that our study aimed to explore.}
\end{itemize}

\subsubsection{Full-text Screening}
We conducted a thorough full-text review of $23$ articles that passed the abstract filtering process. This step involved a detailed examination of each paper's content to ensure its relevance to our research questions. We specifically looked for papers that addressed the intersection of trust-related concepts, SE, and LLMs. Articles were included if they:
\begin{itemize}
    \item Explicitly discussed trust, distrust, or trustworthiness in the context of LLMs for SE tasks.
    \item Provided empirical evidence or theoretical frameworks related to trust in LLMs for SE.
    \item Addressed challenges or proposed solutions to build trust in LLM-assisted SE processes.
\end{itemize}

Papers that only tangentially mentioned trust or LLMs without a substantial focus on SE applications were excluded at this stage.

\subsubsection{Snowballing and Manual Addition of Studies}
To ensure a comprehensive search, we performed backward snowballing by systematically screening the reference lists of all studies that met our inclusion criteria. 
The articles in the references went through the same inclusion criteria as the articles in our primary study. We first filtered them by title and then by abstract. We added $6$ papers from this step.

\subsubsection{Exclusion criteria}

In this step, we applied exclusion criteria to filter out studies that were irrelevant to our research questions. This involved a manual review of all papers to ensure their applicability. Although trust has been widely studied in DL and HCI, we excluded papers that did not explicitly operationalize trust in SE. Our focus is on understanding trust in the context of SE, particularly regarding the integration and use of \llms. Although trust research in DL and HCI offers valuable insights, it does not directly address SE-specific challenges such as workflow integration, developer-user interactions, and task-specific trust calibration. Therefore, we exclude these papers from our main analysis. However, to enrich our analysis, we incorporate relevant findings from broader trust research that, while not explicitly focused on SE, provide conceptual insights applicable to our study (see \ref{subsec:dl-hci} for the methodology for selecting these articles). The finalized corpus contains 18 papers.

\subsubsection{Data Extraction and Analysis}

We iteratively reviewed the articles collected to collect general and specific information relevant to our RQ, following the protocols suggested by ~\cite{Levett_2022} to extract data. We create a dataset with the following data: authors, year of publication, proposed RQs, methodology, study population and characteristics (for user studies), definition of trust, consequences of trust, importance of trust, antecedents of trust, and metric used for trust. 

Both authors independently worked to extract information and fill out the database. The authors discussed any discrepancies with the third author. In the end, we obtained a high Inter-Rater Reliability (IRR), with a Cohen's kappa(k) coefficient of 0.824. 

Based on the extracted information, we derive our taxonomy to answer our RQs. We draw on the framework and vocabulary of the MATCH model~\cite{Liao_2022} to derive our taxonomy to answer \ref{rq:factors} and \ref{rq:metrics}. The MATCH model presents factors that influence a user's trust throughout the AI interaction process, deriving these insights from a comprehensive review of the literature, making it appropriate for our study. The MATCH model breaks down the trust-formation process into three main components:

\begin{enumerate}
\item \textit{Model Trustworthiness Attribute}: These attributes are derived from the ABI framework~\cite{Dietz2006MeasuringTI} which includes three critical qualities. \textbf{Ability} refers to what the model can do, showcasing its competence and performance in SE-related tasks. \textbf{Intention Benevolence} addresses why the model was made, reflecting the intentions behind its development. \textbf{Process Integrity} involves how the model was trained, ensuring adherence to ethical standards, and that the training data are free from biases. 
\item \textit{Trust Affordance}: Trust affordance refers to the exhibited characteristics of a system that indicate how it can be used, guiding users in evaluating the model's trustworthiness. Trust affordance can manifest itself in three ways. \textbf{AI-generated content} is prediction, answers, and other outputs that can directly indicate the AI's ability. \textbf{Transparency} regarding the model’s inner functioning can enhance users' understanding of the benevolence of the model and the integrity of the process. \textbf{Interaction} covers the ways users interact with AI, which can influence trust through the usability and design of the system.
\item \textit{Trust Processing:} Based on users' ability and motivation, they will partake in two different trust judgement making procedures.
\textbf{Systematic Processing} involves a rigorous assessment of information to form rational trust judgments.
\textbf{Heuristic Processing}: uses cognitive shortcuts to make faster trust judgments, sometimes leading to incorrect judgment.
\end{enumerate}

\subsection{Phase\texorpdfstring{$_2$}{2}: Survey study}
\label{subsec: survey_study}
With the approval of our Institutional Review Board (IRB), we conducted a survey study to investigate the perceptions of professionals of concepts related to trust when using LLM4SE tasks. We chose a survey study over interviews or mixed methods to gather a broader and more diverse range of practitioner insights on trust in LLMs for SE tasks. User studies, particularly surveys, enable efficient data collection from larger samples while maintaining consistency for comparisons between participant groups. Unlike interviews, which are time-intensive and prone to interviewer bias, surveys provide anonymity, reducing social desirability bias, and ensuring honest responses. Furthermore, the inclusion of open-ended questions in the survey allowed us to capture qualitative insights alongside quantitative data, offering a balance between depth and scalability suitable for our exploratory research goals. This approach aligns with established practices in trust research, improving the relevance and generalizability of our findings.
In this study, we specifically explore the definitions of trust, its antecedents, and metrics, complementing our systematic review of the literature with insights from experienced and novice practitioners. We employed purposive sampling ~\cite{baltes2021sampling} to recruit practitioners with varying levels of SE and ML expertise, including industrial experts, researchers, students, and software engineers. We reached out to participants using our academic and industrial networks. We also encouraged our network to forward the invitation to relevant colleagues or peers in their networks, thus broadening our reach. Participation was voluntary, no incentives were offered, and participants were informed of the voluntary nature of our study during the request. 

\subsubsection{Survey Structure}
The survey consists of four different sections:  

Section 1 examines the practitioner's familiarity with using \llms for various SE tasks. It qualifies practitioners for the rest of the survey, ensuring valid feedback by excluding those who have never used \llms for subsequent downstream tasks. 

Section 2 combines blocks of questions investigating practitioners' insights into trust-related concepts in Test Case Generation, Code Generation, and Program Repair. The blocks are presented in random order to avoid bias. Each block covers trust definitions, antecedents, and metrics. Only practitioners with experience in using \llms for these tasks answer these questions.

Section 3 captures the general perception of the importance of trust, the desired metric of trust, and whether practitioners value the importance of these topics.

Section 4 collects demographic information about participants, such as sex, job title, and years of experience. The data is used to categorize trust-related concepts by expertise during data analysis. 

\subsubsection{Qualitative Metric} \label{metric} 
We derive survey metrics based on our SLR to complement it with participant's perceptions. We capture trust definitions, antecedents, and metrics. 

\begin{itemize}
    \item Definition:
We ask practitioners with experience using \llms for specific tasks to define trust in code generation, test case generation, and program repair.

\item Antecedents of Trust:
We gather practitioners' views on useful trust antecedents, as explained in ~\ref{antecedents}, for particular downstream SE tasks, which allows us to investigate whether trust antecedents are consistent across downstream tasks. 

\item Trust Metrics:
We collect information on how practitioners measure trust in specific downstream tasks through open-ended questions. This helps us to compare the trust measurement methods of practitioners with those in the literature. We also ask about their code review process for both human-written code and \llms generated code to identify similarities and differences.

\end{itemize}

\subsubsection{Trust Antecedents Selection} \label{antecedents}

The survey asks participants to select trust antecedents for each downstream task. We identified attributes related to the model and the community in the SE, HCI, and ML literature ~\cite{trustworthyAI, EUwebsite, korber_2018, 10.1145/3411764.3445650, 10.1109/ICSE-SEIP58684.2023.00043, ji2024aialignmentcomprehensivesurvey}. An author compiled all documented antecedents and then two authors independently categorized them, resolving disagreements through discussion. We selected the nine most popular antecedents for our survey, allowing participants to add any additional ones that were not originally listed.

\subsubsection{Data Collection}
We reached out to $96$ potential participants with varying expertise and from different backgrounds, who were not involved or aware of the purpose of this work. Of these groups, $48$ participants answered the survey. However, we had to discard the $23$ participants due to the incomplete or poor quality of their responses leaving $25$ valid responses. The study was performed with Qualtrics ~\cite{noauthor_qualtrics_nodate}.

\subsubsection{Survey Participants}
{
The survey included 25 participants representing a diverse range of professional roles in software engineering and research. The majority were identified as students (40\%), followed by researchers (20\%), engineers (16\%), and professors (12\%), with the remaining roles distributed among freelancers, managers, and other titles. Participants demonstrated various levels of experience in software engineering: 44\% reported more than 5 years of experience, 24\% had 3 to 5 years and 28\% had 1 to 2 years, with only one respondent (4\%) having less than a year of experience. In terms of education, the majority had graduate or professional degrees (60\%), while 36\% had bachelor's degrees and 4\% had completed high school or equivalent.}
When asked about their experience using LLM for software engineering tasks, 36\% of the respondents reported 1 to 2 years of experience, another 36\% had less than 1 year, and 16\% had 3 to 5 years of experience. This distribution reflects a relatively early stage of adoption of LLMs within software engineering, aligning with the novelty of the technology in this domain. Please refer to our online Appendix ~\cite{AnonymousRepoTrustSurvey} for more details.

\subsubsection{Survey Validity}
To solidify our survey, we conducted two pilot studies: one with $5$ participants and the other with $2$ participants who were not invited to the main survey. Based on the results of these pilot studies, we eliminated any leading questions and minor errors and improved our selection of antecedents of trust. Our survey study is also based on the findings of our literature survey. In addition, two of the authors individually performed the data analysis to avoid bias and human error.

\subsection{Complementary Analysis of Broader Trust literature}
\label{subsec:dl-hci}

\input{table/new_key_paper_table}

{Our initial review of trust in LLMs in SE yielded only $18$ articles, revealing significant gaps in existing literature in LLM4SE. Thus, to supplement our systematic review and provide additional context for trust in LLM4SE, we performed a complementary analysis of broader trust literature from HCI, DL and Automation. These fields provide well-established definitions, influencing factors, and evaluation metrics that can inform our understanding of trust in LLM4SE. Thus, we extracted trust-related concepts, such as definitions, influencing factors, and evaluation metrics, that could inform our findings within the SE domain. Table \ref{tab:rep-papers} highlights a selection of representative papers that exemplify these concepts and their contributions.

\subsubsection{Selection Process}
The selection process for this complementary analysis began with revisiting the set of studies that had been excluded from our SE-focussed review due to their lack of direct alignment with LLMs, SE, and trust. Rather than discarding these studies entirely, we reassessed them to identify those that provided valuable insights into trust-related concepts, even if they were not specific to SE.

To refine our selection, we applied predefined inclusion criteria. Papers were considered relevant if they offered conceptual definitions of trust, examined the mechanisms of trust formation, proposed methodologies for measuring trust, or explored factors that influence trust formation. Following this process, we selected $70$ papers that provided meaningful contributions to understanding trust-related concepts.

\subsubsection{Data Extraction and Analysis}

{To analyze the selected papers, we extracted key information, including definitions regarding trust-related concepts, factors contributing to trust formation, and methodologies for trust evaluation.}

{Two authors independently coded the extracted information with discrepancies resolved through discussion with the third author. A comparative analysis was conducted between the findings of this broader review and our systematic review focused on SE, highlighting similarities, key differences, and potential lessons that could be applied to trust in LLM-based SE systems.}

{We present our results in a structured format to ensure clarity. For each research question, we first summarize the insights from the SLR, followed by key takeaways from the survey study. Next, we analyze the broader trust literature to contextualize trust concepts even further. We then synthesize insights from all three sources to identify gaps, similarities, and key takeaways. Finally, we outline opportunities for future research, highlighting areas that require further exploration.}

%% file: table/new_key_paper_table.tex
\begin{table}[]
\caption{Representative Papers from Complementary Trust Literature Analysis}
\label{tab:rep-papers}
\resizebox{\columnwidth}{!}{%
\begin{tabular}{p{3cm} p{6cm} p{6cm}} 
\hline
\textbf{Paper} &
  \textbf{Key Contribution} &
  \textbf{Relevance to LLM4SE} \\
\hline
Jakesch et al. 2019 \cite{10.1145/3290605.3300469} &
  Perception of AI authorship significantly impacts perceived trustworthiness, often leading to distrust in mixed human-AI environments. &
  Highlights need to manage perceptions of LLM involvement to prevent unwarranted distrust of correct LLM-generated code. \\
  
Kaur et al. 2022 \cite{10.1145/3491209} &
  Comprehensive review of trustworthy AI attributes (robustness, fairness, security) as intrinsic system properties. &
  Provides foundational understanding for conceptualizing trustworthiness objectively in LLM-assisted SE. \\
  
Lewis et al. 2018 \cite{Lewis2018} &
  Synthesizes factors affecting trust and measurement methods, highlighting dynamic trust and calibration issues. &
  Provides foundational understanding of trust dynamics and measurement challenges for managing trust in LLM-assisted SE. \\
  
Li et al. 2021 \cite{trustworthyAI} &
  Framework for building trustworthy AI across the lifecycle, covering robustness, explainability, accountability. &
  Provides systematic approach for designing, deploying, and governing reliable LLM-based SE tools. \\
  
Kim et al. 2024 \cite{Kim_2024} &
  Shows first-person uncertainty expressions reduce overreliance on incorrect outputs and calibrate trust. &
  Highlights importance of explicit communication of LLM limitations to calibrate user trust in SE. \\
  
Langer et al. 2021 \cite{LANGER2021103473} &
  Conceptual model linking explainability to satisfaction of diverse stakeholder needs via human understanding. &
  Emphasizes considering stakeholder-specific desiderata beyond technical performance when building trust in LLMs. \\
  
Huang et al. 2024 \cite{trustLLM} &
  Principles for eight LLM trustworthiness dimensions and empirical benchmark to evaluate mainstream LLMs. &
  Provides systematic framework for conceptualizing and measuring LLM trustworthiness relevant to SE contexts. \\
  
Lee \& See 2004 \cite{Lee2004TrustIA} &
  Introduces ``appropriate reliance'' and trust calibration aligning user trust with actual system performance. &
  Foundational model for understanding and managing trust levels in human-LLM interactions. \\
  
Vereschak et al. 2021 \cite{10.1145/3476068} &
  Survey of empirical measurement methods highlighting the need for explicit trust definitions in experiments. &
  Guides empirical studies on trust in LLM-assisted SE, emphasizing clear definitions for valid results. \\
  
Ye et al. 2023 \cite{ye2023assessinghiddenrisksllms} &
  TRUSTLLM framework assessing eight trustworthiness dimensions with empirical evaluation. &
  Provides actionable methods for assessing LLM trustworthiness, improving reliability, safety, and ethics in SE tools. \\
\hline
\end{tabular}
}
\end{table}

%% file: text/6_RQ1.tex
\section{\texorpdfstring{$RQ_1$:}: Definition of Trust} \label{subsec:rq1}
In this section, we present our results for \ref{rq:definition} and discuss its implications.

\input{table/table2}

\subsection{Phase\texorpdfstring{$_1$}{1}: SLR Results}

In our corpus, only 27.8\% ($n=5$ out of 18) of the reviewed articles explicitly defined trust, with all definitions borrowed from established concepts rather than introducing new ones~\cite{cheng2023itworktooonline, 10.1145/3630106.3658984, 10.1145/3411764.3445650,borg2024trustcalibrationidespaving,10.1145/3397481.3450656}. The two primary definitions of trust utilized in the corpus are the following:

\begin{itemize}
\item "An attitude that an agent will achieve an individual’s goal in a situation characterized by uncertainty and vulnerability"~\cite{Lee2004TrustIA}. This definition was cited by~\cite{cheng2023itworktooonline, 10.1145/3630106.3658984, 10.1145/3411764.3445650,borg2024trustcalibrationidespaving} ($n=4$ of all articles that defined trust). 
\item "The extent to which a user is confident and willing to act based on the recommendations, actions, and decisions of an artificially intelligent decision aid"~\cite{Madsen2000MeasuringHT}. This definition appeared in ~\cite{10.1145/3397481.3450656} ($n=1$).
\end{itemize}

Both definitions originate from social science and have been adapted to the Human-Machine context. These definitions emphasize that trust is rooted in the trustee's vulnerable position, arising from uncertainty about whether the LLM will meet expectations and deliver desired outcomes. This inherent uncertainty motivates the trustee to place trust in the model~\cite{10.1145/3476068}. Trust is also characterized by the trustee's positive expectation of the tool's ability and intention to achieve its goals. Furthermore, trust is described as an attitude, as how a person views and evaluates a particular situation, and is not a direct determinant of behavior. Although trust can guide behaviors such as LLM trust, this process is subject to individual, organizational, and cultural contexts~\cite{Liao_2022,10.1145/3630106.3658984}.

Trust plays a crucial role in shaping the adoption and continued use of LLMs and software development tools~\cite{10.1145/3664646.3664757}. However, trust alone is insufficient; fostering "appropriate trust" is essential to ensure that users rely on the model in ways that align with its actual capabilities~\cite{9582305,cheng2023itworktooonline,10.1145/3630106.3658984}. Trust is an attitude that stakeholders have towards the model, yet it does not necessarily indicate whether it is reliable. For instance, if a user's trust level does not match the actual soundness of the system. In that case, they might either accept a flawed response or unjustly dismiss accurate output. Neither scenario is desirable. 

One possible mediator between trust and appropriate trust is the design of trustworthiness models~\cite{9582305}. In the context of SE, a model is considered trustworthy if and only if it functions correctly and it is justified for stakeholders to trust the system~\cite{9582305}. This approach can help bridge the gap between user perception and system reliability, ultimately leading to more effective and responsive use of LLMs in SE practices.

\subsection{Phase\texorpdfstring{$_2$}{2}: Survey Study Results}

We analyzed 25 responses to open-ended questions on the definitions of \textit{ trust} in \llms for SE. One of the authors iteratively coded the survey responses, following an inductive approach to identify recurring themes. The other authors collaboratively verified and refined the final coding to ensure consistency and reliability. This analysis resulted in the following categories: 

\begin{itemize}[
leftmargin=0pt, itemindent=32pt,
labelwidth=15pt, labelsep=1pt, listparindent=0.7cm,
align=left]
    \item \textbf{Functional Correctness and Performance:}
 This was the most frequently mentioned theme across all downstream tasks. The participants emphasized the importance of \llms generating code that meets the specified requirements, passes the test cases, and improves performance. For example, a participant noted: \textit{"Fundamentally, I think trust in this sense means that the code, test, or summary produced is accurate. That is if I produce a code snippet, the model hasn't hallucinated any modules and the snippet will accomplish the task specified in the prompt..."}. This theme was particularly significant for the generation of test cases, and 75\% of the participants referred to it in their responses. 

    \item \textbf{Understandability:}
    Understandability emerged as an essential attribute, reflecting the desire of the participants for a result that is clear and easy to understand without extensive verification. This was especially emphasized for code generation tasks, where 26.32\% of the responses mentioned it. One participant remarked \textit{"I would consider trust to mean that I would be comfortable including the LLM-generate code into a code review packet without feeling the need to verify the code in advance."}

    \item \textbf{Security and System Integrity:}
    Participants expressed concern regarding the generated code's security. They emphasized the critical role of ensuring that \llms do not compromise the systems they generate code for. They underscored the importance of minimizing bugs and vulnerabilities. For example, one of the participants defined trust in \llms for code generation as \textit{"Trust is relying on LLMs do not intentionally try to undermine systems they generate code for."} Interestingly, no participants mentioned security and system integrity as essential factors when defining trust in \llms for test case generation.
    
    \item \textbf{Usability and Knowledge Enhancement:}
Participants valued the use of the generated code directly without careful evaluations and the potential of \llms to provide additional knowledge. One of them stated, \textit{"Trust in LLMS means that I can use the LLMS code directly in my study or work occasion."} Interestingly, the practitioner was a novice with minimal experience in SE and \llms, reinforcing the importance of educating practitioners to foster calibrated trust in \llms for SE applications.

    \item \textbf{Reliability and Consistency:}
     This category reflects the emphasis the participants placed on long-term performance and consistent behavior of \llms in SE tasks. It was the least used theme in all the definitions. One participant stated, \textit{"Trust in LLMs for Code Generation could be interpreted as the percentage of accurate code generated within an extended time frame of use."}
\end{itemize}
Our study reveals widespread use of \llms in SE, with 99.8\% of participants using them at least weekly. Despite widespread use, only a moderate proportion trust the models (52.9\% for code generation, 75\% for test case generation, 54.5\% for program repair). Furthermore, while 84\% believe that conceptualizing trust in LLM4SE is essential, 60\% of the participants struggle to differentiate between trust and trustworthiness or fail to recognize the distinction. This highlights the complexity of these concepts and their varying interpretations in public perception.

\subsection{Complementary Analysis for Definition of Trust}

Trust is a widely studied concept across disciplines such as HCI, DL, and automation. In AI research, two dominant perspectives on trust exist: (1) trust as a subjective human attitude and (2) trustworthiness as an intrinsic system property. The first perspective, common in HCI, defines trust as a user's perception of the reliability of an AI system, shaped by factors such as interpretability, confidence, and perceived competence under conditions of vulnerability ~\cite{10.1145/3476068, 10.1145/3290605.3300469, 10.1145/3514094.3534150, trustLLM, 10.1145/3531146.3533202, 10.1145/3544549.3585808}. In the SE literature, trust is often framed in a similar way to an expectation about the ability of an LLM to function under uncertainty~\cite{ghofrani2022trustchallengesreusingopen, 10.1007/978-3-642-40779-6_23}.

The second perspective, more common in the DL, views trustworthiness as an inherent property of the system, defined by attributes such as robustness, fairness, and security ~\cite{10.1145/3491209, trustworthyAI, liu2021trustworthyaicomputationalperspective}. In AI audits, trustworthiness is associated with fairness metrics and bias mitigation ~\cite{10.1145/3686963}, while in automation, trust calibration ensures that user reliance aligns with actual system performance ~\cite{Lee2004TrustIA}. Furthermore, some studies argue that trust is a multifaceted and context-dependent concept, cautioning against overly broad or narrow definitions that may limit its practical applicability ~\cite{10.1007/978-3-540-24747-0_25}.

Despite these established perspectives, the SE literature rarely distinguishes between trust and trustworthiness, leading to conceptual ambiguity. Our analysis also reveals that distrust is often overlooked, despite being a critical component of trust dynamics.

\subsection{Discussion}
Our findings reveal a disconnect between research and practice: the literature defines trust abstractly as a perception of risk , while practitioners define it using concrete technical factors like functional correctness, security, and usability.

To bridge this gap, we propose the following definitions based on our systematic review of the literature, survey study, and insights from broader trust research. Table \ref{tab:tab2_def} shows an overview of our definitions.

 \begin{enumerate}
    \item \textbf{Trust} Trust in LLM4SE refers to the willingness of a practitioner to rely on the recommendations or outputs of an LLM system for software engineering tasks, based on the belief that the system is capable, reliable, and aligned with their objectives under conditions of uncertainty. This trust is shaped by the perceived and demonstrated performance of the system, including correctness, security, and usability. This definition is supported by broader trust literature, where trust is conceptualized as a subjective attitude shaped by user perception ~\cite{10.1145/3476068, 10.1145/3290605.3300469, 10.1145/3514094.3534150}, as well as our SLR findings, where trust is often framed as expectation driven. Furthermore, the results of our survey reveal that practitioners define trust more concretely in terms of functional correctness, interpretability, and usability, focusing on practical reliability over abstract expectations.

    \item \textbf{Distrust} Distrust in LLM4SE is not merely the absence of trust, but signifies practitioners' active expectation of risk, failure, or harm in the system output for a given SE task. This occurs when the model is perceived to be unreliable, unpredictable, or misaligned with ethical and functional goals, including security vulnerabilities and biased or unsafe recommendations. HCI research highlights that distrust can be driven by inconsistencies in AI behavior, misalignment with user needs, or previous negative experiences ~\cite{10.1145/3613905.3650825}. Similarly, our survey results indicate that professionals actively distrust LLMs when they exhibit frequent inaccuracies or security vulnerabilities, particularly in safety-critical SE tasks. The distinction between trust and distrust is thus crucial, as distrust involves an explicit expectation of failure rather than mere uncertainty.

    \item \textbf{Trustworthiness} The trustworthiness in LLM4SE is the intrinsic quality of the LLM system that justifies a user's trust based on its ability to consistently perform tasks accurately, transparently, securely, and ethically. Unlike trust, which is user-dependent, trustworthiness is an objective property that can be evaluated through measurable factors such as robustness, fairness, security, and reliability. This aligns with the DL research, where trustworthiness is defined through system properties such as fairness, robustness, and security ~\cite{10.1145/3491209, trustworthyAI, 10.1145/3614424}. In contrast to trust, which is a user-driven perception, trustworthiness represents an objective measure of the reliability of the AI system. Our SLR found that trustworthiness in SE is rarely explicitly defined, leading to confusion between perceived and actual system reliability. Survey participants also struggled with this distinction, reinforcing the need to establish trustworthiness as a measurable property rather than an assumed user sentiment.  
\end{enumerate}

One key insight from our survey is that trust in LLM4SE is highly task-dependent. While 75\% of respondents expressed trust in LLMs for test case generation, trust was lower for code generation (52.9\%) and program repair (54.5\%). This finding is consistent with previous HCI research, which suggests that trust is dynamic and varies based on task complexity and perceived risk ~\cite{10.1145/3290605.3300469}. It also supports observations in DL and automation research, where trust is shaped by domain-specific considerations ~\cite{Lee2004TrustIA}.

Furthermore, our results indicate that practitioners often equate trust with reliability. Many survey participants described trust as something that builds over time through consistent correctness of outputs, rather than as an abstract attitude toward AI systems. This perspective is consistent with findings from Automation and Robotics, where trust is calibrated through ongoing system performance rather than static ~\cite{Lee2004TrustIA}. Similarly, in SE, the trust in the tools is related to their reputation, verification, and the ability to reduce the need for manual supervision ~\cite{hou2022systematicliteraturereviewtrust, Rutinowski2024BenchmarkingTA}. Our results suggest that trust in LLM4SE follows a similar pattern, reinforcing the need to incorporate subjective and system-oriented factors in trust research.

Finally, security considerations emerged as a crucial component of how practitioners define trust, yet none of the reviewed SLR papers explicitly incorporated security into trust definitions. Several survey respondents associated trust with the model’s ability to produce safe and non-malicious code and prevent vulnerabilities. This gap highlights the need to integrate security into trustworthiness discussions, as emphasized in DL research on robust and privacy-preserving AI~\cite{10.1145/3614424, trustworthyAI}. Furthermore, institutional trust perspectives propose that trust in AI should extend beyond individual system performance to include regulatory oversight and transparency~\cite{10.1145/3442188.3445890}. These insights could inform how trust in LLM4SE is conceptualized, particularly as AI governance frameworks evolve.

\subsection{Opportunities for Future Work}
Our findings suggest several critical avenues for future research. One key area that needs attention is the refinement and validation of concepts related to trust. The definitions that we have proposed are based on insights from the literature review and the perspective of practitioners. However, their applicability in real-world SE tasks requires further empirical assessment. A promising direction for future research could involve conducting longitudinal studies to examine how trust evolves as practitioners interact with LLM for SE tasks. Additionally, task-specific experiments focused on different SE tasks, such as code generation, test case generation, and program repair, could help clarify whether trust in LLMs is highly task dependent or whether users develop overarching heuristics that generalize across diverse SE contexts.

While our corpus has derived a definition of trust from the social sciences, trust is a topic well-studied across a variety of fields. Psychological safety, for example, is similar to trust in our context, as it refers to an individual’s perception that they can take interpersonal risks without fear of negative consequences \cite{doi:10.2307/2666999}. Although this concept aligns closely with our definition of trust, previous papers in LLM4SE have largely overlooked it. Future studies should broaden the exploration of trust by drawing inspiration from other domains.

Another important area of future research is the study of distrust in the context of LLM4SE. Although trust and trustworthiness have been fairly studied, distrust in LLM4SE remains an underexplored phenomenon. Our findings suggest that distrust is not simply the absence of trust, but an active expectation of failure, risk, or potential harm. Future research should investigate how distrust develops and whether it leads to outright rejection of LLM-generated outputs or selective verification based on task complexity. In addition, designing interventions to mitigate distrust, such as improved transparency, explicit explanations of failures, and confidence estimates, could play a crucial role in improving user confidence in LLM. Understanding the dynamics between trust, distrust, and corrective user behavior will be essential for designing AI-assisted development environments that align with practitioner expectations.

Lastly, there is a need to explore the distinction between trust and trustworthiness further. Our findings suggest that this distinction is not well understood by practitioners, as 60\% of the survey respondents struggled to differentiate between the two concepts. By exploring the interactions between trust, trustworthiness, and related constructs such as confidence, belief, and reliability, researchers could contribute to more precise conceptual frameworks for trust in AI. This would help bridge the gap between theoretical understandings of trust and the practical considerations that shape how users interact with LLMs in SE contexts.

\begin{tcolorbox}[colframe=black!50, colback=gray!10, title=Summary of $RQ_1$: Definition of Trust Concepts in LLM4SE]
Our study found that only 27.8\% ($n=5$ out of 18) of reviewed papers explicitly defined trust, borrowing established definitions from social science and AI research rather than introducing new ones. The survey revealed that practitioners conceptualize trust in \llms based on functional correctness, security, understandability, and usability, while the literature primarily frames it as an attitude toward uncertainty and risk. This discrepancy highlights a significant gap between theoretical definitions and real-world expectations. Furthermore, 60\% of the participants struggled to distinguish between trust and trustworthiness, indicating conceptual confusion. The trust in \llms was found to depend on tasks, with higher trust levels for test case generation (75\%) than for code generation (52.9\%) and program repair (54.5\%). We propose distinct definitions: trust as the willingness of a practitioner to rely on \llms based on perceived reliability, distrust as an expectation of failure, and trustworthiness as an intrinsic property of the system that ensures reliability, security, and ethical alignment. Addressing these differences is crucial for fostering the appropriate trust in \llms for software engineering.  
\end{tcolorbox}

%% file: table/table2.tex
\begin{table}[t]
\caption{Trust concepts in LLM4SE}
\label{tab:tab2_def}
\centering
\resizebox{\textwidth}{!}{%
\begin{tabular}{p{2cm}p{4cm}p{4cm}p{5cm}}
\hline
\textbf{Aspect}          & \textbf{Trust}        & \textbf{Distrust}     & \textbf{Trustworthiness} \\ \hline
\textit{\textbf{Definition}} &
  Willingness to rely on LLM outputs under uncertainty, based on positive expectations  &
  Active expectation of failure or risk in LLM outputs or behavior &
  Intrinsic system qualities ensuring reliable, transparent, and ethical performance \\ 
\textit{\textbf{Focus}}  & Positive expectations & Negative expectations & System characteristics   \\ [0.2cm]
\textit{\textbf{Nature}} & User perception       & User perception       & System attribute         \\ [0.2cm]
\textit{\textbf{Example}} &
  Using LLM-generated test cases based on high observed accuracy and interpretability &
  Avoiding code from LLM due to prior output errors or security concerns &
  A system offering explainable outputs and consistent performance aligned with SE standards \\ \hline
\end{tabular}%
}
\end{table}

%% file: text/7_RQ2.tex
\section{\texorpdfstring{$RQ_2$:}: Antecedents of trust}\label{subseq:rq2}

In this section, we present our results for \ref{rq:factors} and discuss its implications. 

\input{table/table1}

\subsection{Phase\texorpdfstring{$_1$}{1}: SLR results}
From the literature, we extracted a large collection of antecedents that impact human trust in the use of LLM4SE. We categorize these antecedents into two main groups: model-specific trustworthiness attributes, which are inherent to the model, and user-centric trust attributes, which are extrinsic to the model. Table \ref{tab:antecedents} provides a comprehensive overview of the taxonomy we developed, constructed using the MATCH model ~\cite{Liao_2022} as a guiding framework. 61\% (n=11 out of 18) of the studies focused on the antecedents inherent to the model during the design phase; Conversely, 22\% (n=4 out of 18) explored how attributes associated with interface interaction help deliver model trustworthiness to stakeholders; 17\% (n=3 out of 18) explored both aspects.

\subsubsection{Model-Specific Attribute} Model-Specific attributes encompass inherent characteristics and capabilities of the model independent of user perception. Understanding how model-specific attributes contribute to trust formation is essential for designing LLMs that foster appropriate trust, ensuring users neither overestimate nor underestimate the model's reliability in SE tasks. In this section, we present results related to model-specific attributes. 
\begin{itemize}[
leftmargin=0pt, itemindent=32pt,
labelwidth=15pt, labelsep=1pt, listparindent=0.7cm,
align=left]
\item\textbf{Ability}: Ability is an overarching term for LLMs' performance or competence~\cite{10.1145/3630106.3658984}. We identified the three most popular performance-related antecedents in our literature: accuracy, efficiency, and reliability. Accuracy is the degree to which the output matches the expected results in terms of accuracy and precision~\cite{key2023trustworthyneuralprogramsynthesis, lo2023trustworthysynergisticartificialintelligence, 10.1109/ICSE-SEIP58684.2023.00043}. Efficiency refers to the ability of the software to use computational resources judiciously and effectively~\cite{lo2023trustworthysynergisticartificialintelligence}. Reliability is the ability of the software to function correctly and consistently under various conditions for an extended period~\cite{10.1109/ICSE-SEIP58684.2023.00043, lo2023trustworthysynergisticartificialintelligence, 10.1145/3641540}. Given the probabilistic nature of \llms, models sometimes generate incorrect or insecure outputs. This inherent unreliability poses a key challenge in establishing trust in these models~\cite {Maninger_2024}. Reliability also includes the model's ability to handle unexpected input gracefully and to have appropriate fault tolerance~\cite{10.1145/3641540}.

\item\textbf{Benevolence}: A trustworthy model should prioritize acting in the best interest of users by aligning its goals with theirs~\cite{10.1145/3630106.3658984}. This alignment can be categorized into two levels: \textbf{short-term goal alignment} and \textbf{long-term goal alignment}. In the short term, models must effectively address users' immediate needs, which can vary among individuals in different situations. These needs may include offering clear advantages in usage~\cite{10.1109/ICSE-SEIP58684.2023.00043}, improving productivity during the coding process~\cite{10.1145/3630106.3658984}, and meeting specific functional expectations~\cite{10.1109/ICSE-SEIP58684.2023.00043}. Although achieving short-term goals is crucial for fostering initial trust, it alone is not sufficient to ensure long-lasting trustworthiness. 

Beyond immediate outputs, models must align with the long-term goals of users. For example, developers often rely on \llms to quickly and accurately achieve their short-term objectives. However, they are also concerned about maintaining and improving their programming skills over time, as excessive dependence on \llms may lead to skill degradation ~\cite{10.1145/3630106.3658984}. To address this concern, \llms should incorporate educational content into their generated outputs, fostering learning and supporting developers in strengthening their expertise~\cite{10.1109/ICSE-SEIP58684.2023.00043}. However, current models lack mechanisms that allow developers to clearly specify both their short-term and long-term goals, leading to a misalignment between the developer's expectations and the results generated by AI~\cite{10.1145/3630106.3658984}.

\item\textbf{Integrity}: This attribute is related to the LLM training process, including that the development practice was safe and secure, according to ethical guidelines~\cite{Maninger_2024, lo2023trustworthysynergisticartificialintelligence, 10.1109/ICSE-SEIP58684.2023.00043}. A study highlights the critical link between training data quality and model trustworthiness, emphasizing how issues such as biases, outdated information, or security vulnerabilities can propagate into model output, directly impacting its integrity~\cite{wang2024trustworthyllmscodedatacentric}. Regarding training data, models are expected to reduce bias, which can be achieved by updating training data for new tasks and curating real-world data sets~\cite{russo2024navigatingcomplexitygenerativeai,lo2023trustworthysynergisticartificialintelligence, Maninger_2024}.
\end{itemize}

\subsubsection{User-Centric Attributes} User-Centric Trust Attributes are trust factors that arise from a user’s interaction with the system, including transparency, explainability, usability and the ability to align the results with user requirements and expectations. In an ideal scenario, model-specific attributes are effectively communicated to users. However, in typical usage, users cannot directly assess a model's trustworthiness; instead, they infer it through various affordances. Designing a trustworthy model involves creating trust affordances that deliver the actual capability of the model, so users can form appropriate trust without overestimating or underestimating the model's capability. In this section, we present results related to user-centric attributes. 

\begin{itemize}[
leftmargin=0pt, itemindent=32pt,
labelwidth=15pt, labelsep=1pt, listparindent=0.7cm,
align=left]
\item\textbf{AI-generated Content}: AI-generated content is directly produced by LLMs. Besides natural language, this content might include SE-related outputs such as code, test cases, and program repairs, based on users' specific needs. As AI-generated content is the primary indicator that users use for the model's ability, the output should be well-delivered to users to receive appropriate trust assessment from them. However, our corpus has shown that the current presentation mediums for the model have difficulty capturing code structure and semantics~\cite{Maninger_2024}. LLM should delve into finding new mediums of representation, such as graphical representations of the generated output, to boost the effectiveness of content delivery~\cite{lo2023trustworthysynergisticartificialintelligence,10.1145/3397481.3450656,noller2022trustenhancementissuesprogram}.

\item\textbf{Transparency}: Transparency is an affordance employed by developers to signal to stakeholders that their trust in the system is well-founded~\cite{9582305}. By offering clear justifications, transparency helps users believe that the system operates as intended. In our analysis, normative metrics and Explainable Artificial Intelligence (XAI) features are the most emphasized factors for achieving transparency.

One way transparency is conveyed is through \textbf{normative metrics}, which refer to standardized benchmarks that provide measurable, quantitative evidence of models' performance. These metrics play a key role in evaluating the model’s ability by offering objective indicators of how well they meet expectations for accuracy, efficiency, and reliability. Importantly, these metrics help align user perceptions of the model with its actual trustworthiness  ~\cite{10.1109/MC.2023.3240730, kowald2024establishingevaluatingtrustworthyai}. For example, accuracy can be assessed through metrics such as the proportion of correct results within the top N suggestions ~\cite{lo2023trustworthysynergisticartificialintelligence}. A high accuracy score suggests that users can confidently rely on the model to provide the correct answer within a few searches. On the other hand, Initial False Alarm (IFA) measures the number of false positives before a true positive is found. A low IFA value is preferred, as it indicates that users do not need to sift through many incorrect results ~\cite{lo2023trustworthysynergisticartificialintelligence, key2023trustworthyneuralprogramsynthesis}.

Efficiency is another key aspect evaluated through normative metrics. Efficiency metrics assess the time taken by the model to execute specific tasks, providing users with clear expectations about response times and helping to prevent surprises. This transparency about response time also builds confidence in the model’s practical usability. Similarly, reliability is often measured by the Mean Time Between Failures (MTBF), assuring users that the model consistently delivers accurate outputs without frequent errors~\cite{10.1145/3411764.3445650}.

Normative metrics are closely tied to transparency because they offer users the data they need to gauge the model’s performance. A well-calibrated model, for instance, provides probabilistic estimates that align closely with actual outcomes, enhancing trust by showing that the model’s predictions are accurate, effective, and reliable ~\cite{key2023trustworthyneuralprogramsynthesis, spiess2024calibrationcorrectnesslanguagemodels}. However, the current models suffer from issues like data duplication, leading to over-optimistic results ~\cite{10.1145/3641540}.

Alongside normative metrics, XAI features serve as another essential mechanism for enhancing transparency. XAI enhances transparency by offering insights into how and why decisions are made, helping users understand the model’s reasoning and inner workings. Relevant documentation and training materials that clarify the rationale behind the model’s outputs can further aid in building trust ~\cite{10.1145/3411764.3445650}. Studies have shown that providing clear comments and specifications in code helps users understand a program’s behavior ~\cite{10.1109/ICSE-SEIP58684.2023.00043}.

Additionally, when a model links outputs to specific source code, methodologies, and reasoning processes, users can trace the model’s logic, ensuring higher levels of provenance transparency. Finally, incorporating formal test cases and realistic test values further enhances transparency, ensuring that the model adheres to ethical guidelines and regulatory standards, which reinforces trust in the model’s governance ~\cite{key2023trustworthyneuralprogramsynthesis, 10.1145/3397481.3450656, noller2022trustenhancementissuesprogram}.

\item\textbf{Interaction}: Interaction plays a pivotal role in how users evaluate the trustworthiness of LLMs. This encompasses direct engagement with the model itself and indirect interaction through the broader user community.

When users engage directly with the model, they often assess its \textbf{customizability}—the extent to which the model adapts to user preferences and feedback. Studies show that users value models that can evolve based on their input, reinforcing perceptions of the model’s ability and responsiveness ~\cite{10.1109/ICSE-SEIP58684.2023.00043,10.1145/3411764.3445650}. Despite these affordances, limitations in \textbf{prompt comprehension} remain a significant barrier. Users who can effectively articulate and refine their prompts tend to achieve better outcomes, while novice users often struggle with the learning curve of prompt engineering ~\cite{Perry_2023}.

Beyond direct engagement, users frequently turn to the broader community to inform their trust judgments. \textbf{Socialization}—interacting with user communities and leveraging collective experiences—provides valuable insights into the model’s strengths and limitations ~\cite{cheng2023itworktooonline}. Trustworthiness perceptions may also differ based on whether the LLM is open-source or proprietary. As observed by Huang et al. \cite{trustLLM}, proprietary models tend to outperform open-source ones in terms of trustworthiness; however, models like LLaMA2 demonstrate that open-source systems can closely rival their proprietary counterparts. These structural differences in model openness further shape how users interpret and validate feedback from the community. Observing community feedback allows users to gain a more comprehensive understanding of the model’s capabilities, often supplementing the limited perspective they might form through personal use~\cite{cheng2023itworktooonline}. Positive examples, shared strategies, and troubleshooting tips from others help build trust by demonstrating the model’s practical value across diverse scenarios~\cite{cheng2023itworktooonline}.

However, this reliance on community signals carries inherent risks. Inaccurate or misleading feedback can skew user expectations, leading to either misplaced trust or unwarranted skepticism. Ensuring that community-driven signals are evidence-based and moderated is critical to maintaining trustworthiness ~\cite{cheng2023itworktooonline,10.1109/ICSE-SEIP58684.2023.00043}.
\end{itemize}

\subsection{Phase\texorpdfstring{$_2$}{2}: Survey Study Results}

In the survey study, we asked respondents to select one or more antecedents that would influence their trust in LLM4SE tasks. These factors can be categorized into the following:
\begin{itemize}[
leftmargin=0pt, itemindent=32pt,
labelwidth=15pt, labelsep=1pt, listparindent=0.7cm,
align=left]
    \item Model attributes: Accuracy, Robustness, Ethicality, Interpretability, Controllability;
    \item HCI attributes: Source Reputation, Workflow Integration, Endorsement, Community Engagement.
\end{itemize}

\begin{figure}[ht]
  \centering
  \includegraphics[width=\linewidth]{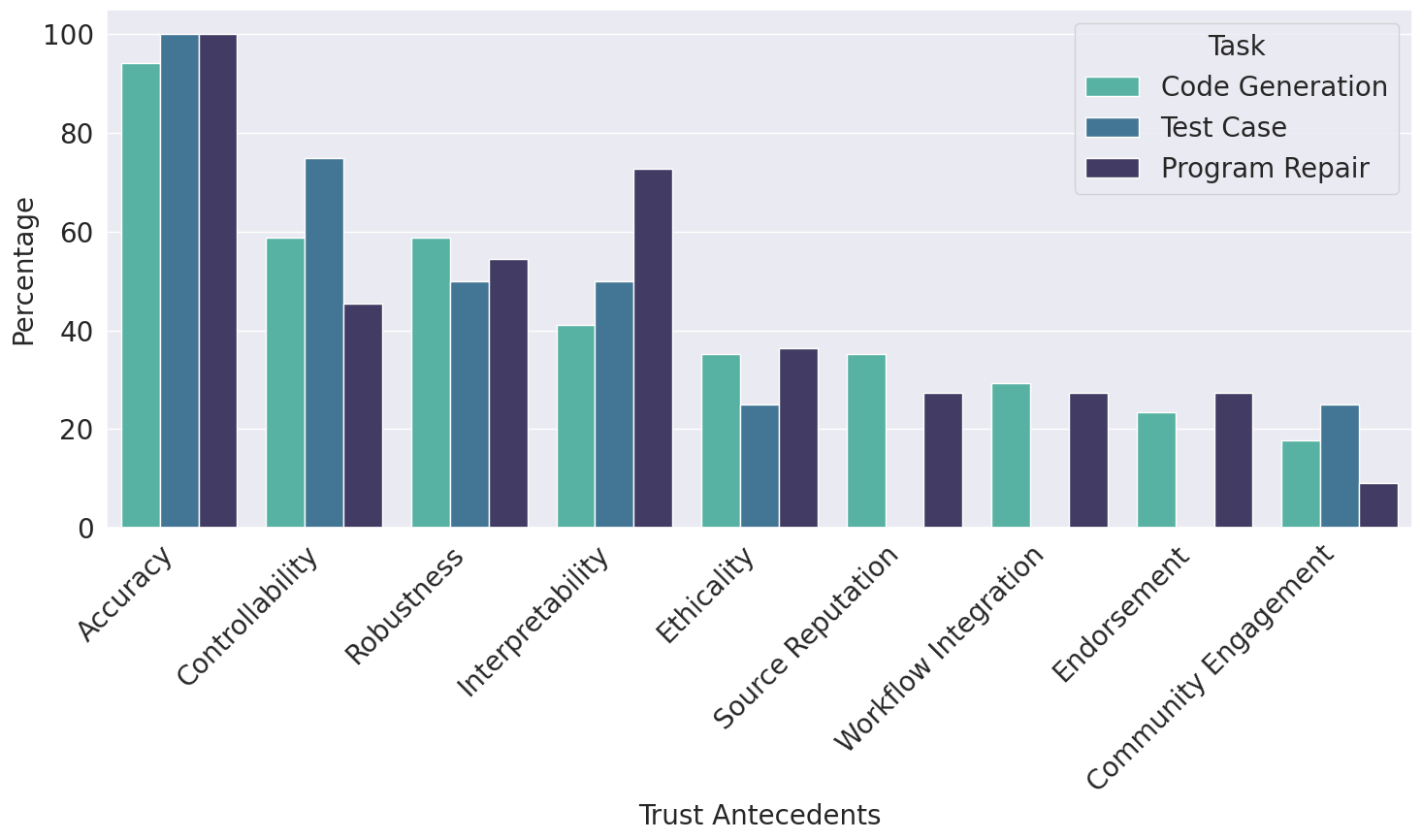}
  \Description[Trust Antecedents]{Trust Antecedents Selection Based on SE Tasks}
  \caption{Trust Antecedents Selection Based on SE Tasks}
   \label{RQ2_task_rundown}  
\end{figure}

Fig.~\ref{RQ2_task_rundown} is a grouped bar chart that visualizes the selection of trust antecedents across three tasks. In general, we observe that model trustworthiness attributes are valued more than human-model interaction attributes, regardless of SE tasks. Accuracy is the most selected trust antecedent in all tasks (94\% for code generation; 100\% for test case generation and program repair). However, preferences for the second most important antecedents vary by task: for code generation, controllability (58.8\%) and robustness (58.8\%) are equally valued; for test-case generation, controllability (75\%) is preferred; and for program repair, interpretability (72.7\%) is valued. 

In the context of program repair, one respondent added to our antecedent collection, expressing concern that their programming skills might degrade if they rely too heavily on the model. This highlights the user's requirement for the model to support their long-term improvement in SE over solving immediate queries. It indicates the user's emphasis on the model's benevolence.

\begin{figure}[ht]
  \centering
  \includegraphics[width=.8\linewidth]{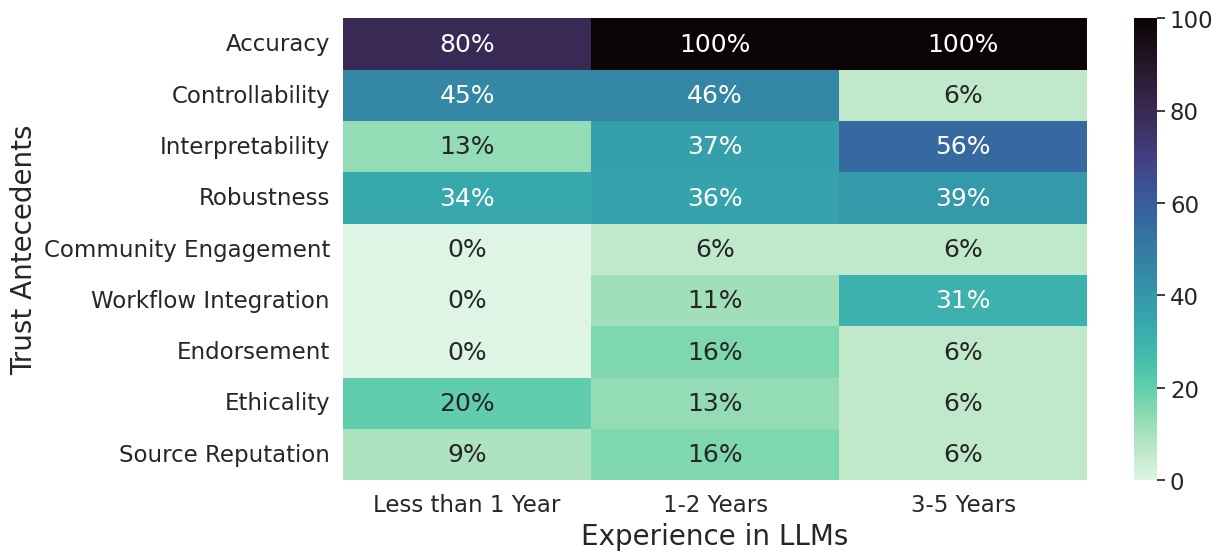}
  \Description[Trust Antecedents]{Trust Antecedents Selection Based on Years of Experience with LLMs}
  \caption{Trust Antecedents Selection Based on Years of Experience with LLMs}
   \label{RQ2_experience_LLM}  
\end{figure}

Fig.~\ref{RQ2_experience_LLM} is a heat map showing the relationship between user experiences using LLM4SE and their trust antecedent selection. The percentage is calculated as the weighted average of respondents in three SE tasks. 

The visualization shows that accuracy and robustness are highly valued across all experience levels. Conversely, community-associated factors (community engagement 0\%, endorsement 0\%, source reputation 0\%) are not valued by users with less than a year of experience, but gain attention with increased experience, though still less than model-related attributes. This suggests that users are more dependent on direct model interaction than on external validation.

A clear trend also emerges based on user experience: experienced users place less emphasis on the model's controllability (6\%) and ethical compliance . This could imply that, with increasing familiarity, users developed a level of trust that mitigates concerns about how to use the tool effectively and the model's benevolence. However, experienced users increasingly value the interpretability (56\%) of the generated output and prefer models that integrate seamlessly into their established procedure.

\begin{figure}[ht]
  \centering
  \includegraphics[width=.85\linewidth]{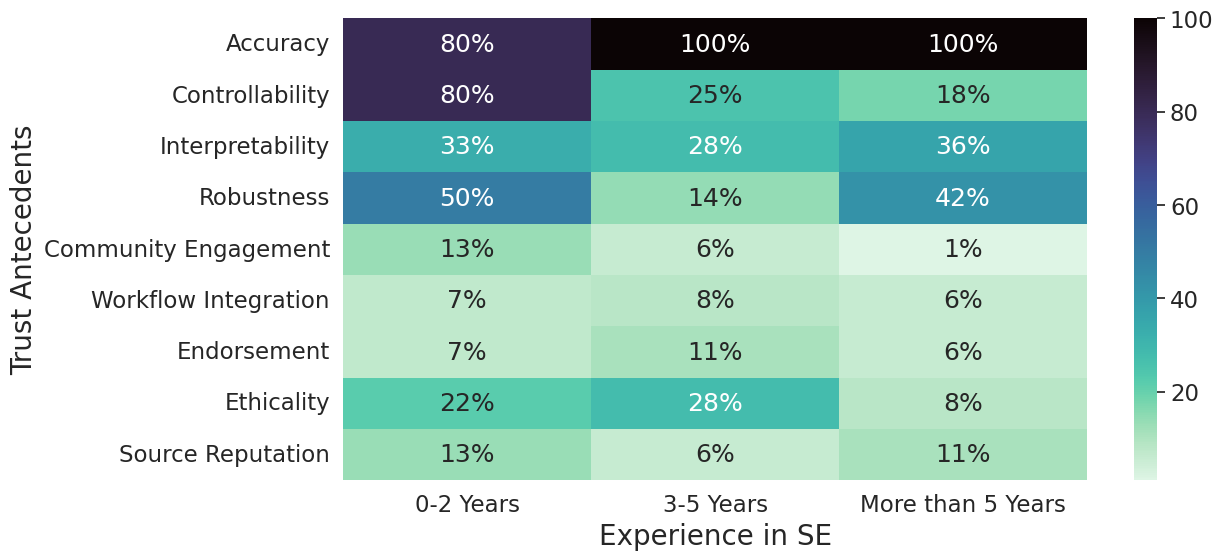}
  \Description[Trust Antecedents]{Trust Antecedents based on years of Experience in SE}
  \caption{Trust Antecedents Selection Based on Years of Experience in SE}
   \label{RQ2_experience_SE}  
\end{figure}

We observe a similar result from Fig.~\ref{RQ2_experience_SE}, a heat map between users' year of expertise in SE and their trust antecedent selection. However, we find that there is no significant relationship between higher levels of knowledge in SE and their selection of community-related factors.
Interestingly, we observe that intermediates pay significantly less emphasis on robustness compared to novices and experts.

\subsection{Complementary Analysis of Trust Antecedents}

Trust in AI systems, including LLM, is shaped by multiple interrelated factors, including technical performance, user perception, and ethical considerations. In AI research, accuracy is widely recognized as a fundamental trust factor in all domains ~\cite{khandelwal2020generalizationmemorizationnearestneighbor, hendrycks2021measuringmassivemultitasklanguage, yin2023largelanguagemodelsknow}. In SE, ensuring accuracy and reliability is essential to prevent security vulnerabilities ~\cite{hou2022systematicliteraturereviewtrust, ahuja2020openingsoftwareengineeringtoolbox, hassan2024rethinkingsoftwareengineeringfoundation, Baldassarre_2024, 9474373, Rutinowski2024BenchmarkingTA}. However, trust is not solely dependent on accuracy; overconfidence in the accuracy of the model can hinder trust by making unreliable results appear convincing ~\cite{huang2023trustgptbenchmarktrustworthyresponsible}. This phenomenon is observed in various AI applications, where users may develop misplaced trust in systems despite their inherent unreliability ~\cite{10.1145/3579460}.

Beyond precision, trust is also influenced by the model’s ability to balance technical performance with inclusivity and multilingual capabilities. Natural Language Processing (NLP) research highlights that ensuring unbiased and culturally aware results remains a challenge when optimizing high performance in different linguistic contexts ~\cite{10.1145/3597512.3599715, zhang2023donttrustchatgptquestion, wang2024decodingtrustcomprehensiveassessmenttrustworthiness}. This trade-off between precision and inclusivity is particularly relevant for SE, where domain-specific terminology may not generalize well across different programming cultures.

Robustness is another critical factor in trust formation. AI models, including LLMs, often struggle to maintain consistent performance on similar inputs. Studies show that even minor perturbations in prompts can lead to significantly different outputs ~\cite{ye2023assessinghiddenrisksllms, 10.1145/3610578}, and evaluations such as PromptBench highlight the sensitivity of LLMs to word-level variations ~\cite{zhu2024promptrobustevaluatingrobustnesslarge}. Similar robustness challenges have been identified in NLP tasks, where adversarial manipulations can undermine the reliability of the model ~\cite{wang2022adversarialgluemultitaskbenchmark, Chen_Hsieh_2023}.

Transparency and accountability further shape trust in AI systems. Research suggests that users develop greater trust when they perceive AI as competent, benevolent, and transparent in their decision-making ~\cite{Baughan_2023, 10.1145/3290605.3300469, 10605304, 10636143}. However, the black-box nature of LLMs and corporate secrecy hinder transparency ~\cite{wu2022aichainstransparentcontrollable, Liao2024AI, LANGER2021103473}. Techniques such as uncertainty expressions (\eg 'I am not sure') can improve trust by signaling limitations ~\cite{Kim_2024}, while XAI approaches offer interpretability at the potential cost of performance ~\cite{wu2022aichainstransparentcontrollable}. The regulatory landscape also plays a role, with frameworks such as the EU AI Act emphasizing responsibility ~\cite{EUwebsite, Jobin2019, Floridi2019Unified, d16c5995650c4bc7ae5ba90b76da1d32, 10430152}, yet leaving questions of liability for AI errors unresolved ~\cite{d16c5995650c4bc7ae5ba90b76da1d32,10.1145/3686963}. The proposed solutions, such as blockchain-based traceability, could improve accountability in AI decision making ~\cite{10.1145/3614424}.

Finally, ethical considerations remain an essential, albeit sometimes secondary, factor in trust formation. Although performance and usability often take priority, fairness in AI is critical to long-term trustworthiness. Studies reveal that LLMs can introduce social biases even in code generation tasks ~\cite{liu2023uncoveringquantifyingsocialbiases}, and research on models such as Codex indicates a trade-off between fairness and performance ~\cite{cui2024fftharmlessnessevaluationanalysis}. Trust is also shaped by concerns about the impact of artificial intelligence on professional development. Similarly to educational AI, where trust is linked to fairness and correctness in recommendations ~\cite{10.1145/3626252.3630842}, SE users worry that overreliance on LLM could degrade their programming skills. However, while not a primary focus for experienced users, ethical concerns remain an important aspect of long-term trust in AI ~\cite{10.1145/3661167.3661214}.

\subsection{Discussion}
Our findings reveal that trust in LLM4SE is shaped by an interplay of model-specific and user-centric attributes, highlighting the complexity of user confidence in LLMs. Accuracy has emerged as a fundamental driver of trust. Our survey study also confirms that practitioners prioritize accuracy in LLM-generated outputs, mirroring observations from a wider trust literature. However, overemphasis on accuracy without mechanisms for uncertainty expression can lead to misplaced trust, a well-documented concern in AI research. Although accuracy is crucial, our findings underscore the need to equip users with tools to critically assess reliability.

A key gap in existing research in LLM4SE is the lack of efforts to establish a priority of antecedent trust. Although efforts have been made to identify trust antecedents by model and user specificity, there has been no systematic investigation of their relative importance to trust. Understanding which trust antecedents are the highest priority for software engineers could enable more targeted improvements in LLM4SE, ensuring that development efforts are aligned with user expectations. This need for prioritization is evident in our survey study, where users consistently emphasized certain factors, such as precision, interpretability, controllability, and robustness over endorsement or source reputation. A lack of clarity on the relative importance of trust antecedents could lead development efforts to focus on less critical aspects while neglecting those that matter most to users. 

Transparency and accountability remain central concerns for trust in LLM4SE. Our survey study indicates that experienced software engineers demand clearer explanations for model decisions, a concern echoed in the broader trust literature. However, the opaque nature of LLMs, exacerbated by corporate secrecy, presents a significant barrier to transparency. Broader AI research suggests that incorporating uncertainty expressions or explainable AI techniques can enhance trust. The importance of interpretability was clearly visible in our survey study, where it was deemed important in all downstream SE tasks. 

Robustness is another critical dimension where expectations in software engineering diverge from broader AI applications. Our findings reveal that experienced software engineers emphasize the need for LLMs to maintain consistency across different programming tasks. This aligns with concerns in the AI literature, where models frequently exhibit instability when faced with adversarial inputs or minor variations in phrasing. While robustness is a universal challenge for AI, software engineering demands particularly high levels of consistency due to the necessity of reproducibility in development workflows.

Ethical considerations, while not the primary concern of experienced software engineers, remain integral to trust formation in LLM4SE. Concerns about social biases in LLM-generated code and potential skill degradation among developers underscore the need for ethical safeguards. The ability of LLMs to handle diverse linguistic and cultural contexts is key to ensuring trust across the global community of Software Developers. Although technical reliability is a prerequisite for trust, long-term confidence in AI systems requires ongoing attention to fairness, professional impact, and user empowerment. 

Ethical considerations, while not always the primary concern of experienced software engineers, remain integral to trust formation in LLM4SE because they directly affect the reliability, fairness, and long-term adoption of these systems. Concerns about social biases in LLM-generated code underscore the crucial need for ethical safeguards, as such biases can lead to discriminatory software that perpetuates societal inequities and erodes public trust \cite{ling2025biasunveiledinvestigatingsocial, du2025faircoderevaluatingsocialbias}. Furthermore, as stated in our user study, the potential for developer skill degradation is a legitimate concern for professional development. When developers become too dependent on LLM, their critical thinking and foundational competencies may erode. This erosion can foster resistance and distrust within the software engineering community, ultimately affecting the perceived value and sustainability of these tools. The importance of transparency in an LLM’s ethical guidelines, especially regarding how it handles sensitive data and makes decisions, is crucial \cite{whytransparancy}. Without clear communication of these principles, even technically proficient systems risk breeding distrust. Research shows that transparency and accountability are the core to fostering trust in AI systems, as they enable stakeholders to understand how decisions are made and provide mechanisms for oversight and redress when needed \cite{du2025faircoderevaluatingsocialbias,whytransparancy}. Therefore, aligning LLM behavior with the principles of fairness, accountability, and transparency is not merely optional; it is a fundamental prerequisite for building enduring trust.

Beyond these ethical dimensions, our study reveals further critical areas where practitioner priorities for trust antecedents diverge from or offer new perspectives compared to established literature. For example, security emerged as an paramount concern for practitioners when defining trust, despite being largely overlooked in traditional SLR definitions of trust. Survey respondents explicitly linked trust to an LLM's ability to produce safe, nonmalicious code and prevent vulnerabilities. This highlights a crucial gap between academic conceptualizations and the practical demands of secure software development using LLMs.

\subsection{Opportunities for Future Work}
The results of this study present several avenues for future research aimed at enhancing the trustworthiness of LLMs in the context of SE tasks. An important avenue of research could be determining the relative importance of trust antecedents to prioritize trustworthy model development. Causal analysis, survey studies, controlled experiments, and A/B testing could help quantify the impact of different factors on trust. Identifying high-priority antecedents will enable targeted development of a trustworthy LLMS for various SE tasks.

As trust in LLMs is highly dependent on their alignment with users’ needs, especially in a dynamic field like SE, future research should explore models that evolve with users’ skills and tasks. This includes integrating learning aids, offering skill-enhancement suggestions, and ensuring that LLMs help maintain or improve users’ programming capabilities over time. Addressing concerns about AI degrading human expertise will be crucial in promoting long-term trust.

Transparency and Interpretability emerged as significant factors for experienced users. Future work should focus on developing explainability and interpretability mechanisms that offer clear, actionable justifications for LLM-generated outputs in SE tasks. This will not only address the black-box nature of LLMs but also improve users’ understanding of how model decisions are made. Effective transparency mechanisms could help bridge the gap between technical complexity and user trust.

A crucial avenue for future work is the development of concrete ethical guidelines and best practices for LLM-assisted SE. Research should focus on creating implementable strategies for practitioners and tool builders to mitigate ethical risks. This includes developing tools to help identify biases and security vulnerabilities in generated code. Future research should also advocate for designing LLMs that are more transparent about their ethical limitations, thereby integrating human oversight and accountability into the development lifecycle.

Another key area is community-driven trust signals; future research should investigate methods for moderating and validating community-driven feedback to ensure its reliability and usefulness. Moreover, incorporating adaptive feedback loops into LLMs could allow them to refine their performance based on real-time user input, fostering greater trust among diverse developer communities.

Lastly, customized trust models could improve usability by adjusting trust mechanisms based on user expertise and task complexity. Novice users might benefit from simplified explanations and clear guidance, while experienced developers may seek more detailed insights into model behavior, particularly for tasks that demand high accuracy and robustness. Adapting trust signals dynamically could improve confidence across different levels of experience.

\begin{tcolorbox}[colframe=black!50, colback=gray!10, title=Summary of $RQ_2$: Antecedents of Trust]

Trust in LLM4SE tasks is shaped by model-specific attributes such as accuracy, robustness, and interpretability, as well as user-centric factors like transparency and usability. Accuracy emerged as the most critical factor, with 94–100\% of respondents emphasizing its importance across tasks like code generation, test case generation, and program repair. Experienced users prioritize interpretability and seamless workflow integration, while novice users focus more on controllability and ethicality. Transparency mechanisms, such as XAI features and normative metrics, are essential for fostering trust but are hindered by the black-box nature of LLMs. Users also express concerns about over-reliance on these models potentially degrading their programming skills, highlighting the need for LLMs to support both immediate task completion and long-term skill development. Community-driven signals influence trust but require moderation to avoid misinformation, as direct interaction with models remains a stronger trust driver than external endorsements.
\end{tcolorbox}

%% file: table/table1.tex
\begin{table*}[t]
\caption{Trust Antecedent Overview}
\label{tab:antecedents}
\centering
\resizebox{\textwidth}{!}{%
\begin{tabular}{llll}
\toprule
\textbf{Construct} &
  \textbf{Factor} &
  \textbf{Survey Example} &
  \textbf{References} \\ \hline
\multicolumn{4}{c}{\textbf{Model-Specific Attributes}} \\ \hline
\multirow{3}{*}{Ability} &
  Accuracy &
  \textit{"The confidence that the LLMs can identify and fix bugs in code correctly."} &
  \cite{key2023trustworthyneuralprogramsynthesis, lo2023trustworthysynergisticartificialintelligence, 10.1109/ICSE-SEIP58684.2023.00043, 10.1007/978-3-031-55486-5_16} \\ 
 &
  Efficiency &
  \textit{"Make the program more efficient, without changing its functionality."} &
  \cite{lo2023trustworthysynergisticartificialintelligence, 10.1007/978-3-031-55486-5_16} \\  
 &
  Reliability &
  \textit{"My trust is based on the tool's ability to consistently and reliably generate correct, understandable outputs."} &
  \cite{10.1109/ICSE-SEIP58684.2023.00043, lo2023trustworthysynergisticartificialintelligence, 10.1145/3641540,Maninger_2024} \\[0.2cm]
\multirow{2}{*}{Benevolence} &
  Intended Use &
  \textit{"The generated code fulfilled my requirements."} &
  \cite{10.1109/ICSE-SEIP58684.2023.00043, 10.1145/3630106.3658984, 10.1007/978-3-031-55486-5_16} \\  
 &
  Long-term Growth &
  \textit{"I don’t want it to take over my job."} &
  \cite{10.1109/ICSE-SEIP58684.2023.00043, 10.1145/3630106.3658984} \\ [0.2cm] 
\multirow{2}{*}{Process Integrity} &
  Decision Process &
  \textit{"My trust relies on LLMs to not intentionally try to undermine systems."} &
  \cite{russo2024navigatingcomplexitygenerativeai,lo2023trustworthysynergisticartificialintelligence, Maninger_2024, wang2024trustworthyllmscodedatacentric} \\ 
 &
  Ethical Compliance &
  \textit{"The LLMs should maintain the privacy of data; it should not use the prompts for its training."} &
  \cite{Maninger_2024, lo2023trustworthysynergisticartificialintelligence, 10.1109/ICSE-SEIP58684.2023.00043, wang2024trustworthyllmscodedatacentric} \\ \hline
\multicolumn{4}{c}{
\textbf{User-centric Attributes}} \\ 
\midrule
\multirow{3}{*}{AI-generated Content} &
  Code &
  \textit{"The result is well-tested and properly takes care of corner cases."} &
  \cite{Maninger_2024,lo2023trustworthysynergisticartificialintelligence} \\  
 &
  Test Case &
  \textit{"It can generate a rich set of tests."} &
  \cite{10.1145/3397481.3450656} \\ 
 &
  Program Repair &
  \textit{"Help me detect the bug and fix the bug, and I can confirm that the bug does exist."} &
  \cite{noller2022trustenhancementissuesprogram} \\ [0.2cm]
\multirow{2}{*}{Transparency} &
  Normative Metrics &
  \textit{"Trust in LLMs could be interpreted as the percentage of accurate code generated within an extended time frame of use."} &
  \cite{key2023trustworthyneuralprogramsynthesis,spiess2024calibrationcorrectnesslanguagemodels, lo2023trustworthysynergisticartificialintelligence, 10.1145/3411764.3445650,10.1145/3641540} \\ 
 &
  Explainable AI Features &
  \textit{"I trust the tools if I can understand the snippet and how it functions for myself."} &
  \cite{10.1145/3411764.3445650,10.1109/ICSE-SEIP58684.2023.00043,key2023trustworthyneuralprogramsynthesis, 10.1145/3397481.3450656, noller2022trustenhancementissuesprogram} \\  [0.2cm]
\multirow{2}{*}{Interaction} &
  Customization &
  \textit{"It's difficult to capture the requirements through a simple prompt."} &
  \cite{Perry_2023,10.1109/ICSE-SEIP58684.2023.00043,10.1145/3411764.3445650} \\ 
 &
  Socialization &
  \textit{"Providing reliable answers which are accepted by the community like stack overflow or reliable static analysis tools."} &
  \cite{cheng2023itworktooonline,10.1109/ICSE-SEIP58684.2023.00043} \\ 
\bottomrule
\end{tabular}%
}
\end{table*}

%% file: text/8_RQ3.tex
\section{\texorpdfstring{$RQ_3$:}: Trust Metrics}\label{subseq_rq3}
In this section, we present our results for \ref{rq:metrics} and discuss its implications.

\subsection{Phase\texorpdfstring{$_1$}{1}: SLR results}
User trust in AI models is shaped by perceived signals and individual experiences, varying between groups~\cite{10.1145/3664646.3664757}. According to Liao et al. ~\cite{Liao_2022}, users can either strategically analyze LLM-generated output or rely on heuristics such as intuition for trust decisions. The goal is to cultivate calibrated trust among stakeholders, avoiding both blind acceptance and unwarranted mistrust, ensuring that trust judgments are well-founded rather than based on flawed intuition.

When users lack the motivation or ability to perform in-depth evaluations, they often make trust judgments based on surface-level cues or general impressions. Heuristics may sometimes be an intentional choice, but trust is often influenced by subtle, difficult-to-articulate factors often linked to the'sixth sense' ~\cite{10.1145/3411764.3445650}. Several heuristic-based trust factors have been identified in the literature.

The scope of the task significantly influences the trust in LLM-generated suggestions. Developers were less likely to accept multiline suggestions for smaller code changes, which are often more straightforward and may require less assistance. In contrast, larger code changes were associated with higher acceptance rates for multiline suggestions~\cite{10.1145/3664646.3664757}. One hypothesis is that smaller changes provide developers with clearer ideas about their intended functionality, reducing the perceived need for AI assistance. Furthermore, developers were significantly less likely to accept suggestions when editing test files. This may be because test code is often restricted to specific goals that AI suggestions may not align with, or because the suggestions might not adhere to local codebase practices, instead favoring global best practices~\cite{10.1145/3664646.3664757}.

On an individual level, a person’s experience in SE and LLMs, along with their disposition or mindset, influences the judgement of trust. Developers with recent experience using an LLM-generated programming language are more likely to trust suggestions in that language ~\cite{10.1145/3664646.3664757}. Interestingly, novice programmers tend to show higher trust in the security of machine-generated code than in human-generated code ~\cite{Perry_2023}. Developers trained to follow specific style guidelines for readability in a programming language are more likely to accept multi-line suggestions generated by LLMs ~\cite{10.1145/3664646.3664757}. Furthermore, positive user experiences with other models can increase their trust in new models, with their general tendency to trust further shaping their judgments ~\cite{9582305}

Sociological aspects of the model and its building also play a role in trust judgment. People pay attention to the reputation of the company or institution that developed the model~\cite{noller2022trustenhancementissuesprogram,cheng2023itworktooonline,10.1109/ICSE-SEIP58684.2023.00043}. The amount of institutional investment is also a factor that users believe indicates the trustworthiness of the model~\cite{10.1145/3411764.3445650}. Furthermore, the population of model users can lead to trust from people, as a larger number of users offer people assurance that the model is under many levels of supervision, encouraging their belief in the provenance of the security of the model~\cite{10.1145/3411764.3445650}.

Several factors associated with functionality are positively valued by users. The model is preferably easy to use~\cite{10.1145/3411764.3445650}. For example, if the model’s features are intuitive or similar to the usage of other popular models, this allows users to operate within their comfort zone and reduces cognitive burden~\cite{10.1145/3411764.3445650}. In addition, the ease of integration of the model with the existing user workflow and alignment with the developer’s coding style directly help build trust~\cite{10.1109/ICSE-SEIP58684.2023.00043}. A polished interface design also leads to positive trust building~\cite{10.1109/ICSE-SEIP58684.2023.00043}.

During interactions, users often employ heuristics to evaluate the model. For example, a sense of control over the model’s outputs and autonomy in decision-making are highly valued. Users who can directly manipulate the code and make choices about the generated results are more likely to trust the model ~\cite{10.1109/ICSE-SEIP58684.2023.00043}. Over time, as users accumulate experience and exposure to the model, they become better equipped to assess its output, further strengthening trust ~\cite{10.1145/3411764.3445650}. The ability of the model to generate predictable and consistent results also plays a crucial role in trust. Although users may be resilient to occasional errors, they tend to lose trust in models that produce unpredictable or surprising results ~\cite{10.1145/3411764.3445650}.

\subsection{Phase\texorpdfstring{$_2$}{2}: Survey Study Results}

An overwhelming number of participants (96\%) believe that multi-item metrics are a more effective indicator of trust than single-item metrics. This confirms our belief that trust is a multidimensional construct that must be captured more comprehensively.


\textbf{Trust Metric and Measurement.} For all three SE tasks, we observed that the precision of the results is a highly valued aspect. Accuracy is the metric most frequently mentioned by our respondents; 38\% of the people who answered the question mentioned accuracy in their responses. The emphasis on accuracy highlights that users expect LLMs to deliver precise and dependable results, as reflected in the participant's quote \textit{ "If the generated output is flawed, there would be no point in using the model"}. This expectation of output correctness aligns with the critical nature of SE, where flawed output can cause significant issues.

Usability cited by 23.8\% of participants is the second most mentioned metric across tasks. It is assessed by the volume of correctly functioning code generated over time with minimal modifications needed. Users gauge trustworthiness by how well the model meets requirements without needing further changes. Usability encompasses the comprehensive fulfillment of user needs, including high performance and alignment with specifications. Additionally, although LLMs are capable of generating large programs over a short period, this needs to be balanced by the \textit{"amount of time lost trying to get a good response or wasted in debugging errors that it introduces."} With increased usability, less time needs to be spent on modifying the code, thus boosting overall productivity.

Another interesting result common across tasks is how people judge their trust in the model. This judgment is often based on the likelihood that the author accepts the generated output without fully understanding or carefully scrutinizing it. The respondents believe that their trust is related to their motivation to review the output. As their trust in LLMs increases, they are less likely to engage in systematic processing to gauge the model's trustworthiness, including its ability, benevolence, and process integrity. Interestingly, these were mostly echoed by novice developers. 

For Code Generation, several ability-related metrics were identified. Comprehensibility refers to the model's ability to form logical reasoning through neural networks. Understandability refers to the ease of understanding the code, and participants note that their trust \textit{"is largely tied to measures of my understanding of the generated code."} Benevolence is another metric mentioned by users, referring to the notion that \textit{"the generated code is not maliciously trying to cause harm."}

For Program Repair, ethical compliance was mentioned as an important aspect by one of the respondents. They emphasized the importance of the generated repair following industry standards. The respondent also mentioned that their trust would be measured by the frequency of usage.

For Test Case Generation, one metric mentioned is comprehensiveness, the ability of the model to generate all possible test cases. Additionally, if the model can verify the user's doubts, it would be deemed more trustworthy. As one respondent stated, \textit{"I would trust the test cases more if they prove that my doubts about correct syntax and calling conventions do not apply."}

\textbf{Reviewing Human versus LLM-Generated Code.} Trust measurement procedures commonly include manual code reviews and running tests. Our survey revealed that 80.6\% of respondents often review generated code, 16.6\% rarely review it, and only 2.8\% never review it. Participants often compared \llms with human developers, suggesting that models acting similarly to human experts are considered more trustworthy. 

For code generation, 40\% of the participants distinguish between reviewing LLM-generated code and human-generated code. LLMs often generate code with unconventional syntax, style, and functionality, which requires more time to comprehend and integrate. One participant reported that LLMs occasionally generate \textit{"meaningless code"}, whereas \textit{"code written by humans is usually functionally done, barring minor bugs"}. However, participants acknowledge that LLM-generated code in smaller chunks can be faster to read and process.

For the generation of test cases, 25\% of participants differentiate their review process between LLM-generated code and human-generated code, with LLM-generated tests requiring stricter scrutiny.

For program repair, 25\% of the participants distinguish between LLM-generated code and human-generated code reviews. Participants mentioned that the repaired \textit{"can often give wildly incorrect approaches (both in function and style)."} Furthermore, the model may produce false positive output, where the justification seems reasonable, but the change does not make sense. These issues require a thorough review of LLM-generated repairs. In contrast, users tend to automatically attribute more credibility to humans, assuming fewer significant flaws in their approach, unless the individual has a reputation for errors.

\subsection{Broader Analysis of Trust Metric}
Measuring trust in LLMs for SE is a complex challenge, as it encompasses both technical and human-centric dimensions. In AI research, trust assessment varies by domain, reflecting different priorities and risks. In SE and machine learning engineering, trust metrics are predominantly technical, emphasizing benchmark-driven evaluations and reproducibility ~\cite{ahuja2020openingsoftwareengineeringtoolbox}. Accuracy is often measured through adversarial testing frameworks such as FFT ~\cite{cui2024fftharmlessnessevaluationanalysis}, while TRUSTGPT assesses factual correctness by analyzing toxicity, bias, and alignment of values ~\cite{huang2023trustgptbenchmarktrustworthyresponsible}. Robustness is evaluated through adversarial success rates ~\cite{wang2022adversarialgluemultitaskbenchmark}, performance degradation from synonym substitutions ~\cite{zhu2024promptrobustevaluatingrobustnesslarge}, and handling of noisy inputs ~\cite{wang2024decodingtrustcomprehensiveassessmenttrustworthiness}.

Accountability metrics in SE focus on tracking model versions, training data, and decision-making processes via immutable logs ~\cite{deLaat2018}, while external audits further enhance reliability ~\cite{deLaat2018}. Fairness, a growing concern in AI, is assessed using adversarial testing to detect biases in outputs ~\cite{cui2024fftharmlessnessevaluationanalysis, liu2023uncoveringquantifyingsocialbiases}, with frameworks such as TRUSTGPT standardizing toxicity and bias evaluations ~\cite{huang2023trustgptbenchmarktrustworthyresponsible}.

In contrast, HCI research prioritizes user-centric trust metrics, relying on qualitative methods such as Likert scale assessments ~\cite{Baughan_2023, 10.1145/3290605.3300469, 10494274} and trust recovery studies ~\cite{Baughan_2023}. Transparency is evaluated through structured frameworks such as POLARIS, which incorporates transparency checks throughout the software development cycle ~\cite{Baldassarre_2024}. Furthermore, methods such as LIME assess the fidelity of local explanations to the overall behavior of the model, ensuring the alignment between interpretability and performance ~\cite{liu2021trustworthyaicomputationalperspective}.

Despite these advancements, existing frameworks rarely integrate user-centered and technical trust metrics ~\cite{Baldassarre_2024}. This gap is problematic, as trust cannot be fully established if AI systems are technically robust but socially distrusted, or user-friendly but functionally unreliable. In addition, trust measurement frameworks often overlook the temporal dynamics of trust, which evolves over time based on user experience, failure rates, and behavioral consistency ~\cite{Lee2004TrustIA, Lewis2018}. In domains such as robotics and education, trust is assessed longitudinally, capturing how user confidence fluctuates with repeated interactions. A similar approach is needed for SE, where trust in LLMs is influenced by ongoing exposure and reliability trends.

\subsection{Discussion}
Our study highlights the need for a comprehensive framework to measure trust in LLM4SE, as existing approaches do not account for its multifaceted nature. Although technical metrics such as accuracy, robustness, and accountability are crucial, our survey results indicate that trust also depends on usability, comprehensibility, and ethical considerations. Developers emphasized that trust is not solely determined by correctness, but also by how seamlessly LLMs integrate into their workflows. The need for minimal debugging and rework reinforces the importance of trust metrics that capture practical usability alongside model performance.

The survey findings also emphasize that the measurement of trust should be multidimensional. A strong preference (96\% of participants) for multi-item trust metrics aligns with broader research indicating that single-item trust measures oversimplify complex trust dynamics. This is particularly critical in SE, where small errors can lead to security vulnerabilities and software failures. A multidimensional approach that incorporates accuracy, usability, ethical compliance, and consistency provides a more holistic assessment of trustworthiness.

Another major gap in current trust measurement frameworks is their inability to capture the temporal evolution of trust. Our findings indicate that trust in LLMs fluctuates over time, shaped by output quality and the extent to which developers feel they can rely on generated suggestions. This aligns with research in robotics and education, where trust is measured longitudinally through failure rates and behavioral consistency ~\cite{Lee2004TrustIA, Lewis2018}. Future SE trust assessment frameworks should go beyond static evaluations and incorporate temporal metrics that track trust development over prolonged interactions with LLMs.

These insights suggest that an effective trust measurement framework for LLM4SE must integrate both technical and user-centric perspectives, consider multidimensional factors, and account for the evolving nature of trust. Addressing these gaps will enable more reliable and user-aligned assessments, ultimately fostering greater confidence in LLMs for software engineering tasks.

\subsection{Opportunities for Future Work}
The results of this study present several avenues for future research aimed at enhancing the metrics of trust in LLM4SE. One crucial avenue of future research based on our survey study is the development of a multi-item trust framework. We propose constructing a multi-item ``LLM4SE Trustworthiness Index'' that integrates objective performance metrics with user-centric evaluations. For example, based on our survey findings, key dimensions like Functional Correctness, Security \& System Integrity, Understandability, and Usability would be prioritized. Operational examples for these metrics could include the following: Functional Correctness: Measured by automated test suite pass rates on LLM-generated code, or the percentage of generated code snippets that achieve desired functionality without modification. Security \& System Integrity: Assessed by the rate of security vulnerabilities (e.g., using static analysis tools on generated code), or the frequency of nonmalicious yet harmful outputs. Understandability/Interpretability: Quantified through user ratings on code clarity, the time required for developers to comprehend and verify LLM-generated solutions for critical tasks or/and through interpretability techniques such as Code Rationale \cite{palacio2025explaininglargelanguagemodels}, Shap \cite{lundberg_unified_nodate}, and attention-based scores\cite{mohankumar_towards_2020}. Usability/Workflow Integration: Evaluated using metrics such as reduction in manual debugging effort, time saved in task completion, or adherence to existing coding standards and project conventions. These individual metrics, drawn from both technical analysis and structured user feedback, would contribute to a composite trustworthiness score. Such an index could be task-specific, with different weightings for each dimension based on the task's criticality (e.g., a higher weight for security in program repair vs. boilerplate generation). Furthermore, a robust framework would track these metrics longitudinally, allowing real-time calibration of user trust based on observed performance over time and across diverse contexts. 

To effectively establish and validate such a comprehensive metric, a key future direction involves conducting causal analysis to determine the precise relative importance and interdependencies of each contributing factor. Understanding how different dimensions of trustworthiness (e.g., accuracy, interpretability, security) causally influence overall user trust will enable the development of more targeted and impactful trust-building interventions.

Trust in LLMs evolves over time, influenced by repeated interactions and exposure to the model. Future research should explore longitudinal studies that track how trust builds or erodes as users gain experience with the model. These studies could examine changes in trust based on output quality, model performance over time, and user feedback. Understanding the temporal dynamics of trust would help in designing models that foster long-term user confidence, making them more effective and reliable for continuous use in SE tasks. Additionally, collecting extensive user-centric data from a larger, more diverse population will be necessary to ensure the generalizability and robustness of these measurements across various SE contexts and practitioner demographics. This scaled data collection will be crucial for validating proposed metrics and understanding how trust dynamics vary across different expertise levels, organizational cultures, and geographical locations.

Context-specific trust metrics are another crucial area for future work. Trust may be prioritized differently in tasks such as code generation, program repair, or test case generation, where each task involves distinct expectations and challenges. For example, code generation can emphasize accuracy, ethical considerations, and contextual relevance, while program repair may require metrics that account for error minimization, maintainability, and adherence to coding standards. Tailored trust metrics will ensure that the measurement framework is aligned with the specific needs of each task and provides more meaningful insights into trust.

An important avenue for future research is trust calibration, which ensures that users' trust levels align with the model's actual performance. Over-reliance on faulty outputs can lead to critical errors, while undertrusting a model may result in missed opportunities and efficiency gains. Future research could investigate feedback mechanisms that help users understand why a model generated certain outputs or offer insights into the decision-making process. Transparency tools, such as model explanations and uncertainty indicators, would allow users to better align their trust with the reliability of the model. This would be especially valuable in high-stakes SE tasks, where the consequences of errors can be significant.

Future work should therefore focus on developing sophisticated feedback mechanisms that empower developers to effectively interpret LLM outputs, moving beyond mere acceptance or rejection. This involves designing tools that provide actionable insights into the LLM's reasoning and confidence levels, allowing practitioners to understand why a particular output was generated and how reliable it is for their specific task. Developing concrete recommendations for SE practitioners and tool builders regarding dynamic trust adjustment is crucial. For practitioners, this implies establishing heuristics for increasing scrutiny (e.g., for outputs in high-risk tasks or when confidence indicators are low) versus increasing reliance (e.g., for boilerplate code or outputs consistently meeting high-performance benchmarks). Tool builders should focus on integrating context-aware transparency tools and uncertainty indicators directly into SE environments, enabling dynamic trust alignment based on observed metrics.

\begin{tcolorbox}[colframe=black!50, colback=gray!10, title=Summary of $RQ_3$: Trust Metrics]
We found that trust in LLM4SE is a multidimensional construct influenced by factors such as accuracy, usability, comprehensibility, and ethical compliance. A key finding from the user study revealed that 96\% of the participants preferred multi-item trust metrics over single-item ones, emphasizing the need to capture trust comprehensively. User trust in LLMs for SE is influenced by both technical and human-centric factors. Accuracy and usability emerge as primary trust metrics, with accuracy being the most frequently cited criterion. Developers also highlighted the importance of comprehensibility and ethical compliance, particularly for tasks like program repair and test case generation. Sociological factors, such as the reputation of the model's developers and its user base, further influence trust perceptions. The findings underscore the necessity of developing multidimensional trust measurement frameworks that take into account both technical performance and user-centric dimensions to better evaluate and foster trust in LLMs.
\end{tcolorbox}

%% file: text/9_threats.tex
\section{Threats to Validity}
We conducted our literature review following the Kithenham et al. guidelines ~\cite{barabaraCit} and our survey study using purposive sampling ~\cite{baltes2021sampling}. However, limitations may exist in our search strategy, data analysis, and participant recruitment.

\textbf{External Threats.} One potential threat to the generalizability of our results is our search strategy. Our search string might have missed some studies. To mitigate this threat, we tried multiple search strategy: 1) \textit{"Large Language Model" OR "Software Engineering" OR "Trust"}, 2) \textit{"Software Engineering" OR "SE" OR "LLM" OR "Large Language Model" OR "Trust"}, 3) \textit{"Large Language Model" AND "Software Enginnering" AND "Trust"}, 4) \textit{"LLM" OR "SE" OR "Trust" OR "Distrust" OR "Trustworthiness"}, 5) \textit{"LLM" OR "Large Language Model" OR "SE" OR "Software Engineering" OR "Trust" OR "Distrust" OR "Trustworthiness"}, 6) \textit{(``trust'' OR ``distrust'' OR ``trustworthiness'') AND (``SE'' OR ``Software Engineering'') AND (``LLM'' OR ``Large Language Model'' OR ``LLMs'' OR ``Large Language Models'' OR ``Deep Learning'' OR ``Machine Learning'')}, 7) \textit{(“trust” OR “distrust” OR “trustworthiness”) AND (“SE” OR “Software Engineering” OR “Software Development” OR “Code Generation” OR “Program Repair” OR "Automatic Program Repair" OR “Software Testing” OR “Test Generation”) AND (“LLM” OR “Large Language Model” OR “LLMs” OR “Large Language Models” OR “Deep Learning” OR “Machine Learning”)}. Search string 7 returned the most results. Although this increased the time required to filter out unrelated papers, it reduced selection bias due to the large pool of results. Another threat might arise from our inclusion and exclusion criteria. To remove biases, we predefined our criteria and independently conducted the methodology by two authors. Furthermore, we also conducted a complementary analysis of $70$ articles that were filtered from our exclusion criteria, but contained trust concepts from the broader literature.

Although our user study consists of $25$ practitioners, it is important to note that this reflects a highly specialized group of individuals with direct experience in LLMs for SE tasks. Given the niche nature of LLM applications in SE, identifying qualified participants is inherently challenging. Similar SE studies have successfully used small targeted samples~\cite{krishna2024disagreementproblemexplainablemachine, cotealllard2019deeplearningelectromyographichand}. Furthermore, filtering questions ensured that participants were qualified, and only qualifying responses were used for analysis. We also published our responses in an online appendix~\cite{AnonymousRepoTrustSurvey}. Although we acknowledge that larger-scale validation would be beneficial, this study provides an essential first step in mapping the trust landscape of LLMs in SE, laying the groundwork for future research. Future studies should aim to conduct larger surveys with more diverse samples, such as testers, developers, and managers, to enable more generalizable insights.

\textbf{Internal Threats.} The internal threat stems from the derived taxonomy characterizing definitions, antecedents, and measurement of trust. To mitigate this, we followed a process inspired by open coding in constructive grounded theory ~\cite{kathy2006}, where every attribute in our taxonomy was reviewed by four authors. Our taxonomy is also available in an online appendix ~\cite{AnonymousRepoTrustSurvey}.

%% file: text/12_conclusion.tex
\section{Conclusion}
We center the motivation for this study around a fundamental question: \textit{Why is trust research essential in LLM4SE?} To answer this, we conducted a comprehensive literature review, collected practitioners' feedback through a survey study, and analyzed the broader trust literature. This approach allowed us to identify key gaps in the way trust is defined, what factors shape it, and how it is measured. Our work provides a structured understanding of these concepts, mapping the current research landscape, and highlighting their critical role in shaping LLM adoption in SE workflows.

Our findings reveal that trust in LLMs is neither absolute nor universal; it is highly task dependent and must be carefully calibrated. Developers who overtrust LLM-generated code risk introducing security vulnerabilities, code smells, and data leaks, while excessive distrust limits the potential benefits of automation and efficiency. We show that trust is influenced by multiple factors, including accuracy, robustness, interpretability, and ethical considerations. However, trust alone is not enough; trustworthiness, as an inherent property of an LLM, must be systematically evaluated through well-defined metrics rather than inferred from user perceptions alone.

By highlighting the need for structured trust assessment frameworks, our study lays the groundwork for future research on trust-aware LLM integration. As LLMs continue to shape modern software development, it will be essential to ensure trust is well calibrated to maximize their benefits while mitigating risks. Moving forward, researchers should focus on developing standardized trust metrics and improving transparency mechanisms to align LLM capabilities with developer expectations, ensuring their responsible and effective adoption in SE.

Our research shows that trust in LLM4SE is a multidisciplinary concept that requires collaborative efforts across fields to be fully understood and effectively operationalized. Insights from HCI can support the design of more intuitive trust affordances and interactive workflows. AI ethics can guide frameworks for fairness, accountability, and transparency in model behavior. Cognitive science can offer a deeper understanding of how trust is formed, calibrated, and disrupted in developer–LLM interactions. Incorporating these perspectives will help establish more robust and human-aligned approaches to trust in LLM-supported software engineering.

%% file: sample-acmsmall.bbl

\begin{thebibliography}{120}


\ifx \showCODEN    \undefined \def \showCODEN     #1{\unskip}     \fi
\ifx \showDOI      \undefined \def \showDOI       #1{#1}\fi
\ifx \showISBNx    \undefined \def \showISBNx     #1{\unskip}     \fi
\ifx \showISBNxiii \undefined \def \showISBNxiii  #1{\unskip}     \fi
\ifx \showISSN     \undefined \def \showISSN      #1{\unskip}     \fi
\ifx \showLCCN     \undefined \def \showLCCN      #1{\unskip}     \fi
\ifx \shownote     \undefined \def \shownote      #1{#1}          \fi
\ifx \showarticletitle \undefined \def \showarticletitle #1{#1}   \fi
\ifx \showURL      \undefined \def \showURL       {\relax}        \fi
\providecommand\bibfield[2]{#2}
\providecommand\bibinfo[2]{#2}
\providecommand\natexlab[1]{#1}
\providecommand\showeprint[2][]{arXiv:#2}

\bibitem[EUw({[n.\,d.]})]%
        {EUwebsite}
 \bibinfo{year}{[n.\,d.]}\natexlab{}.
\newblock
\newblock
\urldef\tempurl%
\url{https://digital-strategy.ec.europa.eu/en/library/ethics-guidelines-trustworthy-ai}
\showURL{%
\tempurl}


\bibitem[noa({[n.\,d.]})]%
        {noauthor_qualtrics_nodate}
 \bibinfo{year}{[n.\,d.]}\natexlab{}.
\newblock \bibinfo{title}{Qualtrics {XM}: {The} {Leading} {Experience} {Management} {Software}}.
\newblock
\newblock
\urldef\tempurl%
\url{https://www.qualtrics.com/}
\showURL{%
\tempurl}


\bibitem[Ahuja et~al\mbox{.}(2020)]%
        {ahuja2020openingsoftwareengineeringtoolbox}
\bibfield{author}{\bibinfo{person}{Mohit~Kumar Ahuja}, \bibinfo{person}{Mohamed-Bachir Belaid}, \bibinfo{person}{Pierre Bernabé}, \bibinfo{person}{Mathieu Collet}, \bibinfo{person}{Arnaud Gotlieb}, \bibinfo{person}{Chhagan Lal}, \bibinfo{person}{Dusica Marijan}, \bibinfo{person}{Sagar Sen}, \bibinfo{person}{Aizaz Sharif}, {and} \bibinfo{person}{Helge Spieker}.} \bibinfo{year}{2020}\natexlab{}.
\newblock \bibinfo{title}{Opening the Software Engineering Toolbox for the Assessment of Trustworthy AI}.
\newblock
\newblock
\showeprint[arxiv]{2007.07768}~[cs.SE]
\urldef\tempurl%
\url{https://arxiv.org/abs/2007.07768}
\showURL{%
\tempurl}


\bibitem[Aljohani et~al\mbox{.}(2025)]%
        {aljohani2025comprehensivesurveytrustworthinesslarge}
\bibfield{author}{\bibinfo{person}{Manar Aljohani}, \bibinfo{person}{Jun Hou}, \bibinfo{person}{Sindhura Kommu}, {and} \bibinfo{person}{Xuan Wang}.} \bibinfo{year}{2025}\natexlab{}.
\newblock \bibinfo{title}{A Comprehensive Survey on the Trustworthiness of Large Language Models in Healthcare}.
\newblock
\newblock
\showeprint[arxiv]{2502.15871}~[cs.CY]
\urldef\tempurl%
\url{https://arxiv.org/abs/2502.15871}
\showURL{%
\tempurl}


\bibitem[Amoozadeh et~al\mbox{.}(2024)]%
        {10.1145/3626252.3630842}
\bibfield{author}{\bibinfo{person}{Matin Amoozadeh}, \bibinfo{person}{David Daniels}, \bibinfo{person}{Daye Nam}, \bibinfo{person}{Aayush Kumar}, \bibinfo{person}{Stella Chen}, \bibinfo{person}{Michael Hilton}, \bibinfo{person}{Sruti Srinivasa~Ragavan}, {and} \bibinfo{person}{Mohammad~Amin Alipour}.} \bibinfo{year}{2024}\natexlab{}.
\newblock \showarticletitle{Trust in Generative AI among Students: An exploratory study}. In \bibinfo{booktitle}{\emph{Proceedings of the 55th ACM Technical Symposium on Computer Science Education V. 1}} (Portland, OR, USA) \emph{(\bibinfo{series}{SIGCSE 2024})}. \bibinfo{publisher}{Association for Computing Machinery}, \bibinfo{address}{New York, NY, USA}, \bibinfo{pages}{67–73}.
\newblock
\showISBNx{9798400704239}
\urldef\tempurl%
\url{https://doi.org/10.1145/3626252.3630842}
\showDOI{\tempurl}


\bibitem[AnonymousRepoTrustRepo(2024)]%
        {AnonymousRepoTrustSurvey}
\bibfield{author}{\bibinfo{person}{AnonymousRepoTrustRepo}.} \bibinfo{year}{2024}\natexlab{}.
\newblock \bibinfo{title}{Trust in LLM in SE}.
\newblock
\newblock
\urldef\tempurl%
\url{https://anonymous.4open.science/r/Mapping-the-Trust-Terrain-LLMs-in-Software-Engineering-Insights-and-Perspectives-A667/README.md}
\showURL{%
\tempurl}
\newblock
\shownote{[Accessed 08-02-2024]}.


\bibitem[Balayn et~al\mbox{.}(2024)]%
        {balayn2024empiricalexplorationtrustdynamics}
\bibfield{author}{\bibinfo{person}{Agathe Balayn}, \bibinfo{person}{Mireia Yurrita}, \bibinfo{person}{Fanny Rancourt}, \bibinfo{person}{Fabio Casati}, {and} \bibinfo{person}{Ujwal Gadiraju}.} \bibinfo{year}{2024}\natexlab{}.
\newblock \bibinfo{title}{An Empirical Exploration of Trust Dynamics in LLM Supply Chains}.
\newblock
\newblock
\showeprint[arxiv]{2405.16310}~[cs.HC]
\urldef\tempurl%
\url{https://arxiv.org/abs/2405.16310}
\showURL{%
\tempurl}


\bibitem[Baldassarre et~al\mbox{.}(2024a)]%
        {Baldassarre_2024}
\bibfield{author}{\bibinfo{person}{Maria~Teresa Baldassarre}, \bibinfo{person}{Domenico Gigante}, \bibinfo{person}{Marcos Kalinowski}, {and} \bibinfo{person}{Azzurra Ragone}.} \bibinfo{year}{2024}\natexlab{a}.
\newblock \showarticletitle{POLARIS: A Framework to Guide the Development of Trustworthy AI Systems}. In \bibinfo{booktitle}{\emph{Proceedings of the IEEE/ACM 3rd International Conference on AI Engineering - Software Engineering for AI}} \emph{(\bibinfo{series}{CAIN 2024})}. \bibinfo{publisher}{ACM}, \bibinfo{pages}{200–210}.
\newblock
\urldef\tempurl%
\url{https://doi.org/10.1145/3644815.3644947}
\showDOI{\tempurl}


\bibitem[Baldassarre et~al\mbox{.}(2024b)]%
        {10.1145/3661167.3661214}
\bibfield{author}{\bibinfo{person}{Maria~Teresa Baldassarre}, \bibinfo{person}{Domenico Gigante}, \bibinfo{person}{Marcos Kalinowski}, \bibinfo{person}{Azzurra Ragone}, {and} \bibinfo{person}{Sara Tibid\`{o}}.} \bibinfo{year}{2024}\natexlab{b}.
\newblock \showarticletitle{Trustworthy AI in practice: an analysis of practitioners' needs and challenges}. In \bibinfo{booktitle}{\emph{Proceedings of the 28th International Conference on Evaluation and Assessment in Software Engineering}} (Salerno, Italy) \emph{(\bibinfo{series}{EASE '24})}. \bibinfo{publisher}{Association for Computing Machinery}, \bibinfo{address}{New York, NY, USA}, \bibinfo{pages}{293–302}.
\newblock
\showISBNx{9798400717017}
\urldef\tempurl%
\url{https://doi.org/10.1145/3661167.3661214}
\showDOI{\tempurl}


\bibitem[Baltes and Ralph(2021)]%
        {baltes2021sampling}
\bibfield{author}{\bibinfo{person}{Sebastian Baltes} {and} \bibinfo{person}{Paul Ralph}.} \bibinfo{year}{2021}\natexlab{}.
\newblock \bibinfo{title}{Sampling in Software Engineering Research: A Critical Review and Guidelines}.
\newblock
\newblock
\showeprint[arxiv]{2002.07764}


\bibitem[Banovic et~al\mbox{.}(2023)]%
        {10.1145/3579460}
\bibfield{author}{\bibinfo{person}{Nikola Banovic}, \bibinfo{person}{Zhuoran Yang}, \bibinfo{person}{Aditya Ramesh}, {and} \bibinfo{person}{Alice Liu}.} \bibinfo{year}{2023}\natexlab{}.
\newblock \showarticletitle{Being Trustworthy is Not Enough: How Untrustworthy Artificial Intelligence (AI) Can Deceive the End-Users and Gain Their Trust}.
\newblock \bibinfo{journal}{\emph{Proc. ACM Hum.-Comput. Interact.}} \bibinfo{volume}{7}, \bibinfo{number}{CSCW1}, Article \bibinfo{articleno}{27} (\bibinfo{date}{April} \bibinfo{year}{2023}), \bibinfo{numpages}{17}~pages.
\newblock
\urldef\tempurl%
\url{https://doi.org/10.1145/3579460}
\showDOI{\tempurl}


\bibitem[Baughan et~al\mbox{.}(2023)]%
        {Baughan_2023}
\bibfield{author}{\bibinfo{person}{Amanda Baughan}, \bibinfo{person}{Xuezhi Wang}, \bibinfo{person}{Ariel Liu}, \bibinfo{person}{Allison Mercurio}, \bibinfo{person}{Jilin Chen}, {and} \bibinfo{person}{Xiao Ma}.} \bibinfo{year}{2023}\natexlab{}.
\newblock \showarticletitle{A Mixed-Methods Approach to Understanding User Trust after Voice Assistant Failures}. In \bibinfo{booktitle}{\emph{Proceedings of the 2023 CHI Conference on Human Factors in Computing Systems}} \emph{(\bibinfo{series}{CHI ’23})}. \bibinfo{publisher}{ACM}, \bibinfo{pages}{1–16}.
\newblock
\urldef\tempurl%
\url{https://doi.org/10.1145/3544548.3581152}
\showDOI{\tempurl}


\bibitem[Bengio et~al\mbox{.}(2000)]%
        {NIPS2000_728f206c}
\bibfield{author}{\bibinfo{person}{Yoshua Bengio}, \bibinfo{person}{R\'{e}jean Ducharme}, {and} \bibinfo{person}{Pascal Vincent}.} \bibinfo{year}{2000}\natexlab{}.
\newblock \showarticletitle{A Neural Probabilistic Language Model}. In \bibinfo{booktitle}{\emph{Advances in Neural Information Processing Systems}}, \bibfield{editor}{\bibinfo{person}{T.~Leen}, \bibinfo{person}{T.~Dietterich}, {and} \bibinfo{person}{V.~Tresp}} (Eds.), Vol.~\bibinfo{volume}{13}. \bibinfo{publisher}{MIT Press}.
\newblock
\urldef\tempurl%
\url{https://proceedings.neurips.cc/paper_files/paper/2000/file/728f206c2a01bf572b5940d7d9a8fa4c-Paper.pdf}
\showURL{%
\tempurl}


\bibitem[Borg(2024)]%
        {borg2024trustcalibrationidespaving}
\bibfield{author}{\bibinfo{person}{Markus Borg}.} \bibinfo{year}{2024}\natexlab{}.
\newblock \bibinfo{title}{Trust Calibration in IDEs: Paving the Way for Widespread Adoption of AI Refactoring}.
\newblock
\newblock
\showeprint[arxiv]{2412.15948}~[cs.SE]
\urldef\tempurl%
\url{https://arxiv.org/abs/2412.15948}
\showURL{%
\tempurl}


\bibitem[Brown et~al\mbox{.}(2024)]%
        {10.1145/3664646.3664757}
\bibfield{author}{\bibinfo{person}{Adam Brown}, \bibinfo{person}{Sarah D'Angelo}, \bibinfo{person}{Ambar Murillo}, \bibinfo{person}{Ciera Jaspan}, {and} \bibinfo{person}{Collin Green}.} \bibinfo{year}{2024}\natexlab{}.
\newblock \showarticletitle{Identifying the Factors That Influence Trust in AI Code Completion}. In \bibinfo{booktitle}{\emph{Proceedings of the 1st ACM International Conference on AI-Powered Software}} (Porto de Galinhas, Brazil) \emph{(\bibinfo{series}{AIware 2024})}. \bibinfo{publisher}{Association for Computing Machinery}, \bibinfo{address}{New York, NY, USA}, \bibinfo{pages}{1–9}.
\newblock
\showISBNx{9798400706851}
\urldef\tempurl%
\url{https://doi.org/10.1145/3664646.3664757}
\showDOI{\tempurl}


\bibitem[Butler et~al\mbox{.}(2004)]%
        {10.1007/978-3-540-24747-0_25}
\bibfield{author}{\bibinfo{person}{Michael Butler}, \bibinfo{person}{Michael Leuschel}, \bibinfo{person}{St{\'e}phane~Lo Presti}, {and} \bibinfo{person}{Phillip Turner}.} \bibinfo{year}{2004}\natexlab{}.
\newblock \showarticletitle{The Use of Formal Methods in the Analysis of Trust (Position Paper)}. In \bibinfo{booktitle}{\emph{Trust Management}}, \bibfield{editor}{\bibinfo{person}{Christian Jensen}, \bibinfo{person}{Stefan Poslad}, {and} \bibinfo{person}{Theo Dimitrakos}} (Eds.). \bibinfo{publisher}{Springer Berlin Heidelberg}, \bibinfo{address}{Berlin, Heidelberg}, \bibinfo{pages}{333--339}.
\newblock
\showISBNx{978-3-540-24747-0}


\bibitem[Cabrero-Daniel and Sanagust\'{\i}n~Cabrero(2023)]%
        {10.1145/3597512.3599715}
\bibfield{author}{\bibinfo{person}{Beatriz Cabrero-Daniel} {and} \bibinfo{person}{Andrea Sanagust\'{\i}n~Cabrero}.} \bibinfo{year}{2023}\natexlab{}.
\newblock \showarticletitle{Perceived Trustworthiness of Natural Language Generators}. In \bibinfo{booktitle}{\emph{Proceedings of the First International Symposium on Trustworthy Autonomous Systems}} (Edinburgh, United Kingdom) \emph{(\bibinfo{series}{TAS '23})}. \bibinfo{publisher}{Association for Computing Machinery}, \bibinfo{address}{New York, NY, USA}, Article \bibinfo{articleno}{23}, \bibinfo{numpages}{9}~pages.
\newblock
\showISBNx{9798400707346}
\urldef\tempurl%
\url{https://doi.org/10.1145/3597512.3599715}
\showDOI{\tempurl}


\bibitem[Charmaz(2006)]%
        {kathy2006}
\bibfield{author}{\bibinfo{person}{Kathy Charmaz}.} \bibinfo{year}{2006}\natexlab{}.
\newblock \bibinfo{booktitle}{\emph{Constructing Grounded Theory: A Practical Guide through Qualitative Analysis}}.
\newblock \bibinfo{publisher}{SAGE Publications Inc.}
\newblock


\bibitem[Chen and Hsieh(2023)]%
        {Chen_Hsieh_2023}
\bibfield{author}{\bibinfo{person}{Pin-Yu Chen} {and} \bibinfo{person}{Cho-Jui Hsieh}.} \bibinfo{year}{2023}\natexlab{}.
\newblock \bibinfo{booktitle}{\emph{Adversarial robustness for machine learning}}.
\newblock \bibinfo{publisher}{Academic Press, an imprint of Elsevier}.
\newblock


\bibitem[Cheng et~al\mbox{.}(2023)]%
        {cheng2023itworktooonline}
\bibfield{author}{\bibinfo{person}{Ruijia Cheng}, \bibinfo{person}{Ruotong Wang}, \bibinfo{person}{Thomas Zimmermann}, {and} \bibinfo{person}{Denae Ford}.} \bibinfo{year}{2023}\natexlab{}.
\newblock \bibinfo{title}{"It would work for me too": How Online Communities Shape Software Developers' Trust in AI-Powered Code Generation Tools}.
\newblock
\newblock
\showeprint[arxiv]{2212.03491}~[cs.HC]
\urldef\tempurl%
\url{https://arxiv.org/abs/2212.03491}
\showURL{%
\tempurl}


\bibitem[Choung et~al\mbox{.}(2022)]%
        {Choung_2022}
\bibfield{author}{\bibinfo{person}{Hyesun Choung}, \bibinfo{person}{Prabu David}, {and} \bibinfo{person}{Arun Ross}.} \bibinfo{year}{2022}\natexlab{}.
\newblock \showarticletitle{Trust in AI and Its Role in the Acceptance of AI Technologies}.
\newblock \bibinfo{journal}{\emph{International Journal of Human–Computer Interaction}} \bibinfo{volume}{39}, \bibinfo{number}{9} (\bibinfo{date}{April} \bibinfo{year}{2022}), \bibinfo{pages}{1727–1739}.
\newblock
\showISSN{1532-7590}
\urldef\tempurl%
\url{https://doi.org/10.1080/10447318.2022.2050543}
\showDOI{\tempurl}


\bibitem[Corley et~al\mbox{.}(2015)]%
        {feature_location}
\bibfield{author}{\bibinfo{person}{Christopher~S. Corley}, \bibinfo{person}{Kostadin Damevski}, {and} \bibinfo{person}{Nicholas~A. Kraft}.} \bibinfo{year}{2015}\natexlab{}.
\newblock \showarticletitle{Exploring the use of deep learning for feature location}. In \bibinfo{booktitle}{\emph{2015 IEEE International Conference on Software Maintenance and Evolution (ICSME)}}. \bibinfo{pages}{556--560}.
\newblock
\urldef\tempurl%
\url{https://doi.org/10.1109/ICSM.2015.7332513}
\showDOI{\tempurl}


\bibitem[Cui et~al\mbox{.}(2024)]%
        {cui2024fftharmlessnessevaluationanalysis}
\bibfield{author}{\bibinfo{person}{Shiyao Cui}, \bibinfo{person}{Zhenyu Zhang}, \bibinfo{person}{Yilong Chen}, \bibinfo{person}{Wenyuan Zhang}, \bibinfo{person}{Tianyun Liu}, \bibinfo{person}{Siqi Wang}, {and} \bibinfo{person}{Tingwen Liu}.} \bibinfo{year}{2024}\natexlab{}.
\newblock \bibinfo{title}{FFT: Towards Harmlessness Evaluation and Analysis for LLMs with Factuality, Fairness, Toxicity}.
\newblock
\newblock
\showeprint[arxiv]{2311.18580}~[cs.CL]
\urldef\tempurl%
\url{https://arxiv.org/abs/2311.18580}
\showURL{%
\tempurl}


\bibitem[Côté-Allard et~al\mbox{.}(2019)]%
        {cotealllard2019deeplearningelectromyographichand}
\bibfield{author}{\bibinfo{person}{Ulysse Côté-Allard}, \bibinfo{person}{Cheikh~Latyr Fall}, \bibinfo{person}{Alexandre Drouin}, \bibinfo{person}{Alexandre Campeau-Lecours}, \bibinfo{person}{Clément Gosselin}, \bibinfo{person}{Kyrre Glette}, \bibinfo{person}{François Laviolette}, {and} \bibinfo{person}{Benoit Gosselin}.} \bibinfo{year}{2019}\natexlab{}.
\newblock \bibinfo{title}{Deep Learning for Electromyographic Hand Gesture Signal Classification Using Transfer Learning}.
\newblock
\newblock
\showeprint[arxiv]{1801.07756}~[cs.LG]
\urldef\tempurl%
\url{https://arxiv.org/abs/1801.07756}
\showURL{%
\tempurl}


\bibitem[de~Laat(2018)]%
        {deLaat2018}
\bibfield{author}{\bibinfo{person}{Paul~B. de Laat}.} \bibinfo{year}{2018}\natexlab{}.
\newblock \showarticletitle{Algorithmic Decision-Making Based on Machine Learning from Big Data: Can Transparency Restore Accountability?}
\newblock \bibinfo{journal}{\emph{Philosophy \& Technology}}  \bibinfo{volume}{31} (\bibinfo{year}{2018}), \bibinfo{pages}{525--541}.
\newblock
\urldef\tempurl%
\url{https://doi.org/10.1007/s13347-017-0293-z}
\showDOI{\tempurl}


\bibitem[Degachi et~al\mbox{.}(2024)]%
        {10.1145/3613905.3650825}
\bibfield{author}{\bibinfo{person}{Chadha Degachi}, \bibinfo{person}{Siddharth Mehrotra}, \bibinfo{person}{Mireia Yurrita}, \bibinfo{person}{Evangelos Niforatos}, {and} \bibinfo{person}{Myrthe~Lotte Tielman}.} \bibinfo{year}{2024}\natexlab{}.
\newblock \showarticletitle{Practising Appropriate Trust in Human-Centred AI Design}. In \bibinfo{booktitle}{\emph{Extended Abstracts of the CHI Conference on Human Factors in Computing Systems}} (Honolulu, HI, USA) \emph{(\bibinfo{series}{CHI EA '24})}. \bibinfo{publisher}{Association for Computing Machinery}, \bibinfo{address}{New York, NY, USA}, Article \bibinfo{articleno}{269}, \bibinfo{numpages}{8}~pages.
\newblock
\showISBNx{9798400703317}
\urldef\tempurl%
\url{https://doi.org/10.1145/3613905.3650825}
\showDOI{\tempurl}


\bibitem[Dietz and Hartog(2006)]%
        {Dietz2006MeasuringTI}
\bibfield{author}{\bibinfo{person}{Graham Dietz} {and} \bibinfo{person}{Deanne N.~Den Hartog}.} \bibinfo{year}{2006}\natexlab{}.
\newblock \showarticletitle{Measuring trust inside organisations}.
\newblock \bibinfo{journal}{\emph{Personnel Review}}  \bibinfo{volume}{35} (\bibinfo{year}{2006}), \bibinfo{pages}{557--588}.
\newblock
\urldef\tempurl%
\url{https://api.semanticscholar.org/CorpusID:59142486}
\showURL{%
\tempurl}


\bibitem[Du et~al\mbox{.}(2025)]%
        {du2025faircoderevaluatingsocialbias}
\bibfield{author}{\bibinfo{person}{Yongkang Du}, \bibinfo{person}{Jen tse Huang}, \bibinfo{person}{Jieyu Zhao}, {and} \bibinfo{person}{Lu Lin}.} \bibinfo{year}{2025}\natexlab{}.
\newblock \bibinfo{title}{FairCoder: Evaluating Social Bias of LLMs in Code Generation}.
\newblock
\newblock
\showeprint[arxiv]{2501.05396}~[cs.CL]
\urldef\tempurl%
\url{https://arxiv.org/abs/2501.05396}
\showURL{%
\tempurl}


\bibitem[Edmondson(1999)]%
        {doi:10.2307/2666999}
\bibfield{author}{\bibinfo{person}{Amy Edmondson}.} \bibinfo{year}{1999}\natexlab{}.
\newblock \showarticletitle{Psychological Safety and Learning Behavior in Work Teams}.
\newblock \bibinfo{journal}{\emph{Administrative Science Quarterly}} \bibinfo{volume}{44}, \bibinfo{number}{2} (\bibinfo{year}{1999}), \bibinfo{pages}{350--383}.
\newblock
\urldef\tempurl%
\url{https://doi.org/10.2307/2666999}
\showDOI{\tempurl}
\showeprint{https://journals.sagepub.com/doi/pdf/10.2307/2666999}


\bibitem[Ferrario and Loi(2022)]%
        {10.1145/3531146.3533202}
\bibfield{author}{\bibinfo{person}{Andrea Ferrario} {and} \bibinfo{person}{Michele Loi}.} \bibinfo{year}{2022}\natexlab{}.
\newblock \showarticletitle{How Explainability Contributes to Trust in AI}. In \bibinfo{booktitle}{\emph{Proceedings of the 2022 ACM Conference on Fairness, Accountability, and Transparency}} (Seoul, Republic of Korea) \emph{(\bibinfo{series}{FAccT '22})}. \bibinfo{publisher}{Association for Computing Machinery}, \bibinfo{address}{New York, NY, USA}, \bibinfo{pages}{1457–1466}.
\newblock
\showISBNx{9781450393522}
\urldef\tempurl%
\url{https://doi.org/10.1145/3531146.3533202}
\showDOI{\tempurl}


\bibitem[Floridi and Cowls(2019)]%
        {Floridi2019Unified}
\bibfield{author}{\bibinfo{person}{Luciano Floridi} {and} \bibinfo{person}{Josh Cowls}.} \bibinfo{year}{2019}\natexlab{}.
\newblock \showarticletitle{A {Unified} {Framework} of {Five} {Principles} for {AI} in {Society}}.
\newblock \bibinfo{journal}{\emph{Harvard Data Science Review}} \bibinfo{volume}{1}, \bibinfo{number}{1} (\bibinfo{date}{jul 1} \bibinfo{year}{2019}).
\newblock
\newblock
\shownote{https://hdsr.mitpress.mit.edu/pub/l0jsh9d1}.


\bibitem[Gambo et~al\mbox{.}(2024)]%
        {10636143}
\bibfield{author}{\bibinfo{person}{Ishaya Gambo}, \bibinfo{person}{Rhodes Massenon}, \bibinfo{person}{Chia-Chen Lin}, \bibinfo{person}{Roseline~Oluwaseun Ogundokun}, \bibinfo{person}{Saurabh Agarwal}, {and} \bibinfo{person}{Wooguil Pak}.} \bibinfo{year}{2024}\natexlab{}.
\newblock \showarticletitle{Enhancing User Trust and Interpretability in AI-Driven Feature Request Detection for Mobile App Reviews: An Explainable Approach}.
\newblock \bibinfo{journal}{\emph{IEEE Access}}  \bibinfo{volume}{12} (\bibinfo{year}{2024}), \bibinfo{pages}{114023--114045}.
\newblock
\urldef\tempurl%
\url{https://doi.org/10.1109/ACCESS.2024.3443527}
\showDOI{\tempurl}


\bibitem[Ghofrani et~al\mbox{.}(2022)]%
        {ghofrani2022trustchallengesreusingopen}
\bibfield{author}{\bibinfo{person}{Javad Ghofrani}, \bibinfo{person}{Paria Heravi}, \bibinfo{person}{Kambiz~A. Babaei}, {and} \bibinfo{person}{Mohammad Soorati}.} \bibinfo{year}{2022}\natexlab{}.
\newblock \bibinfo{title}{Trust Challenges in Reusing Open Source Software: An Interview-based Initial Study}.
\newblock
\newblock
\showeprint[arxiv]{2208.01137}~[cs.SE]
\urldef\tempurl%
\url{https://arxiv.org/abs/2208.01137}
\showURL{%
\tempurl}


\bibitem[Glikson and Woolley(2020)]%
        {Glikson2020HumanTI}
\bibfield{author}{\bibinfo{person}{Ella Glikson} {and} \bibinfo{person}{Anita~Williams Woolley}.} \bibinfo{year}{2020}\natexlab{}.
\newblock \showarticletitle{Human Trust in Artificial Intelligence: Review of Empirical Research}.
\newblock \bibinfo{journal}{\emph{Academy of Management Annals}} (\bibinfo{year}{2020}).
\newblock
\urldef\tempurl%
\url{https://api.semanticscholar.org/CorpusID:216198731}
\showURL{%
\tempurl}


\bibitem[Gu et~al\mbox{.}(2018)]%
        {deep_code_search}
\bibfield{author}{\bibinfo{person}{Xiaodong Gu}, \bibinfo{person}{Hongyu Zhang}, {and} \bibinfo{person}{Sunghun Kim}.} \bibinfo{year}{2018}\natexlab{}.
\newblock \showarticletitle{Deep code search}. In \bibinfo{booktitle}{\emph{Proceedings of the 40th International Conference on Software Engineering}} (Gothenburg, Sweden) \emph{(\bibinfo{series}{ICSE '18})}. \bibinfo{publisher}{Association for Computing Machinery}, \bibinfo{address}{New York, NY, USA}, \bibinfo{pages}{933–944}.
\newblock
\showISBNx{9781450356381}
\urldef\tempurl%
\url{https://doi.org/10.1145/3180155.3180167}
\showDOI{\tempurl}


\bibitem[Guo et~al\mbox{.}(2017)]%
        {traceability_2017}
\bibfield{author}{\bibinfo{person}{Jin Guo}, \bibinfo{person}{Jinghui Cheng}, {and} \bibinfo{person}{Jane Cleland-Huang}.} \bibinfo{year}{2017}\natexlab{}.
\newblock \showarticletitle{Semantically Enhanced Software Traceability Using Deep Learning Techniques}. In \bibinfo{booktitle}{\emph{2017 IEEE/ACM 39th International Conference on Software Engineering (ICSE)}}. \bibinfo{pages}{3--14}.
\newblock
\urldef\tempurl%
\url{https://doi.org/10.1109/ICSE.2017.9}
\showDOI{\tempurl}


\bibitem[Harjit~Sekhon and Devlin(2014)]%
        {trustor_trustee}
\bibfield{author}{\bibinfo{person}{Husni~Kharouf Harjit~Sekhon, Christine~Ennew} {and} \bibinfo{person}{James Devlin}.} \bibinfo{year}{2014}\natexlab{}.
\newblock \showarticletitle{Trustworthiness and trust: influences and implications}.
\newblock \bibinfo{journal}{\emph{Journal of Marketing Management}} \bibinfo{volume}{30}, \bibinfo{number}{3-4} (\bibinfo{year}{2014}), \bibinfo{pages}{409--430}.
\newblock
\urldef\tempurl%
\url{https://doi.org/10.1080/0267257X.2013.842609}
\showDOI{\tempurl}
\showeprint{https://doi.org/10.1080/0267257X.2013.842609}


\bibitem[Hassan et~al\mbox{.}(2024)]%
        {hassan2024rethinkingsoftwareengineeringfoundation}
\bibfield{author}{\bibinfo{person}{Ahmed~E. Hassan}, \bibinfo{person}{Dayi Lin}, \bibinfo{person}{Gopi~Krishnan Rajbahadur}, \bibinfo{person}{Keheliya Gallaba}, \bibinfo{person}{Filipe~R. Cogo}, \bibinfo{person}{Boyuan Chen}, \bibinfo{person}{Haoxiang Zhang}, \bibinfo{person}{Kishanthan Thangarajah}, \bibinfo{person}{Gustavo~Ansaldi Oliva}, \bibinfo{person}{Jiahuei Lin}, \bibinfo{person}{Wali~Mohammad Abdullah}, {and} \bibinfo{person}{Zhen~Ming Jiang}.} \bibinfo{year}{2024}\natexlab{}.
\newblock \bibinfo{title}{Rethinking Software Engineering in the Foundation Model Era: A Curated Catalogue of Challenges in the Development of Trustworthy FMware}.
\newblock
\newblock
\showeprint[arxiv]{2402.15943}~[cs.SE]
\urldef\tempurl%
\url{https://arxiv.org/abs/2402.15943}
\showURL{%
\tempurl}


\bibitem[Hendrycks et~al\mbox{.}(2021)]%
        {hendrycks2021measuringmassivemultitasklanguage}
\bibfield{author}{\bibinfo{person}{Dan Hendrycks}, \bibinfo{person}{Collin Burns}, \bibinfo{person}{Steven Basart}, \bibinfo{person}{Andy Zou}, \bibinfo{person}{Mantas Mazeika}, \bibinfo{person}{Dawn Song}, {and} \bibinfo{person}{Jacob Steinhardt}.} \bibinfo{year}{2021}\natexlab{}.
\newblock \bibinfo{title}{Measuring Massive Multitask Language Understanding}.
\newblock
\newblock
\showeprint[arxiv]{2009.03300}~[cs.CY]
\urldef\tempurl%
\url{https://arxiv.org/abs/2009.03300}
\showURL{%
\tempurl}


\bibitem[Hou and Jansen(2022)]%
        {hou2022systematicliteraturereviewtrust}
\bibfield{author}{\bibinfo{person}{Fang Hou} {and} \bibinfo{person}{Slinger Jansen}.} \bibinfo{year}{2022}\natexlab{}.
\newblock \bibinfo{title}{A Systematic Literature Review on Trust in the Software Ecosystem}.
\newblock
\newblock
\showeprint[arxiv]{2203.05678}~[cs.SE]
\urldef\tempurl%
\url{https://arxiv.org/abs/2203.05678}
\showURL{%
\tempurl}


\bibitem[Huang et~al\mbox{.}(2024)]%
        {trustLLM}
\bibfield{author}{\bibinfo{person}{Yue Huang}, \bibinfo{person}{Lichao Sun}, \bibinfo{person}{Haoran Wang}, \bibinfo{person}{Siyuan Wu}, \bibinfo{person}{Qihui Zhang}, \bibinfo{person}{Yuan Li}, \bibinfo{person}{Chujie Gao}, \bibinfo{person}{Yixin Huang}, \bibinfo{person}{Wenhan Lyu}, \bibinfo{person}{Yixuan Zhang}, \bibinfo{person}{Xiner Li}, \bibinfo{person}{Hanchi Sun}, \bibinfo{person}{Zhengliang Liu}, \bibinfo{person}{Yixin Liu}, \bibinfo{person}{Yijue Wang}, \bibinfo{person}{Zhikun Zhang}, \bibinfo{person}{Bertie Vidgen}, \bibinfo{person}{Bhavya Kailkhura}, \bibinfo{person}{Caiming Xiong}, \bibinfo{person}{Chaowei Xiao}, \bibinfo{person}{Chunyuan Li}, \bibinfo{person}{Eric~P. Xing}, \bibinfo{person}{Furong Huang}, \bibinfo{person}{Hao Liu}, \bibinfo{person}{Heng Ji}, \bibinfo{person}{Hongyi Wang}, \bibinfo{person}{Huan Zhang}, \bibinfo{person}{Huaxiu Yao}, \bibinfo{person}{Manolis Kellis}, \bibinfo{person}{Marinka Zitnik}, \bibinfo{person}{Meng Jiang}, \bibinfo{person}{Mohit Bansal},
  \bibinfo{person}{James Zou}, \bibinfo{person}{Jian Pei}, \bibinfo{person}{Jian Liu}, \bibinfo{person}{Jianfeng Gao}, \bibinfo{person}{Jiawei Han}, \bibinfo{person}{Jieyu Zhao}, \bibinfo{person}{Jiliang Tang}, \bibinfo{person}{Jindong Wang}, \bibinfo{person}{Joaquin Vanschoren}, \bibinfo{person}{John Mitchell}, \bibinfo{person}{Kai Shu}, \bibinfo{person}{Kaidi Xu}, \bibinfo{person}{Kai-Wei Chang}, \bibinfo{person}{Lifang He}, \bibinfo{person}{Lifu Huang}, \bibinfo{person}{Michael Backes}, \bibinfo{person}{Neil~Zhenqiang Gong}, \bibinfo{person}{Philip~S. Yu}, \bibinfo{person}{Pin-Yu Chen}, \bibinfo{person}{Quanquan Gu}, \bibinfo{person}{Ran Xu}, \bibinfo{person}{Rex Ying}, \bibinfo{person}{Shuiwang Ji}, \bibinfo{person}{Suman Jana}, \bibinfo{person}{Tianlong Chen}, \bibinfo{person}{Tianming Liu}, \bibinfo{person}{Tianyi Zhou}, \bibinfo{person}{William~Yang Wang}, \bibinfo{person}{Xiang Li}, \bibinfo{person}{Xiangliang Zhang}, \bibinfo{person}{Xiao Wang}, \bibinfo{person}{Xing Xie}, \bibinfo{person}{Xun Chen},
  \bibinfo{person}{Xuyu Wang}, \bibinfo{person}{Yan Liu}, \bibinfo{person}{Yanfang Ye}, \bibinfo{person}{Yinzhi Cao}, \bibinfo{person}{Yong Chen}, {and} \bibinfo{person}{Yue Zhao}.} \bibinfo{year}{2024}\natexlab{}.
\newblock \showarticletitle{Position: {T}rust{LLM}: Trustworthiness in Large Language Models}. In \bibinfo{booktitle}{\emph{Proceedings of the 41st International Conference on Machine Learning}} \emph{(\bibinfo{series}{Proceedings of Machine Learning Research}, Vol.~\bibinfo{volume}{235})}, \bibfield{editor}{\bibinfo{person}{Ruslan Salakhutdinov}, \bibinfo{person}{Zico Kolter}, \bibinfo{person}{Katherine Heller}, \bibinfo{person}{Adrian Weller}, \bibinfo{person}{Nuria Oliver}, \bibinfo{person}{Jonathan Scarlett}, {and} \bibinfo{person}{Felix Berkenkamp}} (Eds.). \bibinfo{publisher}{PMLR}, \bibinfo{pages}{20166--20270}.
\newblock
\urldef\tempurl%
\url{https://proceedings.mlr.press/v235/huang24x.html}
\showURL{%
\tempurl}


\bibitem[Huang et~al\mbox{.}(2023)]%
        {huang2023trustgptbenchmarktrustworthyresponsible}
\bibfield{author}{\bibinfo{person}{Yue Huang}, \bibinfo{person}{Qihui Zhang}, \bibinfo{person}{Philip~S. Y}, {and} \bibinfo{person}{Lichao Sun}.} \bibinfo{year}{2023}\natexlab{}.
\newblock \bibinfo{title}{TrustGPT: A Benchmark for Trustworthy and Responsible Large Language Models}.
\newblock
\newblock
\showeprint[arxiv]{2306.11507}~[cs.CL]
\urldef\tempurl%
\url{https://arxiv.org/abs/2306.11507}
\showURL{%
\tempurl}


\bibitem[Jakesch et~al\mbox{.}(2019)]%
        {10.1145/3290605.3300469}
\bibfield{author}{\bibinfo{person}{Maurice Jakesch}, \bibinfo{person}{Megan French}, \bibinfo{person}{Xiao Ma}, \bibinfo{person}{Jeffrey~T. Hancock}, {and} \bibinfo{person}{Mor Naaman}.} \bibinfo{year}{2019}\natexlab{}.
\newblock \showarticletitle{AI-Mediated Communication: How the Perception that Profile Text was Written by AI Affects Trustworthiness}. In \bibinfo{booktitle}{\emph{Proceedings of the 2019 CHI Conference on Human Factors in Computing Systems}} (Glasgow, Scotland Uk) \emph{(\bibinfo{series}{CHI '19})}. \bibinfo{publisher}{Association for Computing Machinery}, \bibinfo{address}{New York, NY, USA}, \bibinfo{pages}{1–13}.
\newblock
\showISBNx{9781450359702}
\urldef\tempurl%
\url{https://doi.org/10.1145/3290605.3300469}
\showDOI{\tempurl}


\bibitem[Ji et~al\mbox{.}(2024)]%
        {ji2024aialignmentcomprehensivesurvey}
\bibfield{author}{\bibinfo{person}{Jiaming Ji}, \bibinfo{person}{Tianyi Qiu}, \bibinfo{person}{Boyuan Chen}, \bibinfo{person}{Borong Zhang}, \bibinfo{person}{Hantao Lou}, \bibinfo{person}{Kaile Wang}, \bibinfo{person}{Yawen Duan}, \bibinfo{person}{Zhonghao He}, \bibinfo{person}{Jiayi Zhou}, \bibinfo{person}{Zhaowei Zhang}, \bibinfo{person}{Fanzhi Zeng}, \bibinfo{person}{Kwan~Yee Ng}, \bibinfo{person}{Juntao Dai}, \bibinfo{person}{Xuehai Pan}, \bibinfo{person}{Aidan O'Gara}, \bibinfo{person}{Yingshan Lei}, \bibinfo{person}{Hua Xu}, \bibinfo{person}{Brian Tse}, \bibinfo{person}{Jie Fu}, \bibinfo{person}{Stephen McAleer}, \bibinfo{person}{Yaodong Yang}, \bibinfo{person}{Yizhou Wang}, \bibinfo{person}{Song-Chun Zhu}, \bibinfo{person}{Yike Guo}, {and} \bibinfo{person}{Wen Gao}.} \bibinfo{year}{2024}\natexlab{}.
\newblock \bibinfo{title}{AI Alignment: A Comprehensive Survey}.
\newblock
\newblock
\showeprint[arxiv]{2310.19852}~[cs.AI]
\urldef\tempurl%
\url{https://arxiv.org/abs/2310.19852}
\showURL{%
\tempurl}


\bibitem[Jobin et~al\mbox{.}(2019)]%
        {Jobin2019}
\bibfield{author}{\bibinfo{person}{Anna Jobin}, \bibinfo{person}{Marcello Ienca}, {and} \bibinfo{person}{Effy Vayena}.} \bibinfo{year}{2019}\natexlab{}.
\newblock \showarticletitle{The global landscape of AI ethics guidelines}.
\newblock \bibinfo{journal}{\emph{Nature Machine Intelligence}}  \bibinfo{volume}{1} (\bibinfo{year}{2019}), \bibinfo{pages}{389--399}.
\newblock
\urldef\tempurl%
\url{https://doi.org/10.1038/s42256-019-0088-2}
\showDOI{\tempurl}


\bibitem[Johnson et~al\mbox{.}(2023)]%
        {10.1109/ICSE-SEIP58684.2023.00043}
\bibfield{author}{\bibinfo{person}{Brittany Johnson}, \bibinfo{person}{Christian Bird}, \bibinfo{person}{Denae Ford}, \bibinfo{person}{Nicole Forsgren}, {and} \bibinfo{person}{Thomas Zimmermann}.} \bibinfo{year}{2023}\natexlab{}.
\newblock \showarticletitle{Make Your Tools Sparkle with Trust: The PICSE Framework for Trust in Software Tools}. In \bibinfo{booktitle}{\emph{Proceedings of the 45th International Conference on Software Engineering: Software Engineering in Practice}} (Melbourne, Australia) \emph{(\bibinfo{series}{ICSE-SEIP '23})}. \bibinfo{publisher}{IEEE Press}, \bibinfo{pages}{409–419}.
\newblock
\showISBNx{9798350300376}
\urldef\tempurl%
\url{https://doi.org/10.1109/ICSE-SEIP58684.2023.00043}
\showDOI{\tempurl}


\bibitem[Kaur et~al\mbox{.}(2022)]%
        {10.1145/3491209}
\bibfield{author}{\bibinfo{person}{Davinder Kaur}, \bibinfo{person}{Suleyman Uslu}, \bibinfo{person}{Kaley~J. Rittichier}, {and} \bibinfo{person}{Arjan Durresi}.} \bibinfo{year}{2022}\natexlab{}.
\newblock \showarticletitle{Trustworthy Artificial Intelligence: A Review}.
\newblock \bibinfo{journal}{\emph{ACM Comput. Surv.}} \bibinfo{volume}{55}, \bibinfo{number}{2}, Article \bibinfo{articleno}{39} (\bibinfo{date}{Jan.} \bibinfo{year}{2022}), \bibinfo{numpages}{38}~pages.
\newblock
\showISSN{0360-0300}
\urldef\tempurl%
\url{https://doi.org/10.1145/3491209}
\showDOI{\tempurl}


\bibitem[Key et~al\mbox{.}(2023)]%
        {key2023trustworthyneuralprogramsynthesis}
\bibfield{author}{\bibinfo{person}{Darren Key}, \bibinfo{person}{Wen-Ding Li}, {and} \bibinfo{person}{Kevin Ellis}.} \bibinfo{year}{2023}\natexlab{}.
\newblock \bibinfo{title}{Toward Trustworthy Neural Program Synthesis}.
\newblock
\newblock
\showeprint[arxiv]{2210.00848}~[cs.SE]
\urldef\tempurl%
\url{https://arxiv.org/abs/2210.00848}
\showURL{%
\tempurl}


\bibitem[Khandelwal et~al\mbox{.}(2020)]%
        {khandelwal2020generalizationmemorizationnearestneighbor}
\bibfield{author}{\bibinfo{person}{Urvashi Khandelwal}, \bibinfo{person}{Omer Levy}, \bibinfo{person}{Dan Jurafsky}, \bibinfo{person}{Luke Zettlemoyer}, {and} \bibinfo{person}{Mike Lewis}.} \bibinfo{year}{2020}\natexlab{}.
\newblock \bibinfo{title}{Generalization through Memorization: Nearest Neighbor Language Models}.
\newblock
\newblock
\showeprint[arxiv]{1911.00172}~[cs.CL]
\urldef\tempurl%
\url{https://arxiv.org/abs/1911.00172}
\showURL{%
\tempurl}


\bibitem[Kim et~al\mbox{.}(2024)]%
        {Kim_2024}
\bibfield{author}{\bibinfo{person}{Sunnie S.~Y. Kim}, \bibinfo{person}{Q.~Vera Liao}, \bibinfo{person}{Mihaela Vorvoreanu}, \bibinfo{person}{Stephanie Ballard}, {and} \bibinfo{person}{Jennifer~Wortman Vaughan}.} \bibinfo{year}{2024}\natexlab{}.
\newblock \showarticletitle{“I’m Not Sure, But...”: Examining the Impact of Large Language Models’ Uncertainty Expression on User Reliance and Trust}. In \bibinfo{booktitle}{\emph{The 2024 ACM Conference on Fairness, Accountability, and Transparency}} \emph{(\bibinfo{series}{FAccT ’24})}. \bibinfo{publisher}{ACM}, \bibinfo{pages}{822–835}.
\newblock
\urldef\tempurl%
\url{https://doi.org/10.1145/3630106.3658941}
\showDOI{\tempurl}


\bibitem[Kitchenham(2012)]%
        {barabaraCit}
\bibfield{author}{\bibinfo{person}{Barbara~A. Kitchenham}.} \bibinfo{year}{2012}\natexlab{}.
\newblock \showarticletitle{Systematic review in software engineering: where we are and where we should be going}. In \bibinfo{booktitle}{\emph{Proceedings of the 2nd International Workshop on Evidential Assessment of Software Technologies}} (Lund, Sweden) \emph{(\bibinfo{series}{EAST '12})}. \bibinfo{publisher}{Association for Computing Machinery}, \bibinfo{address}{New York, NY, USA}, \bibinfo{pages}{1–2}.
\newblock
\showISBNx{9781450315098}
\urldef\tempurl%
\url{https://doi.org/10.1145/2372233.2372235}
\showDOI{\tempurl}


\bibitem[Knowles and Richards(2021)]%
        {10.1145/3442188.3445890}
\bibfield{author}{\bibinfo{person}{Bran Knowles} {and} \bibinfo{person}{John~T. Richards}.} \bibinfo{year}{2021}\natexlab{}.
\newblock \showarticletitle{The Sanction of Authority: Promoting Public Trust in AI}. In \bibinfo{booktitle}{\emph{Proceedings of the 2021 ACM Conference on Fairness, Accountability, and Transparency}} (Virtual Event, Canada) \emph{(\bibinfo{series}{FAccT '21})}. \bibinfo{publisher}{Association for Computing Machinery}, \bibinfo{address}{New York, NY, USA}, \bibinfo{pages}{262–271}.
\newblock
\showISBNx{9781450383097}
\urldef\tempurl%
\url{https://doi.org/10.1145/3442188.3445890}
\showDOI{\tempurl}


\bibitem[Kowald et~al\mbox{.}(2024)]%
        {kowald2024establishingevaluatingtrustworthyai}
\bibfield{author}{\bibinfo{person}{Dominik Kowald}, \bibinfo{person}{Sebastian Scher}, \bibinfo{person}{Viktoria Pammer-Schindler}, \bibinfo{person}{Peter Müllner}, \bibinfo{person}{Kerstin Waxnegger}, \bibinfo{person}{Lea Demelius}, \bibinfo{person}{Angela Fessl}, \bibinfo{person}{Maximilian Toller}, \bibinfo{person}{Inti Gabriel~Mendoza Estrada}, \bibinfo{person}{Ilija Simic}, \bibinfo{person}{Vedran Sabol}, \bibinfo{person}{Andreas Truegler}, \bibinfo{person}{Eduardo Veas}, \bibinfo{person}{Roman Kern}, \bibinfo{person}{Tomislav Nad}, {and} \bibinfo{person}{Simone Kopeinik}.} \bibinfo{year}{2024}\natexlab{}.
\newblock \bibinfo{title}{Establishing and Evaluating Trustworthy AI: Overview and Research Challenges}.
\newblock
\newblock
\showeprint[arxiv]{2411.09973}~[cs.LG]
\urldef\tempurl%
\url{https://arxiv.org/abs/2411.09973}
\showURL{%
\tempurl}


\bibitem[Krishna et~al\mbox{.}(2024)]%
        {krishna2024disagreementproblemexplainablemachine}
\bibfield{author}{\bibinfo{person}{Satyapriya Krishna}, \bibinfo{person}{Tessa Han}, \bibinfo{person}{Alex Gu}, \bibinfo{person}{Steven Wu}, \bibinfo{person}{Shahin Jabbari}, {and} \bibinfo{person}{Himabindu Lakkaraju}.} \bibinfo{year}{2024}\natexlab{}.
\newblock \bibinfo{title}{The Disagreement Problem in Explainable Machine Learning: A Practitioner's Perspective}.
\newblock
\newblock
\showeprint[arxiv]{2202.01602}~[cs.LG]
\urldef\tempurl%
\url{https://arxiv.org/abs/2202.01602}
\showURL{%
\tempurl}


\bibitem[Kästner et~al\mbox{.}(2021)]%
        {9582305}
\bibfield{author}{\bibinfo{person}{Lena Kästner}, \bibinfo{person}{Markus Langer}, \bibinfo{person}{Veronika Lazar}, \bibinfo{person}{Astrid Schomäcker}, \bibinfo{person}{Timo Speith}, {and} \bibinfo{person}{Sarah Sterz}.} \bibinfo{year}{2021}\natexlab{}.
\newblock \showarticletitle{On the Relation of Trust and Explainability: Why to Engineer for Trustworthiness}. In \bibinfo{booktitle}{\emph{2021 IEEE 29th International Requirements Engineering Conference Workshops (REW)}}. \bibinfo{pages}{169--175}.
\newblock
\urldef\tempurl%
\url{https://doi.org/10.1109/REW53955.2021.00031}
\showDOI{\tempurl}


\bibitem[Körber(2018)]%
        {korber_2018}
\bibfield{author}{\bibinfo{person}{Moritz Körber}.} \bibinfo{year}{2018}\natexlab{}.
\newblock \bibinfo{title}{Theoretical considerations and development of a questionnaire to measure trust in automation}.
\newblock
\newblock
\urldef\tempurl%
\url{https://doi.org/10.31234/osf.io/nfc45}
\showDOI{\tempurl}


\bibitem[Langer et~al\mbox{.}(2021)]%
        {LANGER2021103473}
\bibfield{author}{\bibinfo{person}{Markus Langer}, \bibinfo{person}{Daniel Oster}, \bibinfo{person}{Timo Speith}, \bibinfo{person}{Holger Hermanns}, \bibinfo{person}{Lena Kästner}, \bibinfo{person}{Eva Schmidt}, \bibinfo{person}{Andreas Sesing}, {and} \bibinfo{person}{Kevin Baum}.} \bibinfo{year}{2021}\natexlab{}.
\newblock \showarticletitle{What do we want from Explainable Artificial Intelligence (XAI)? – A stakeholder perspective on XAI and a conceptual model guiding interdisciplinary XAI research}.
\newblock \bibinfo{journal}{\emph{Artificial Intelligence}}  \bibinfo{volume}{296} (\bibinfo{year}{2021}), \bibinfo{pages}{103473}.
\newblock
\showISSN{0004-3702}
\urldef\tempurl%
\url{https://doi.org/10.1016/j.artint.2021.103473}
\showDOI{\tempurl}


\bibitem[Lassiter and Fleischmann(2024)]%
        {10.1145/3686963}
\bibfield{author}{\bibinfo{person}{Tina~B. Lassiter} {and} \bibinfo{person}{Kenneth~R. Fleischmann}.} \bibinfo{year}{2024}\natexlab{}.
\newblock \showarticletitle{"Something Fast and Cheap" or "A Core Element of Building Trust"? - AI Auditing Professionals' Perspectives on Trust in AI}.
\newblock \bibinfo{journal}{\emph{Proc. ACM Hum.-Comput. Interact.}} \bibinfo{volume}{8}, \bibinfo{number}{CSCW2}, Article \bibinfo{articleno}{424} (\bibinfo{date}{Nov.} \bibinfo{year}{2024}), \bibinfo{numpages}{22}~pages.
\newblock
\urldef\tempurl%
\url{https://doi.org/10.1145/3686963}
\showDOI{\tempurl}


\bibitem[Lee and See(2004)]%
        {Lee2004TrustIA}
\bibfield{author}{\bibinfo{person}{John~D. Lee} {and} \bibinfo{person}{Katrina~A. See}.} \bibinfo{year}{2004}\natexlab{}.
\newblock \showarticletitle{Trust in Automation: Designing for Appropriate Reliance}.
\newblock \bibinfo{journal}{\emph{Human Factors: The Journal of Human Factors and Ergonomics Society}}  \bibinfo{volume}{46} (\bibinfo{year}{2004}), \bibinfo{pages}{50 -- 80}.
\newblock
\urldef\tempurl%
\url{https://api.semanticscholar.org/CorpusID:5210390}
\showURL{%
\tempurl}


\bibitem[Levett(2022)]%
        {Levett_2022}
\bibfield{author}{\bibinfo{person}{Paul Levett}.} \bibinfo{year}{2022}\natexlab{}.
\newblock \bibinfo{title}{Research guides: Systematic reviews: Data Extraction/coding/study characteristics/results}.
\newblock
\newblock
\urldef\tempurl%
\url{https://guides.himmelfarb.gwu.edu/systematic_review/data-extraction}
\showURL{%
\tempurl}


\bibitem[Lewis et~al\mbox{.}(2018)]%
        {Lewis2018}
\bibfield{author}{\bibinfo{person}{Michael Lewis}, \bibinfo{person}{Katia Sycara}, {and} \bibinfo{person}{Phillip Walker}.} \bibinfo{year}{2018}\natexlab{}.
\newblock \showarticletitle{The Role of Trust in Human-Robot Interaction}.
\newblock In \bibinfo{booktitle}{\emph{Foundations of Trusted Autonomy}}, \bibfield{editor}{\bibinfo{person}{Hussein Abbass}, \bibinfo{person}{Johannes Scholz}, {and} \bibinfo{person}{David Reid}} (Eds.). \bibinfo{series}{Studies in Systems, Decision and Control}, Vol.~\bibinfo{volume}{117}. \bibinfo{publisher}{Springer, Cham}, \bibinfo{pages}{135--159}.
\newblock
\urldef\tempurl%
\url{https://doi.org/10.1007/978-3-319-64816-3_8}
\showDOI{\tempurl}


\bibitem[Li et~al\mbox{.}(2023b)]%
        {trustworthyAI}
\bibfield{author}{\bibinfo{person}{Bo Li}, \bibinfo{person}{Peng Qi}, \bibinfo{person}{Bo Liu}, \bibinfo{person}{Shuai Di}, \bibinfo{person}{Jingen Liu}, \bibinfo{person}{Jiquan Pei}, \bibinfo{person}{Jinfeng Yi}, {and} \bibinfo{person}{Bowen Zhou}.} \bibinfo{year}{2023}\natexlab{b}.
\newblock \showarticletitle{Trustworthy AI: From Principles to Practices}.
\newblock \bibinfo{journal}{\emph{ACM Comput. Surv.}} \bibinfo{volume}{55}, \bibinfo{number}{9}, Article \bibinfo{articleno}{177} (\bibinfo{date}{jan} \bibinfo{year}{2023}), \bibinfo{numpages}{46}~pages.
\newblock
\showISSN{0360-0300}
\urldef\tempurl%
\url{https://doi.org/10.1145/3555803}
\showDOI{\tempurl}


\bibitem[Li et~al\mbox{.}(2023a)]%
        {10.1109/MC.2023.3240730}
\bibfield{author}{\bibinfo{person}{Gongyuan Li}, \bibinfo{person}{Bohan Liu}, {and} \bibinfo{person}{He Zhang}.} \bibinfo{year}{2023}\natexlab{a}.
\newblock \showarticletitle{Quality Attributes of Trustworthy Artificial Intelligence in Normative Documents and Secondary Studies: A Preliminary Review}.
\newblock \bibinfo{journal}{\emph{Computer}} \bibinfo{volume}{56}, \bibinfo{number}{4} (\bibinfo{date}{April} \bibinfo{year}{2023}), \bibinfo{pages}{28–37}.
\newblock
\showISSN{0018-9162}
\urldef\tempurl%
\url{https://doi.org/10.1109/MC.2023.3240730}
\showDOI{\tempurl}


\bibitem[Liao and Sundar(2022)]%
        {Liao_2022}
\bibfield{author}{\bibinfo{person}{Q.Vera Liao} {and} \bibinfo{person}{S.~Shyam Sundar}.} \bibinfo{year}{2022}\natexlab{}.
\newblock \showarticletitle{Designing for Responsible Trust in AI Systems: A Communication Perspective}. In \bibinfo{booktitle}{\emph{2022 ACM Conference on Fairness, Accountability, and Transparency}} \emph{(\bibinfo{series}{FAccT ’22})}. \bibinfo{publisher}{ACM}.
\newblock
\urldef\tempurl%
\url{https://doi.org/10.1145/3531146.3533182}
\showDOI{\tempurl}


\bibitem[Liao and Wortman~Vaughan(2024)]%
        {Liao2024AI}
\bibfield{author}{\bibinfo{person}{Q.~Vera Liao} {and} \bibinfo{person}{Jennifer Wortman~Vaughan}.} \bibinfo{year}{2024}\natexlab{}.
\newblock \showarticletitle{{AI} {Transparency} in the {Age} of {LLMs}: A {Human}-{Centered} {Research} {Roadmap}}.
\newblock \bibinfo{journal}{\emph{Harvard Data Science Review}} \bibinfo{number}{Special Issue 5} (\bibinfo{date}{may 31} \bibinfo{year}{2024}).
\newblock
\newblock
\shownote{https://hdsr.mitpress.mit.edu/pub/aelql9qy}.


\bibitem[Ling et~al\mbox{.}(2025)]%
        {ling2025biasunveiledinvestigatingsocial}
\bibfield{author}{\bibinfo{person}{Lin Ling}, \bibinfo{person}{Fazle Rabbi}, \bibinfo{person}{Song Wang}, {and} \bibinfo{person}{Jinqiu Yang}.} \bibinfo{year}{2025}\natexlab{}.
\newblock \bibinfo{title}{Bias Unveiled: Investigating Social Bias in LLM-Generated Code}.
\newblock
\newblock
\showeprint[arxiv]{2411.10351}~[cs.SE]
\urldef\tempurl%
\url{https://arxiv.org/abs/2411.10351}
\showURL{%
\tempurl}


\bibitem[Liu et~al\mbox{.}(2024a)]%
        {liu2024empiricalstudypotentialllms}
\bibfield{author}{\bibinfo{person}{Bo Liu}, \bibinfo{person}{Yanjie Jiang}, \bibinfo{person}{Yuxia Zhang}, \bibinfo{person}{Nan Niu}, \bibinfo{person}{Guangjie Li}, {and} \bibinfo{person}{Hui Liu}.} \bibinfo{year}{2024}\natexlab{a}.
\newblock \bibinfo{title}{An Empirical Study on the Potential of LLMs in Automated Software Refactoring}.
\newblock
\newblock
\showeprint[arxiv]{2411.04444}~[cs.SE]
\urldef\tempurl%
\url{https://arxiv.org/abs/2411.04444}
\showURL{%
\tempurl}


\bibitem[Liu et~al\mbox{.}(2021)]%
        {liu2021trustworthyaicomputationalperspective}
\bibfield{author}{\bibinfo{person}{Haochen Liu}, \bibinfo{person}{Yiqi Wang}, \bibinfo{person}{Wenqi Fan}, \bibinfo{person}{Xiaorui Liu}, \bibinfo{person}{Yaxin Li}, \bibinfo{person}{Shaili Jain}, \bibinfo{person}{Yunhao Liu}, \bibinfo{person}{Anil~K. Jain}, {and} \bibinfo{person}{Jiliang Tang}.} \bibinfo{year}{2021}\natexlab{}.
\newblock \bibinfo{title}{Trustworthy AI: A Computational Perspective}.
\newblock
\newblock
\showeprint[arxiv]{2107.06641}~[cs.AI]
\urldef\tempurl%
\url{https://arxiv.org/abs/2107.06641}
\showURL{%
\tempurl}


\bibitem[Liu et~al\mbox{.}(2018)]%
        {feature_envy_2018}
\bibfield{author}{\bibinfo{person}{Hui Liu}, \bibinfo{person}{Zhifeng Xu}, {and} \bibinfo{person}{Yanzhen Zou}.} \bibinfo{year}{2018}\natexlab{}.
\newblock \showarticletitle{Deep learning based feature envy detection}. In \bibinfo{booktitle}{\emph{Proceedings of the 33rd ACM/IEEE International Conference on Automated Software Engineering}} (Montpellier, France) \emph{(\bibinfo{series}{ASE '18})}. \bibinfo{publisher}{Association for Computing Machinery}, \bibinfo{address}{New York, NY, USA}, \bibinfo{pages}{385–396}.
\newblock
\showISBNx{9781450359375}
\urldef\tempurl%
\url{https://doi.org/10.1145/3238147.3238166}
\showDOI{\tempurl}


\bibitem[Liu et~al\mbox{.}(2017)]%
        {input_generation}
\bibfield{author}{\bibinfo{person}{Peng Liu}, \bibinfo{person}{Xiangyu Zhang}, \bibinfo{person}{Marco Pistoia}, \bibinfo{person}{Yunhui Zheng}, \bibinfo{person}{Manoel Marques}, {and} \bibinfo{person}{Lingfei Zeng}.} \bibinfo{year}{2017}\natexlab{}.
\newblock \showarticletitle{Automatic Text Input Generation for Mobile Testing}. In \bibinfo{booktitle}{\emph{2017 IEEE/ACM 39th International Conference on Software Engineering (ICSE)}}. \bibinfo{pages}{643--653}.
\newblock
\urldef\tempurl%
\url{https://doi.org/10.1109/ICSE.2017.65}
\showDOI{\tempurl}


\bibitem[Liu et~al\mbox{.}(2023)]%
        {liu2023uncoveringquantifyingsocialbiases}
\bibfield{author}{\bibinfo{person}{Yan Liu}, \bibinfo{person}{Xiaokang Chen}, \bibinfo{person}{Yan Gao}, \bibinfo{person}{Zhe Su}, \bibinfo{person}{Fengji Zhang}, \bibinfo{person}{Daoguang Zan}, \bibinfo{person}{Jian-Guang Lou}, \bibinfo{person}{Pin-Yu Chen}, {and} \bibinfo{person}{Tsung-Yi Ho}.} \bibinfo{year}{2023}\natexlab{}.
\newblock \bibinfo{title}{Uncovering and Quantifying Social Biases in Code Generation}.
\newblock
\newblock
\showeprint[arxiv]{2305.15377}~[cs.CL]
\urldef\tempurl%
\url{https://arxiv.org/abs/2305.15377}
\showURL{%
\tempurl}


\bibitem[Liu et~al\mbox{.}(2024b)]%
        {10.1145/3641540}
\bibfield{author}{\bibinfo{person}{Yue Liu}, \bibinfo{person}{Chakkrit Tantithamthavorn}, \bibinfo{person}{Yonghui Liu}, {and} \bibinfo{person}{Li Li}.} \bibinfo{year}{2024}\natexlab{b}.
\newblock \showarticletitle{On the Reliability and Explainability of Language Models for Program Generation}.
\newblock \bibinfo{journal}{\emph{ACM Trans. Softw. Eng. Methodol.}} \bibinfo{volume}{33}, \bibinfo{number}{5}, Article \bibinfo{articleno}{126} (\bibinfo{date}{jun} \bibinfo{year}{2024}), \bibinfo{numpages}{26}~pages.
\newblock
\showISSN{1049-331X}
\urldef\tempurl%
\url{https://doi.org/10.1145/3641540}
\showDOI{\tempurl}


\bibitem[Liu et~al\mbox{.}(2024c)]%
        {liu2024trustworthyllmssurveyguideline}
\bibfield{author}{\bibinfo{person}{Yang Liu}, \bibinfo{person}{Yuanshun Yao}, \bibinfo{person}{Jean-Francois Ton}, \bibinfo{person}{Xiaoying Zhang}, \bibinfo{person}{Ruocheng Guo}, \bibinfo{person}{Hao Cheng}, \bibinfo{person}{Yegor Klochkov}, \bibinfo{person}{Muhammad~Faaiz Taufiq}, {and} \bibinfo{person}{Hang Li}.} \bibinfo{year}{2024}\natexlab{c}.
\newblock \bibinfo{title}{Trustworthy LLMs: a Survey and Guideline for Evaluating Large Language Models' Alignment}.
\newblock
\newblock
\showeprint[arxiv]{2308.05374}~[cs.AI]
\urldef\tempurl%
\url{https://arxiv.org/abs/2308.05374}
\showURL{%
\tempurl}


\bibitem[Lo(2023)]%
        {lo2023trustworthysynergisticartificialintelligence}
\bibfield{author}{\bibinfo{person}{David Lo}.} \bibinfo{year}{2023}\natexlab{}.
\newblock \bibinfo{title}{Trustworthy and Synergistic Artificial Intelligence for Software Engineering: Vision and Roadmaps}.
\newblock
\newblock
\showeprint[arxiv]{2309.04142}~[cs.SE]
\urldef\tempurl%
\url{https://arxiv.org/abs/2309.04142}
\showURL{%
\tempurl}


\bibitem[Lundberg and Lee({[n.\,d.]})]%
        {lundberg_unified_nodate}
\bibfield{author}{\bibinfo{person}{Scott~M Lundberg} {and} \bibinfo{person}{Su-In Lee}.} \bibinfo{year}{[n.\,d.]}\natexlab{}.
\newblock \showarticletitle{A {Unified} {Approach} to {Interpreting} {Model} {Predictions}}.
\newblock  (\bibinfo{year}{[n.\,d.]}).
\newblock


\bibitem[Madsen and Gregor(2000)]%
        {Madsen2000MeasuringHT}
\bibfield{author}{\bibinfo{person}{Maria Madsen} {and} \bibinfo{person}{Shirley~D Gregor}.} \bibinfo{year}{2000}\natexlab{}.
\newblock \showarticletitle{Measuring Human-Computer Trust}.
\newblock
\urldef\tempurl%
\url{https://api.semanticscholar.org/CorpusID:18821611}
\showURL{%
\tempurl}


\bibitem[Maninger et~al\mbox{.}(2024)]%
        {Maninger_2024}
\bibfield{author}{\bibinfo{person}{Daniel Maninger}, \bibinfo{person}{Krishna Narasimhan}, {and} \bibinfo{person}{Mira Mezini}.} \bibinfo{year}{2024}\natexlab{}.
\newblock \showarticletitle{Towards Trustworthy AI Software Development Assistance}. In \bibinfo{booktitle}{\emph{Proceedings of the 2024 ACM/IEEE 44th International Conference on Software Engineering: New Ideas and Emerging Results}} \emph{(\bibinfo{series}{ICSE-NIER’24})}. \bibinfo{publisher}{ACM}.
\newblock
\urldef\tempurl%
\url{https://doi.org/10.1145/3639476.3639770}
\showDOI{\tempurl}


\bibitem[Mehrotra et~al\mbox{.}(2024)]%
        {10.1145/3610578}
\bibfield{author}{\bibinfo{person}{Siddharth Mehrotra}, \bibinfo{person}{Carolina~Centeio Jorge}, \bibinfo{person}{Catholijn~M. Jonker}, {and} \bibinfo{person}{Myrthe~L. Tielman}.} \bibinfo{year}{2024}\natexlab{}.
\newblock \showarticletitle{Integrity-based Explanations for Fostering Appropriate Trust in AI Agents}.
\newblock \bibinfo{journal}{\emph{ACM Trans. Interact. Intell. Syst.}} \bibinfo{volume}{14}, \bibinfo{number}{1}, Article \bibinfo{articleno}{4} (\bibinfo{date}{Jan.} \bibinfo{year}{2024}), \bibinfo{numpages}{36}~pages.
\newblock
\showISSN{2160-6455}
\urldef\tempurl%
\url{https://doi.org/10.1145/3610578}
\showDOI{\tempurl}


\bibitem[Mohankumar et~al\mbox{.}(2020)]%
        {mohankumar_towards_2020}
\bibfield{author}{\bibinfo{person}{Akash~Kumar Mohankumar}, \bibinfo{person}{Preksha Nema}, \bibinfo{person}{Sharan Narasimhan}, \bibinfo{person}{Mitesh~M. Khapra}, \bibinfo{person}{Balaji~Vasan Srinivasan}, {and} \bibinfo{person}{Balaraman Ravindran}.} \bibinfo{year}{2020}\natexlab{}.
\newblock \showarticletitle{Towards {Transparent} and {Explainable} {Attention} {Models}}. In \bibinfo{booktitle}{\emph{Proceedings of the 58th {Annual} {Meeting} of the {Association} for {Computational} {Linguistics}}}. \bibinfo{publisher}{Association for Computational Linguistics}, \bibinfo{address}{Online}, \bibinfo{pages}{4206--4216}.
\newblock
\urldef\tempurl%
\url{https://doi.org/10.18653/v1/2020.acl-main.387}
\showDOI{\tempurl}


\bibitem[Mohsin et~al\mbox{.}(2024)]%
        {mohsin2024trustlargelanguagemodels}
\bibfield{author}{\bibinfo{person}{Ahmad Mohsin}, \bibinfo{person}{Helge Janicke}, \bibinfo{person}{Adrian Wood}, \bibinfo{person}{Iqbal~H. Sarker}, \bibinfo{person}{Leandros Maglaras}, {and} \bibinfo{person}{Naeem Janjua}.} \bibinfo{year}{2024}\natexlab{}.
\newblock \bibinfo{title}{Can We Trust Large Language Models Generated Code? A Framework for In-Context Learning, Security Patterns, and Code Evaluations Across Diverse LLMs}.
\newblock
\newblock
\showeprint[arxiv]{2406.12513}~[cs.CR]
\urldef\tempurl%
\url{https://arxiv.org/abs/2406.12513}
\showURL{%
\tempurl}


\bibitem[Nissenbaum(1996)]%
        {d16c5995650c4bc7ae5ba90b76da1d32}
\bibfield{author}{\bibinfo{person}{Helen Nissenbaum}.} \bibinfo{year}{1996}\natexlab{}.
\newblock \showarticletitle{Accountability in a computerized society}.
\newblock \bibinfo{journal}{\emph{Science and engineering ethics}} \bibinfo{volume}{2}, \bibinfo{number}{1} (\bibinfo{year}{1996}), \bibinfo{pages}{25--42}.
\newblock
\showISSN{1353-3452}
\urldef\tempurl%
\url{https://doi.org/10.1007/BF02639315}
\showDOI{\tempurl}


\bibitem[Noller et~al\mbox{.}(2022)]%
        {noller2022trustenhancementissuesprogram}
\bibfield{author}{\bibinfo{person}{Yannic Noller}, \bibinfo{person}{Ridwan Shariffdeen}, \bibinfo{person}{Xiang Gao}, {and} \bibinfo{person}{Abhik Roychoudhury}.} \bibinfo{year}{2022}\natexlab{}.
\newblock \bibinfo{title}{Trust Enhancement Issues in Program Repair}.
\newblock
\newblock
\showeprint[arxiv]{2108.13064}~[cs.SE]
\urldef\tempurl%
\url{https://arxiv.org/abs/2108.13064}
\showURL{%
\tempurl}


\bibitem[Page et~al\mbox{.}(2021)]%
        {Prisma}
\bibfield{author}{\bibinfo{person}{Matthew~J Page}, \bibinfo{person}{Joanne~E McKenzie}, \bibinfo{person}{Patrick~M Bossuyt}, \bibinfo{person}{Isabelle Boutron}, \bibinfo{person}{Tammy~C Hoffmann}, \bibinfo{person}{Cynthia~D Mulrow}, \bibinfo{person}{Larissa Shamseer}, \bibinfo{person}{Jennifer~M Tetzlaff}, \bibinfo{person}{Elie~A Akl}, \bibinfo{person}{Sue~E Brennan}, {and} \bibinfo{person}{et al.}} \bibinfo{year}{2021}\natexlab{}.
\newblock \showarticletitle{The Prisma 2020 statement: An updated guideline for reporting systematic reviews}.
\newblock \bibinfo{journal}{\emph{BMJ}} (\bibinfo{date}{Mar} \bibinfo{year}{2021}).
\newblock
\urldef\tempurl%
\url{https://doi.org/10.1136/bmj.n71}
\showDOI{\tempurl}


\bibitem[Palacio et~al\mbox{.}(2025)]%
        {palacio2025explaininglargelanguagemodels}
\bibfield{author}{\bibinfo{person}{David~N. Palacio}, \bibinfo{person}{Dipin Khati}, \bibinfo{person}{Daniel Rodriguez-Cardenas}, \bibinfo{person}{Alejandro Velasco}, {and} \bibinfo{person}{Denys Poshyvanyk}.} \bibinfo{year}{2025}\natexlab{}.
\newblock \bibinfo{title}{On Explaining (Large) Language Models For Code Using Global Code-Based Explanations}.
\newblock
\newblock
\showeprint[arxiv]{2503.16771}~[cs.SE]
\urldef\tempurl%
\url{https://arxiv.org/abs/2503.16771}
\showURL{%
\tempurl}


\bibitem[Paulus et~al\mbox{.}(2013)]%
        {10.1007/978-3-642-40779-6_23}
\bibfield{author}{\bibinfo{person}{Sachar Paulus}, \bibinfo{person}{Nazila~Gol Mohammadi}, {and} \bibinfo{person}{Thorsten Weyer}.} \bibinfo{year}{2013}\natexlab{}.
\newblock \showarticletitle{Trustworthy Software Development}. In \bibinfo{booktitle}{\emph{Communications and Multimedia Security}}, \bibfield{editor}{\bibinfo{person}{Bart De~Decker}, \bibinfo{person}{Jana Dittmann}, \bibinfo{person}{Christian Kraetzer}, {and} \bibinfo{person}{Claus Vielhauer}} (Eds.). \bibinfo{publisher}{Springer Berlin Heidelberg}, \bibinfo{address}{Berlin, Heidelberg}, \bibinfo{pages}{233--247}.
\newblock
\showISBNx{978-3-642-40779-6}


\bibitem[Pearce et~al\mbox{.}(2025)]%
        {pearce2022copilot}
\bibfield{author}{\bibinfo{person}{Hammond Pearce}, \bibinfo{person}{Baleegh Ahmad}, \bibinfo{person}{Benjamin Tan}, \bibinfo{person}{Brendan Dolan-Gavitt}, {and} \bibinfo{person}{Ramesh Karri}.} \bibinfo{year}{2025}\natexlab{}.
\newblock \showarticletitle{Asleep at the Keyboard? Assessing the Security of GitHub Copilot’s Code Contributions}.
\newblock \bibinfo{journal}{\emph{Commun. ACM}} \bibinfo{volume}{68}, \bibinfo{number}{2} (\bibinfo{date}{Jan.} \bibinfo{year}{2025}), \bibinfo{pages}{96–105}.
\newblock
\showISSN{0001-0782}
\urldef\tempurl%
\url{https://doi.org/10.1145/3610721}
\showDOI{\tempurl}


\bibitem[Peng et~al\mbox{.}(2023)]%
        {peng2023impactaideveloperproductivity}
\bibfield{author}{\bibinfo{person}{Sida Peng}, \bibinfo{person}{Eirini Kalliamvakou}, \bibinfo{person}{Peter Cihon}, {and} \bibinfo{person}{Mert Demirer}.} \bibinfo{year}{2023}\natexlab{}.
\newblock \bibinfo{title}{The Impact of AI on Developer Productivity: Evidence from GitHub Copilot}.
\newblock
\newblock
\showeprint[arxiv]{2302.06590}~[cs.SE]
\urldef\tempurl%
\url{https://arxiv.org/abs/2302.06590}
\showURL{%
\tempurl}


\bibitem[Perrig et~al\mbox{.}(2023)]%
        {10.1145/3544549.3585808}
\bibfield{author}{\bibinfo{person}{Sebastian A.~C. Perrig}, \bibinfo{person}{Nicolas Scharowski}, {and} \bibinfo{person}{Florian Br\"{u}hlmann}.} \bibinfo{year}{2023}\natexlab{}.
\newblock \showarticletitle{Trust Issues with Trust Scales: Examining the Psychometric Quality of Trust Measures in the Context of AI}. In \bibinfo{booktitle}{\emph{Extended Abstracts of the 2023 CHI Conference on Human Factors in Computing Systems}} (Hamburg, Germany) \emph{(\bibinfo{series}{CHI EA '23})}. \bibinfo{publisher}{Association for Computing Machinery}, \bibinfo{address}{New York, NY, USA}, Article \bibinfo{articleno}{297}, \bibinfo{numpages}{7}~pages.
\newblock
\showISBNx{9781450394222}
\urldef\tempurl%
\url{https://doi.org/10.1145/3544549.3585808}
\showDOI{\tempurl}


\bibitem[Perry et~al\mbox{.}(2023)]%
        {Perry_2023}
\bibfield{author}{\bibinfo{person}{Neil Perry}, \bibinfo{person}{Megha Srivastava}, \bibinfo{person}{Deepak Kumar}, {and} \bibinfo{person}{Dan Boneh}.} \bibinfo{year}{2023}\natexlab{}.
\newblock \showarticletitle{Do Users Write More Insecure Code with AI Assistants?}. In \bibinfo{booktitle}{\emph{Proceedings of the 2023 ACM SIGSAC Conference on Computer and Communications Security}} \emph{(\bibinfo{series}{CCS ’23})}. \bibinfo{publisher}{ACM}.
\newblock
\urldef\tempurl%
\url{https://doi.org/10.1145/3576915.3623157}
\showDOI{\tempurl}


\bibitem[Rabani et~al\mbox{.}(2024)]%
        {10.1007/978-3-031-55486-5_16}
\bibfield{author}{\bibinfo{person}{Zeinab~Sadat Rabani}, \bibinfo{person}{Hanieh Khorashadizadeh}, \bibinfo{person}{Shirin Abdollahzade}, \bibinfo{person}{Sven Groppe}, {and} \bibinfo{person}{Javad Ghofrani}.} \bibinfo{year}{2024}\natexlab{}.
\newblock \showarticletitle{Developers' Perspective on Trustworthiness of Code Generated by ChatGPT: Insights from Interviews}. In \bibinfo{booktitle}{\emph{Applied Machine Learning and Data Analytics}}, \bibfield{editor}{\bibinfo{person}{M.~A. Jabbar}, \bibinfo{person}{Sanju Tiwari}, \bibinfo{person}{Fernando Ortiz-Rodr{\'i}guez}, \bibinfo{person}{Sven Groppe}, {and} \bibinfo{person}{Tasneem Bano~Rehman}} (Eds.). \bibinfo{publisher}{Springer Nature Switzerland}, \bibinfo{address}{Cham}, \bibinfo{pages}{215--229}.
\newblock
\showISBNx{978-3-031-55486-5}


\bibitem[Roychoudhury et~al\mbox{.}(2025)]%
        {roychoudhury2025aisoftwareengineerprogramming}
\bibfield{author}{\bibinfo{person}{Abhik Roychoudhury}, \bibinfo{person}{Corina Pasareanu}, \bibinfo{person}{Michael Pradel}, {and} \bibinfo{person}{Baishakhi Ray}.} \bibinfo{year}{2025}\natexlab{}.
\newblock \bibinfo{title}{AI Software Engineer: Programming with Trust}.
\newblock
\newblock
\showeprint[arxiv]{2502.13767}~[cs.SE]
\urldef\tempurl%
\url{https://arxiv.org/abs/2502.13767}
\showURL{%
\tempurl}


\bibitem[Russo(2024)]%
        {russo2024navigatingcomplexitygenerativeai}
\bibfield{author}{\bibinfo{person}{Daniel Russo}.} \bibinfo{year}{2024}\natexlab{}.
\newblock \bibinfo{title}{Navigating the Complexity of Generative AI Adoption in Software Engineering}.
\newblock
\newblock
\showeprint[arxiv]{2307.06081}~[cs.SE]
\urldef\tempurl%
\url{https://arxiv.org/abs/2307.06081}
\showURL{%
\tempurl}


\bibitem[Rutinowski et~al\mbox{.}(2024)]%
        {Rutinowski2024BenchmarkingTA}
\bibfield{author}{\bibinfo{person}{J{\'e}r{\^o}me Rutinowski}, \bibinfo{person}{Simon Kl{\"u}ttermann}, \bibinfo{person}{Jan Endendyk}, \bibinfo{person}{Christopher Reining}, {and} \bibinfo{person}{Emmanuel M{\"u}ller}.} \bibinfo{year}{2024}\natexlab{}.
\newblock \showarticletitle{Benchmarking Trust: A Metric for Trustworthy Machine Learning}. In \bibinfo{booktitle}{\emph{xAI}}.
\newblock
\urldef\tempurl%
\url{https://api.semanticscholar.org/CorpusID:273496081}
\showURL{%
\tempurl}


\bibitem[Sandoval et~al\mbox{.}(2023)]%
        {sandoval2024security}
\bibfield{author}{\bibinfo{person}{Gustavo Sandoval}, \bibinfo{person}{Hammond Pearce}, \bibinfo{person}{Teo Nys}, \bibinfo{person}{Ramesh Karri}, \bibinfo{person}{Siddharth Garg}, {and} \bibinfo{person}{Brendan Dolan-Gavitt}.} \bibinfo{year}{2023}\natexlab{}.
\newblock \showarticletitle{Lost at C: A User Study on the Security Implications of Large Language Model Code Assistants}. In \bibinfo{booktitle}{\emph{32nd USENIX Security Symposium (USENIX Security 23)}}. \bibinfo{publisher}{USENIX Association}, \bibinfo{address}{Anaheim, CA}, \bibinfo{pages}{2205--2222}.
\newblock
\showISBNx{978-1-939133-37-3}
\urldef\tempurl%
\url{https://www.usenix.org/conference/usenixsecurity23/presentation/sandoval}
\showURL{%
\tempurl}


\bibitem[Serban et~al\mbox{.}(2021)]%
        {9474373}
\bibfield{author}{\bibinfo{person}{Alex Serban}, \bibinfo{person}{Koen van~der Blom}, \bibinfo{person}{Holger Hoos}, {and} \bibinfo{person}{Joost Visser}.} \bibinfo{year}{2021}\natexlab{}.
\newblock \showarticletitle{Practices for Engineering Trustworthy Machine Learning Applications}. In \bibinfo{booktitle}{\emph{2021 IEEE/ACM 1st Workshop on AI Engineering - Software Engineering for AI (WAIN)}}. \bibinfo{pages}{97--100}.
\newblock
\urldef\tempurl%
\url{https://doi.org/10.1109/WAIN52551.2021.00021}
\showDOI{\tempurl}


\bibitem[Sharma et~al\mbox{.}(2024)]%
        {sharma2024suggestthathumantrust}
\bibfield{author}{\bibinfo{person}{Manasi Sharma}, \bibinfo{person}{Ho~Chit Siu}, \bibinfo{person}{Rohan Paleja}, {and} \bibinfo{person}{Jaime~D. Peña}.} \bibinfo{year}{2024}\natexlab{}.
\newblock \bibinfo{title}{Why Would You Suggest That? Human Trust in Language Model Responses}.
\newblock
\newblock
\showeprint[arxiv]{2406.02018}~[cs.CL]
\urldef\tempurl%
\url{https://arxiv.org/abs/2406.02018}
\showURL{%
\tempurl}


\bibitem[Spiess et~al\mbox{.}(2024)]%
        {spiess2024calibrationcorrectnesslanguagemodels}
\bibfield{author}{\bibinfo{person}{Claudio Spiess}, \bibinfo{person}{David Gros}, \bibinfo{person}{Kunal~Suresh Pai}, \bibinfo{person}{Michael Pradel}, \bibinfo{person}{Md~Rafiqul~Islam Rabin}, \bibinfo{person}{Amin Alipour}, \bibinfo{person}{Susmit Jha}, \bibinfo{person}{Prem Devanbu}, {and} \bibinfo{person}{Toufique Ahmed}.} \bibinfo{year}{2024}\natexlab{}.
\newblock \bibinfo{title}{Calibration and Correctness of Language Models for Code}.
\newblock
\newblock
\showeprint[arxiv]{2402.02047}~[cs.SE]
\urldef\tempurl%
\url{https://arxiv.org/abs/2402.02047}
\showURL{%
\tempurl}


\bibitem[Stettinger et~al\mbox{.}(2024)]%
        {10430152}
\bibfield{author}{\bibinfo{person}{Georg Stettinger}, \bibinfo{person}{Patrick Weissensteiner}, {and} \bibinfo{person}{Siddartha Khastgir}.} \bibinfo{year}{2024}\natexlab{}.
\newblock \showarticletitle{Trustworthiness Assurance Assessment for High-Risk AI-Based Systems}.
\newblock \bibinfo{journal}{\emph{IEEE Access}}  \bibinfo{volume}{12} (\bibinfo{year}{2024}), \bibinfo{pages}{22718--22745}.
\newblock
\urldef\tempurl%
\url{https://doi.org/10.1109/ACCESS.2024.3364387}
\showDOI{\tempurl}


\bibitem[Thorne(2024)]%
        {Thorne_2024}
\bibfield{author}{\bibinfo{person}{Simon Thorne}.} \bibinfo{year}{2024}\natexlab{}.
\newblock \showarticletitle{Understanding the interplay between trust, reliability, and human factors in the age of Generative AI}.
\newblock \bibinfo{journal}{\emph{International journal of simulation: systems, science \&amp; technology}} (\bibinfo{date}{May} \bibinfo{year}{2024}).
\newblock
\urldef\tempurl%
\url{https://doi.org/10.5013/ijssst.a.25.01.10}
\showDOI{\tempurl}


\bibitem[Tufano et~al\mbox{.}(2018)]%
        {bug_fixing_tufano}
\bibfield{author}{\bibinfo{person}{Michele Tufano}, \bibinfo{person}{Cody Watson}, \bibinfo{person}{Gabriele Bavota}, \bibinfo{person}{Massimiliano Di~Penta}, \bibinfo{person}{Martin White}, {and} \bibinfo{person}{Denys Poshyvanyk}.} \bibinfo{year}{2018}\natexlab{}.
\newblock \showarticletitle{An empirical investigation into learning bug-fixing patches in the wild via neural machine translation}. In \bibinfo{booktitle}{\emph{Proceedings of the 33rd ACM/IEEE International Conference on Automated Software Engineering}} (Montpellier, France) \emph{(\bibinfo{series}{ASE '18})}. \bibinfo{publisher}{Association for Computing Machinery}, \bibinfo{address}{New York, NY, USA}, \bibinfo{pages}{832–837}.
\newblock
\showISBNx{9781450359375}
\urldef\tempurl%
\url{https://doi.org/10.1145/3238147.3240732}
\showDOI{\tempurl}


\bibitem[Turner et~al\mbox{.}(2024)]%
        {Turner2024}
\bibfield{author}{\bibinfo{person}{Amy Turner}, \bibinfo{person}{Meena Kaushik}, \bibinfo{person}{Mu-Ti Huang}, {and} \bibinfo{person}{Srikar Varanasi}.} \bibinfo{year}{2024}\natexlab{}.
\newblock \bibinfo{booktitle}{\emph{Calibrating Trust in AI-Assisted Decision Making}}.
\newblock \bibinfo{type}{{T}echnical {R}eport}. \bibinfo{institution}{UC Berkeley School of Information}.
\newblock
\urldef\tempurl%
\url{https://www.ischool.berkeley.edu/sites/default/files/sproject_attachments/humanai_capstonereport-final.pdf}
\showURL{%
\tempurl}


\bibitem[Vereschak et~al\mbox{.}(2021)]%
        {10.1145/3476068}
\bibfield{author}{\bibinfo{person}{Oleksandra Vereschak}, \bibinfo{person}{Gilles Bailly}, {and} \bibinfo{person}{Baptiste Caramiaux}.} \bibinfo{year}{2021}\natexlab{}.
\newblock \showarticletitle{How to Evaluate Trust in AI-Assisted Decision Making? A Survey of Empirical Methodologies}.
\newblock \bibinfo{journal}{\emph{Proc. ACM Hum.-Comput. Interact.}} \bibinfo{volume}{5}, \bibinfo{number}{CSCW2}, Article \bibinfo{articleno}{327} (\bibinfo{date}{oct} \bibinfo{year}{2021}), \bibinfo{numpages}{39}~pages.
\newblock
\urldef\tempurl%
\url{https://doi.org/10.1145/3476068}
\showDOI{\tempurl}


\bibitem[Vodrahalli et~al\mbox{.}(2022)]%
        {10.1145/3514094.3534150}
\bibfield{author}{\bibinfo{person}{Kailas Vodrahalli}, \bibinfo{person}{Roxana Daneshjou}, \bibinfo{person}{Tobias Gerstenberg}, {and} \bibinfo{person}{James Zou}.} \bibinfo{year}{2022}\natexlab{}.
\newblock \showarticletitle{Do Humans Trust Advice More if it Comes from AI? An Analysis of Human-AI Interactions}. In \bibinfo{booktitle}{\emph{Proceedings of the 2022 AAAI/ACM Conference on AI, Ethics, and Society}} (Oxford, United Kingdom) \emph{(\bibinfo{series}{AIES '22})}. \bibinfo{publisher}{Association for Computing Machinery}, \bibinfo{address}{New York, NY, USA}, \bibinfo{pages}{763–777}.
\newblock
\showISBNx{9781450392471}
\urldef\tempurl%
\url{https://doi.org/10.1145/3514094.3534150}
\showDOI{\tempurl}


\bibitem[Wang et~al\mbox{.}(2024b)]%
        {wang2024decodingtrustcomprehensiveassessmenttrustworthiness}
\bibfield{author}{\bibinfo{person}{Boxin Wang}, \bibinfo{person}{Weixin Chen}, \bibinfo{person}{Hengzhi Pei}, \bibinfo{person}{Chulin Xie}, \bibinfo{person}{Mintong Kang}, \bibinfo{person}{Chenhui Zhang}, \bibinfo{person}{Chejian Xu}, \bibinfo{person}{Zidi Xiong}, \bibinfo{person}{Ritik Dutta}, \bibinfo{person}{Rylan Schaeffer}, \bibinfo{person}{Sang~T. Truong}, \bibinfo{person}{Simran Arora}, \bibinfo{person}{Mantas Mazeika}, \bibinfo{person}{Dan Hendrycks}, \bibinfo{person}{Zinan Lin}, \bibinfo{person}{Yu Cheng}, \bibinfo{person}{Sanmi Koyejo}, \bibinfo{person}{Dawn Song}, {and} \bibinfo{person}{Bo Li}.} \bibinfo{year}{2024}\natexlab{b}.
\newblock \bibinfo{title}{DecodingTrust: A Comprehensive Assessment of Trustworthiness in GPT Models}.
\newblock
\newblock
\showeprint[arxiv]{2306.11698}~[cs.CL]
\urldef\tempurl%
\url{https://arxiv.org/abs/2306.11698}
\showURL{%
\tempurl}


\bibitem[Wang et~al\mbox{.}(2022)]%
        {wang2022adversarialgluemultitaskbenchmark}
\bibfield{author}{\bibinfo{person}{Boxin Wang}, \bibinfo{person}{Chejian Xu}, \bibinfo{person}{Shuohang Wang}, \bibinfo{person}{Zhe Gan}, \bibinfo{person}{Yu Cheng}, \bibinfo{person}{Jianfeng Gao}, \bibinfo{person}{Ahmed~Hassan Awadallah}, {and} \bibinfo{person}{Bo Li}.} \bibinfo{year}{2022}\natexlab{}.
\newblock \bibinfo{title}{Adversarial GLUE: A Multi-Task Benchmark for Robustness Evaluation of Language Models}.
\newblock
\newblock
\showeprint[arxiv]{2111.02840}~[cs.CL]
\urldef\tempurl%
\url{https://arxiv.org/abs/2111.02840}
\showURL{%
\tempurl}


\bibitem[Wang et~al\mbox{.}(2024a)]%
        {wang2024trustworthyllmscodedatacentric}
\bibfield{author}{\bibinfo{person}{Chong Wang}, \bibinfo{person}{Zhenpeng Chen}, \bibinfo{person}{Tianlin Li}, \bibinfo{person}{Yilun Zhao}, {and} \bibinfo{person}{Yang Liu}.} \bibinfo{year}{2024}\natexlab{a}.
\newblock \bibinfo{title}{Towards Trustworthy LLMs for Code: A Data-Centric Synergistic Auditing Framework}.
\newblock
\newblock
\showeprint[arxiv]{2410.09048}~[cs.SE]
\urldef\tempurl%
\url{https://arxiv.org/abs/2410.09048}
\showURL{%
\tempurl}


\bibitem[Wang et~al\mbox{.}(2024c)]%
        {10.1145/3630106.3658984}
\bibfield{author}{\bibinfo{person}{Ruotong Wang}, \bibinfo{person}{Ruijia Cheng}, \bibinfo{person}{Denae Ford}, {and} \bibinfo{person}{Thomas Zimmermann}.} \bibinfo{year}{2024}\natexlab{c}.
\newblock \showarticletitle{Investigating and Designing for Trust in AI-powered Code Generation Tools}. In \bibinfo{booktitle}{\emph{Proceedings of the 2024 ACM Conference on Fairness, Accountability, and Transparency}} (Rio de Janeiro, Brazil) \emph{(\bibinfo{series}{FAccT '24})}. \bibinfo{publisher}{Association for Computing Machinery}, \bibinfo{address}{New York, NY, USA}, \bibinfo{pages}{1475–1493}.
\newblock
\showISBNx{9798400704505}
\urldef\tempurl%
\url{https://doi.org/10.1145/3630106.3658984}
\showDOI{\tempurl}


\bibitem[Wang et~al\mbox{.}(2016)]%
        {defect_prediction}
\bibfield{author}{\bibinfo{person}{Song Wang}, \bibinfo{person}{Taiyue Liu}, {and} \bibinfo{person}{Lin Tan}.} \bibinfo{year}{2016}\natexlab{}.
\newblock \showarticletitle{Automatically learning semantic features for defect prediction}. In \bibinfo{booktitle}{\emph{Proceedings of the 38th International Conference on Software Engineering}} (Austin, Texas) \emph{(\bibinfo{series}{ICSE '16})}. \bibinfo{publisher}{Association for Computing Machinery}, \bibinfo{address}{New York, NY, USA}, \bibinfo{pages}{297–308}.
\newblock
\showISBNx{9781450339001}
\urldef\tempurl%
\url{https://doi.org/10.1145/2884781.2884804}
\showDOI{\tempurl}


\bibitem[Watson et~al\mbox{.}(2021)]%
        {watson2021systematic}
\bibfield{author}{\bibinfo{person}{Cody Watson}, \bibinfo{person}{Nathan Cooper}, \bibinfo{person}{David~Nader Palacio}, \bibinfo{person}{Kevin Moran}, {and} \bibinfo{person}{Denys Poshyvanyk}.} \bibinfo{year}{2021}\natexlab{}.
\newblock \bibinfo{title}{A Systematic Literature Review on the Use of Deep Learning in Software Engineering Research}.
\newblock
\newblock
\showeprint[arxiv]{2009.06520}~[cs.SE]


\bibitem[Weisz et~al\mbox{.}(2021)]%
        {10.1145/3397481.3450656}
\bibfield{author}{\bibinfo{person}{Justin~D. Weisz}, \bibinfo{person}{Michael Muller}, \bibinfo{person}{Stephanie Houde}, \bibinfo{person}{John Richards}, \bibinfo{person}{Steven~I. Ross}, \bibinfo{person}{Fernando Martinez}, \bibinfo{person}{Mayank Agarwal}, {and} \bibinfo{person}{Kartik Talamadupula}.} \bibinfo{year}{2021}\natexlab{}.
\newblock \showarticletitle{Perfection Not Required? Human-AI Partnerships in Code Translation}. In \bibinfo{booktitle}{\emph{Proceedings of the 26th International Conference on Intelligent User Interfaces}} (College Station, TX, USA) \emph{(\bibinfo{series}{IUI '21})}. \bibinfo{publisher}{Association for Computing Machinery}, \bibinfo{address}{New York, NY, USA}, \bibinfo{pages}{402–412}.
\newblock
\showISBNx{9781450380171}
\urldef\tempurl%
\url{https://doi.org/10.1145/3397481.3450656}
\showDOI{\tempurl}


\bibitem[Widder et~al\mbox{.}(2021)]%
        {10.1145/3411764.3445650}
\bibfield{author}{\bibinfo{person}{David~Gray Widder}, \bibinfo{person}{Laura Dabbish}, \bibinfo{person}{James~D. Herbsleb}, \bibinfo{person}{Alexandra Holloway}, {and} \bibinfo{person}{Scott Davidoff}.} \bibinfo{year}{2021}\natexlab{}.
\newblock \showarticletitle{Trust in Collaborative Automation in High Stakes Software Engineering Work: A Case Study at NASA}. In \bibinfo{booktitle}{\emph{Proceedings of the 2021 CHI Conference on Human Factors in Computing Systems}} (Yokohama, Japan) \emph{(\bibinfo{series}{CHI '21})}. \bibinfo{publisher}{Association for Computing Machinery}, \bibinfo{address}{New York, NY, USA}, Article \bibinfo{articleno}{184}, \bibinfo{numpages}{13}~pages.
\newblock
\showISBNx{9781450380966}
\urldef\tempurl%
\url{https://doi.org/10.1145/3411764.3445650}
\showDOI{\tempurl}


\bibitem[Wu et~al\mbox{.}(2022)]%
        {wu2022aichainstransparentcontrollable}
\bibfield{author}{\bibinfo{person}{Tongshuang Wu}, \bibinfo{person}{Michael Terry}, {and} \bibinfo{person}{Carrie~J. Cai}.} \bibinfo{year}{2022}\natexlab{}.
\newblock \bibinfo{title}{AI Chains: Transparent and Controllable Human-AI Interaction by Chaining Large Language Model Prompts}.
\newblock
\newblock
\showeprint[arxiv]{2110.01691}~[cs.HC]
\urldef\tempurl%
\url{https://arxiv.org/abs/2110.01691}
\showURL{%
\tempurl}


\bibitem[Wu et~al\mbox{.}(2023)]%
        {10494274}
\bibfield{author}{\bibinfo{person}{Wennan Wu}, \bibinfo{person}{Ruisi Liu}, {and} \bibinfo{person}{Junjie Chu}.} \bibinfo{year}{2023}\natexlab{}.
\newblock \showarticletitle{How important is Trust: Exploring the Factors Influencing College Students' Use of Chat GPT as a Learning Aid}. In \bibinfo{booktitle}{\emph{2023 16th International Symposium on Computational Intelligence and Design (ISCID)}}. \bibinfo{pages}{67--70}.
\newblock
\urldef\tempurl%
\url{https://doi.org/10.1109/ISCID59865.2023.00024}
\showDOI{\tempurl}


\bibitem[Ye et~al\mbox{.}(2023)]%
        {ye2023assessinghiddenrisksllms}
\bibfield{author}{\bibinfo{person}{Wentao Ye}, \bibinfo{person}{Mingfeng Ou}, \bibinfo{person}{Tianyi Li}, \bibinfo{person}{Yipeng chen}, \bibinfo{person}{Xuetao Ma}, \bibinfo{person}{Yifan Yanggong}, \bibinfo{person}{Sai Wu}, \bibinfo{person}{Jie Fu}, \bibinfo{person}{Gang Chen}, \bibinfo{person}{Haobo Wang}, {and} \bibinfo{person}{Junbo Zhao}.} \bibinfo{year}{2023}\natexlab{}.
\newblock \bibinfo{title}{Assessing Hidden Risks of LLMs: An Empirical Study on Robustness, Consistency, and Credibility}.
\newblock
\newblock
\showeprint[arxiv]{2305.10235}~[cs.LG]
\urldef\tempurl%
\url{https://arxiv.org/abs/2305.10235}
\showURL{%
\tempurl}


\bibitem[Yin et~al\mbox{.}(2023)]%
        {yin2023largelanguagemodelsknow}
\bibfield{author}{\bibinfo{person}{Zhangyue Yin}, \bibinfo{person}{Qiushi Sun}, \bibinfo{person}{Qipeng Guo}, \bibinfo{person}{Jiawen Wu}, \bibinfo{person}{Xipeng Qiu}, {and} \bibinfo{person}{Xuanjing Huang}.} \bibinfo{year}{2023}\natexlab{}.
\newblock \bibinfo{title}{Do Large Language Models Know What They Don't Know?}
\newblock
\newblock
\showeprint[arxiv]{2305.18153}~[cs.CL]
\urldef\tempurl%
\url{https://arxiv.org/abs/2305.18153}
\showURL{%
\tempurl}


\bibitem[Zerick et~al\mbox{.}(2024)]%
        {10605304}
\bibfield{author}{\bibinfo{person}{Juliette Zerick}, \bibinfo{person}{Zachary Kaufman}, \bibinfo{person}{Jonathan Ott}, \bibinfo{person}{Janki Kuber}, \bibinfo{person}{Ember Chow}, \bibinfo{person}{Shyama Shah}, {and} \bibinfo{person}{Gregory Lewis}.} \bibinfo{year}{2024}\natexlab{}.
\newblock \showarticletitle{It Takes Two to Trust: Mediating Human-AI Trust for Resilience and Reliability}. In \bibinfo{booktitle}{\emph{2024 IEEE Conference on Artificial Intelligence (CAI)}}. \bibinfo{pages}{755--761}.
\newblock
\urldef\tempurl%
\url{https://doi.org/10.1109/CAI59869.2024.00145}
\showDOI{\tempurl}


\bibitem[Zerilli et~al\mbox{.}(2022)]%
        {whytransparancy}
\bibfield{author}{\bibinfo{person}{John Zerilli}, \bibinfo{person}{Umang Bhatt}, {and} \bibinfo{person}{Adrian Weller}.} \bibinfo{year}{2022}\natexlab{}.
\newblock \showarticletitle{How transparency modulates trust in artificial intelligence}.
\newblock \bibinfo{journal}{\emph{Patterns}} \bibinfo{volume}{3}, \bibinfo{number}{4} (\bibinfo{year}{2022}), \bibinfo{pages}{100455}.
\newblock
\showISSN{2666-3899}
\urldef\tempurl%
\url{https://doi.org/10.1016/j.patter.2022.100455}
\showDOI{\tempurl}


\bibitem[Zhang et~al\mbox{.}(2024)]%
        {10.1145/3614424}
\bibfield{author}{\bibinfo{person}{Peiyun Zhang}, \bibinfo{person}{Song Ding}, {and} \bibinfo{person}{Qinglin Zhao}.} \bibinfo{year}{2024}\natexlab{}.
\newblock \showarticletitle{Exploiting Blockchain to Make AI Trustworthy: A Software Development Lifecycle View}.
\newblock \bibinfo{journal}{\emph{ACM Comput. Surv.}} \bibinfo{volume}{56}, \bibinfo{number}{7}, Article \bibinfo{articleno}{163} (\bibinfo{date}{April} \bibinfo{year}{2024}), \bibinfo{numpages}{31}~pages.
\newblock
\showISSN{0360-0300}
\urldef\tempurl%
\url{https://doi.org/10.1145/3614424}
\showDOI{\tempurl}


\bibitem[Zhang et~al\mbox{.}(2023)]%
        {zhang2023donttrustchatgptquestion}
\bibfield{author}{\bibinfo{person}{Xiang Zhang}, \bibinfo{person}{Senyu Li}, \bibinfo{person}{Bradley Hauer}, \bibinfo{person}{Ning Shi}, {and} \bibinfo{person}{Grzegorz Kondrak}.} \bibinfo{year}{2023}\natexlab{}.
\newblock \bibinfo{title}{Don't Trust ChatGPT when Your Question is not in English: A Study of Multilingual Abilities and Types of LLMs}.
\newblock
\newblock
\showeprint[arxiv]{2305.16339}~[cs.CL]
\urldef\tempurl%
\url{https://arxiv.org/abs/2305.16339}
\showURL{%
\tempurl}


\bibitem[Zhu et~al\mbox{.}(2024)]%
        {zhu2024promptrobustevaluatingrobustnesslarge}
\bibfield{author}{\bibinfo{person}{Kaijie Zhu}, \bibinfo{person}{Jindong Wang}, \bibinfo{person}{Jiaheng Zhou}, \bibinfo{person}{Zichen Wang}, \bibinfo{person}{Hao Chen}, \bibinfo{person}{Yidong Wang}, \bibinfo{person}{Linyi Yang}, \bibinfo{person}{Wei Ye}, \bibinfo{person}{Yue Zhang}, \bibinfo{person}{Neil~Zhenqiang Gong}, {and} \bibinfo{person}{Xing Xie}.} \bibinfo{year}{2024}\natexlab{}.
\newblock \bibinfo{title}{PromptRobust: Towards Evaluating the Robustness of Large Language Models on Adversarial Prompts}.
\newblock
\newblock
\showeprint[arxiv]{2306.04528}~[cs.CL]
\urldef\tempurl%
\url{https://arxiv.org/abs/2306.04528}
\showURL{%
\tempurl}


\end{thebibliography}
